\documentclass[12pt]{article}
\usepackage{%epsf,
amsfonts,amssymb,epsfig,amsmath}
\usepackage{color}
\usepackage{hyperref,hep}
\usepackage{cancel}

\renewcommand{\text}[1]{#1}

\newcommand{\be}{\begin{equation}}
\newcommand{\ee}{\end{equation}}
\newcommand{\ben}{\begin{displaymath}}
\newcommand{\een}{\end{displaymath}}
\newcommand{\bea}{\begin{eqnarray}}
\newcommand{\eea}{\end{eqnarray}}
\newcommand{\bean}{\begin{eqnarray*}}
\newcommand{\eean}{\end{eqnarray*}}
\newcommand{\nn}{\nonumber \\}
\newcommand{\ba}{\begin{array}}
\newcommand{\ea}{\end{array}}
\newcommand{\bi}{\begin{itemize}}
\newcommand{\ei}{\end{itemize}}

\newcommand{\reef}[1]{(\ref{#1})}

\def\dblone{\hbox{$1\hskip -1.2pt\vrule depth 0pt height 1.6ex width 0.7pt
                  \vrule depth 0pt height 0.3pt width 0.12em$}}

\newcommand{\bbR}{{\mathbb{R}}}
\newcommand{\bbZ}{{\mathbb{Z}}}

\DeclareMathOperator{\im}{Im}

\DeclareMathOperator{\diag}{diag}

\newcommand{\dd}{\mathrm{d}}

\newcommand{\id}{{1}}

\begin{document}

\makeatletter
\renewcommand{\theequation}{\thesection.\arabic{equation}}
\@addtoreset{equation}{section}
\makeatother

\baselineskip 18pt

\begin{titlepage}

\vfill

\begin{flushright}
Imperial/TP/2011/JG/04\\
\end{flushright}

\vfill

\begin{center}
   \baselineskip=16pt
   \begin{Large}\textbf{
        Spectral function of the\\*[5pt] supersymmetry current}
        %\\*[5pt] and AdS/CMT}
   \end{Large}
   \vskip 1.5cm
Jerome P. Gauntlett$^1$,  Julian Sonner$^{1,2}$ and Daniel Waldram$^{1}$ \\
   \vskip .6cm
     \begin{small}
  \textit{$^1$Blackett Laboratory,
        Imperial College\\ London, SW7 2AZ, U.K.}
        %E-mail: j.gauntlett, d.waldram@imperial.ac.uk}
        \end{small}\\*[.4cm]
      %  \\*[.4cm]½
    \begin{small}
    \textit{$^2$D.A.M.T.P.,
        University of Cambridge\\ Cambridge, CB3 0WA, U.K.}
        %E-mail: j.gauntlett, d.waldram@imperial.ac.uk}
        \end{small}\\*[.4cm]
      %  \\*[.4cm]
 %      \begin{small}
   % \textit{$^3$Trinity College,
     %   University of Cambridge\\ Cambridge, CB2 1TQ, U.K.}
        %E-mail: j.gauntlett, d.waldram@imperial.ac.uk}
       % \end{small}\\*[.4cm]
      %  \\*[.4cm]

   \end{center}

\vfill

\begin{center}
\textbf{Abstract}
\end{center}

\begin{quote}
We continue our study of the retarded Green's function of the universal fermionic supersymmetry current
(``supercurrent")
for the most general class of $d=3$ $N=2$ SCFTs with $D=10$ or $D=11$ supergravity duals by
studying the 
propagation of the Dirac gravitino in the electrically charged AdS-Reissner-Nordstr\"om black-brane
background of $N=2$ minimal gauged supergravity in $D=4$. We expand upon results presented
in a companion paper, including the absence of a Fermi surface and the appearance of a soft power-law gap
at zero temperature. We also present the analytic solution of the gravitino equation in the $AdS_2\times \mathbb{R}^2$
background which arises as the near-horizon limit at zero temperature. In addition we determine the quasinormal mode
spectrum.

\end{quote}

\vfill

\end{titlepage}

\setcounter{equation}{0}

%%%%%%%%%%%%%%%%%%%%%%%%%%%%%%%%%%%%%%%%%%%%%%%%%%%%%%%%%%%%%%%%%%%%%%%
\tableofcontents
%%%%%%%%%%%%%%%%%%%%%%%%%%%%%

%%%%%%%%%%%%%%%%%%%%%%%%%%%%%%%%%%%%%%%%%%%%%%%%%%%%%%%%%%%%%%%%%%%%%%%
\section{Introduction}

All $N=2$ SCFTs in $d=3$ dimensions have a universal current supermultiplet consisting of
the energy-momentum tensor, the abelian global $R$-symmetry current, and the fermionic
supersymmetry current or ``supercurrent". We will be interested in the general class of such SCFTs 
that have weakly coupled supergravity duals in either $D=10$ or $D=11$ dimensions.
We will carry out a comprehensive investigation of the Green's function of the supercurrent for this universal class when held at finite temperature and
non-zero chemical potential with respect to the global $R$-symmetry.

A significant motivation for this paper, which technically clarifies and conceptually extends the companion paper 
\cite{Gauntlett:2011mf}, is the extensive work on fermion spectral functions that has been 
carried out using gauge-gravity duality 
starting with \cite{Lee:2008xf,Liu:2009dm,Cubrovic:2009ye,Faulkner:2009wj}
and then further developed in
\cite{Iqbal:2009fd,Albash:2009wz,Basu:2009qz,Denef:2009yy,Kachru:2009xf,Hung:2009qk,Maity:2009zz,Chen:2009pt,Gubser:2009qt,Faulkner:2009am,Gubser:2009dt,Hartnoll:2009kk,
Hartnoll:2009ns, Albash:2010yr,Faulkner:2010tq,Gubser:2010dm,Albash:2010dr,Ammon:2010pg,Faulkner:2010da,Hartman:2010fk,Mross:2010rd,Cai:2010tr,Benini:2010qc,Larsen:2010jt,Hung:2010te,Sachdev:2010um,Gubankova:2010ny,Vegh:2010fc,Benini:2010pr,Hartnoll:2010gu,Kachru:2010dk,Yamamoto:2010yp,Edalati:2010ww,Gulotta:2010cu,Arsiwalla:2010bt,Gubankova:2010rc,Parente:2010fs,Edalati:2010ge,Cubrovic:2010bf,Faulkner:2011tm,Guarrera:2011my,Wu:2011bx,Huijse:2011hp,Iizuka:2011hg,Jensen:2011su,Hartnoll:2011dm,Iqbal:2011in,Cubrovic:2011xm,Edalati:2011yv,Rangamani:2011ae,Hoyos:2011us}.  Remarkably, all of these works have, essentially, been in a bottom-up context, 
with the exception of \cite{Ammon:2010pg} which studied the spectral functions in the context of probe-brane $p$-wave superconductors. 
It is therefore of considerable interest to carry out more detailed top-down analyses to see which features can actually 
be realised in a string/M-theory setting. An appealing feature of the analysis that we carry out is that it is applicable
to a universal class of SCFTs. Note that after the appearance of \cite{Gauntlett:2011mf} and as this paper was being written up, 
a related study appeared in \cite{Belliard:2011qq}.

Consider, then, the most general $AdS_4\times M_6$ and $AdS_4\times M_7$ solutions of
$D=10$ and $D=11$ supergravity, respectively, that are dual to $N=2$ SCFTs in $d=3$.
There is very strong evidence that there is always a consistent 
KK truncation of the $D=10$ or $D=11$ supergravity theory on $M_6$ or $M_7$, respectively,
to minimal $N=2$ gauged supergravity in $D=4$ \cite{Gauntlett:2007ma}. This means,
by definition,
that any solution
of the latter theory can be uplifted on any of the $M_6$ or $M_7$ to obtain an exact solution
of $D=10$ or $D=11$ supergravity, respectively. Indeed for the infinite class of $D=11$ solutions when 
$M_7$ is a seven-dimensional Sasaki-Einstein manifold, $SE_7$,
and also for another general class of $D=11$ solutions with magnetic four-form flux,
this was proven at the level of the bosonic fields in \cite{Gauntlett:2007ma} and, furthermore, consistency with
the supersymmetry variations was shown. For the $SE_7$
case it has also been shown that the truncation is consistent in the fermionic fields, at the quadratic level, in
\cite{gswunpub,Bah:2010yt}.

The bosonic fields of minimal $N=2$ $D=4$ gauged supergravity consist of the metric and an abelian gauge field, which are
dual to the energy-momentum tensor and the R-symmetry current in the $N=2$ $d=3$ SCFT, respectively. 
The $D=4$ supergravity also has
two Majorana gravitini, or equivalently a single Dirac gravitino, which is dual to the supercurrent in the SCFT.
An important feature of minimal $N=2$ $D=4$ gauged supergravity is that the bosonic Lagrangian is simply 
Einstein-Maxwell theory with a negative cosmological constant. In particular, 
the electrically charged AdS-Reissner-Nordstr\"om
(AdS-RN) black-brane solution solves the equations of motion. Thus, the results of \cite{Gauntlett:2007ma}
imply that this solution is relevant as a dual description for {\it all}\footnote{Of course the electrically charged
AdS-RN black-brane can also be relevant 
for certain non-supersymmetric CFTs. A prominent example is the ``skew-whiffed"
$AdS_4\times SE_7$ solutions \cite{Denef:2009tp}.}
 of the $d=3$ $N=2$ SCFTs with gravity duals
at finite temperature $T$ and non-zero chemical potential $\mu$ with respect to the 
R-symmetry.
The strategy for calculating the Green's function for the supercurrent is to solve the linearised gravitino 
equation in the AdS-RN background and then examine the asymptotic behaviour
at the boundary. To calculate the retarded Green's function, which will
be our main objective, we need to impose ingoing boundary conditions on the gravitino solutions
at the black-brane horizon.

The Green's function that we obtain is valid for the entire class of $N=2$ SCFTs at least for large values of 
the dimensionless temperature 
$T/\mu$. However, as $T/\mu$ is lowered it is possible that the field theory undergoes one or more phase transitions. The
corresponding phases
will 
be described by new black-brane solutions which will involve KK modes lying outside the truncation to minimal gauged supergravity
and hence will depend on the details of the specific SCFT that is being considered.
For example, in the $SE_7$ class of SCFTs it is known that there
are superfluid instabilities which spontaneously break the R-symmetry \cite{Denef:2009tp}. 
An analysis for the case of $S^7$ can be found in \cite{Donos:2011ut} and for another class of SCFTs in \cite{Donos:2010ax}. 
It would certainly be very interesting to know the precise nature of all of the phase transitions all the way down to their
ultimate zero temperature ground states, even for specific examples of $M_6$ or $M_7$ (for some discussion see \cite{Donos:2011ut}).  
One could then aim to extend the calculations of this paper and \cite{Gauntlett:2011mf}
by solving the gravitino equation in the thermodynamically preferred black-brane background for any given temperature.
On the other hand, it is possible that for at least some $N=2$ SCFTs the AdS-RN black-brane is valid all the way down to zero temperatures.

The paper is structured as follows. In section \ref{sec:corrfn} we begin with some general comments concerning the
Green's function of the supercurrent  including a discussion of discrete symmetries.
In section \ref{n2gs} we introduce $N=2$ $D=4$ gauged supergravity while in section
\ref{twodform} we derive a convenient form of the gravitino equations of motion.
In section \ref{asym} we analyse the asymptotic behaviour of the gravitino equations at the AdS boundary and
derive a useful formula for the Green's function.

We will calculate the retarded Green's function and hence we must impose
ingoing boundary conditions on the gravitino equations at the black-brane horizon. The details of how to
do this for non-zero and zero temperature are different and so are treated separately in sections \ref{nonzeroT} and \ref{zeroT}, respectively.
In both cases we solve the equations numerically.
We will see that the spectral function, as a function of frequency $\omega$ and wave-number $k\equiv |{\bf k}|$
has a ``phonino" pole at $\omega+\mu=0$ and $k=0$.
This is a collective fermionic excitation that any supersymmetric medium at finite temperature 
has. This long-wavelength mode is important for the hydrodynamical description and
has been studied, for zero chemical potential, 
in \cite{Lebedev:1989rz,Kovtun:2003vj,Kratzert:2003cr}. Furthermore 
the dispersion relation of the pole was obtained holographically in \cite{Policastro:2008cx}. Our work extends
these investigations to non-zero chemical potential. We will see that the pole persists at $T=0$ and we
will also extract the dispersion relation. It is worth emphasising that in our conventions $\omega=0$ corresponds to the Fermi energy and hence this phonino pole is not gapless\footnote{This is reminiscent of the ferromagnetic magnons discussed in \cite{Iqbal:2010eh}.} when $\mu\ne 0$.

Another important feature of the Green's function for $T\ne 0$ is 
a depletion of spectral weight near $\omega= 0$.  At $T=0$ the spectral weight 
vanishes at $\omega=0$ with, for each $k$, a power-law
behaviour. This latter feature is controlled by the locally quantum critical theory dual to the
$AdS_2\times\mathbb{R}^2$ background which arises as the near-horizon limit of the
AdS-RN solution at $T=0$. To see this in detail,
in section \ref{ads2appr} we first obtain the exact solution for
the gravitino equations in the $AdS_2\times\mathbb{R}^2$ background. 
Then in section \ref{matching} we show how the non-analytic structure of the Green's function
for $\omega\approx 0$, as well as the power-law scaling, is governed by the exact results in the $AdS_2\times\mathbb{R}^2$ background. 

In section \ref{leaver} we calculate the spectrum of quasinormal modes for $T\ne 0$
finding results consistent with some of the main features of the
spectral functions.

We conclude in section \ref{conclude}, summarising some of our main results. 
It is worth emphasising that 
we find no  Fermi surface for the supercurrent correlation function in
our supergravity calculation. We suspect that in the top-down models that we are
considering a Fermi surface may be seen in other fermion correlators and/or at higher loop order.
Another interesting feature is that we will see the conformal dimension of the operators in the one-dimensional conformal field theory in the far IR, dual to the $AdS_2$ region, are all real. This means that the log-oscillatory behaviour see in the bottom-up models of \cite{Liu:2009dm,Faulkner:2009wj} is absent. It is also a manifestation of the fixed charge over mass ratio dictated by the string/M-theory compactifications we consider. In physical terms this charge to mass ratio is never high enough to make pair production of charged Fermions energetically favourable \cite{Faulkner:2009wj}. This leads us to speculate that the system displays no instability towards forming an electron star 
\cite{Hartnoll:2009ns,Hartnoll:2010gu,Cubrovic:2011xm}, a state which can be viewed as the result of occupying the corresponding modes 
in the bulk.

Some technical material is presented in four appendices.
Appendix \ref{spconv} contains our spinor conventions. In appendix \ref{sec:spectral-fn} we discuss
the positivity properties of the spectral function.
Appendix \ref{gb} contains a calculation for the asymptotic behaviour of
more general Rarita-Schwinger equations in arbitrary spacetime dimensions, but propagating in pure AdS space.
Although not strictly needed for the main text we found the calculations enlightening. In appendix \ref{app:numerics} we summarise some
aspects of the numerical methods that we employed to obtain our results. Appendix \ref{app:asymptotics}
contains some technical
material that we use in calculating the Green's function in the $AdS_2\times\bbR^2$ background.

\section{The supercurrent correlation function}
\label{sec:corrfn}

\subsection{The form of the correlation function}
\def\tdgamma{\gamma}
\def\tdrho{\gamma}
Let $S_{\alpha}$ be the conserved supercurrent operator of the dual $d=3$ $N=2$ SCFT in $\mathbb{R}^{1,2}$. 
It is a complex vector-spinor: the index $\alpha$, with $\alpha=t,x,y$, is the vectorial index and we have suppressed the spinor index. 
Since the supersymmetric theory is conformal, the supercurrent is gamma-traceless.
We aim to calculate the retarded correlation function
$G_{{\alpha}{\beta}}(p)=
\big<S_{\alpha}(p)\bar{S}_{\beta}(0)\big>_{Ret}$ at temperature $T$ and chemical potential $\mu$ with
respect to the abelian $R$-symmetry. 
%Let us first consider the expectation value of the
%supercurrent in the presence of a vector-spinor source $a_\alpha$:
%\begin{align}\label{source}
%\left\langle S_\alpha\right\rangle_a=\left\langle S_\alpha{\rm exp}[i\int d^3x(\bar S^\alpha a_\alpha+\bar a^\alpha S_\alpha)]\right\rangle_{SCFT}\, .
%\end{align}
%Linearising in $a$ we obtain
We will do this by exploiting the fact that
the expectation value of the supercurrent in the presence of a vector-spinor source $a_\alpha$, at linearised
order in the source, is given by
\begin{align}\label{form}
\langle S_\alpha\rangle= iG_{\alpha\beta}a^\beta\, .
\end{align}
The gamma-tracelessness and conservation of the supercurrent implies that
\begin{equation}
\label{susycond}
   {\tdgamma}^{{\alpha}}\langle S_{\alpha}\rangle = 0 \ , \qquad
   p^{\alpha}\langle S_{\alpha}\rangle=0\, ,
\end{equation}
where $\tdgamma^{{\alpha}}$ are $d=3$ gamma-matrices. Our conventions for $d=3$ and $D=4$ spinors can
be found in appendix A.
Since we are considering the
SCFT at finite $\mu$, which can be viewed as weakly gauging the $R$-symmetry, we have
\begin{equation}\label{pexpression}
   p^{\alpha} = (\tilde\omega,{\bf k}),\qquad \tilde \omega\equiv \omega+\mu\, .
\end{equation}
Note that $\omega=0$ corresponds to the Fermi energy.
Since the source couples to the supercurrent by a term in the Lagrangian of the form
$\int d^3x(\bar S^\alpha a_\alpha+\bar a^\alpha S_\alpha)$,
the source in %\eqref{source},
\eqref{form} can be taken to satisfy 
\begin{align}\label{sceconds}
\gamma^\alpha a_\alpha=0,\qquad 
\delta a_\alpha=\left(
   \delta_\alpha^\beta-\tfrac{1}{3}\gamma_\alpha \gamma^\beta
   \right)p_\beta\epsilon\, ,
\end{align}
where the second equation in \reef{sceconds} arises from the weak gauging of the supersymmetry. Of course
the supercurrent itself, and hence its expectation value, is gauge invariant. 
In general, since $S_\alpha$ has two independent components, $G_{\alpha\beta}$ will have four.

It will be convenient to extract the four components of $G_{\alpha\beta}$ by 
introducing a basis of 3d vector-spinors $e^{(i)}_{\alpha}$, $i=1,2$,
satisfying $ \tdrho^{\alpha}e^{(i)}_{\alpha}=p^{\alpha}e^{(i)}_{\alpha}=0$ and the orthogonality condition
$\bar{e}^{(i)}_\alpha e^{{(j)}\alpha}=-2p^2\epsilon^{ij}$
. We can then write
\begin{equation}\label{gdefij}
   G_{\alpha\beta} 
      = t_{ij}  e^{(i)}_{\alpha}\bar{ {e}}^{(j)}_{\beta}\, ,
\end{equation}
where the four functions $t_{ij}$ of $p^\alpha$ are the four independent components of $G_{\alpha\beta}$.
Note that the mass dimensions of $G_{\alpha\beta}(p)$ is 2 and hence $t_{ij}$ is dimensionless. We
also observe that the expectation value, $\langle S_\alpha(p)\rangle$, and the source, $a_\alpha(p)$,
have mass dimensions $-1/2$ and $-5/2$, respectively.

An $N=2$ SCFT in $\mathbb{R}^{1,2}$ at finite temperature and chemical potential 
will preserve $O(2)$ rotations (and reflections) in the $x-y$ plane and this strongly
constrains the $t_{ij}$ as we now show. We first rotate so that
\begin{align}
p^{\alpha}=(\tilde\omega,k,0)\, ,
\end{align}
with $k\equiv|\bf k|$
and take the following explicit basis
\begin{equation}
\begin{aligned}\label{explicitbb}
   e^{(1)}_{\alpha}(\tilde\omega,k) &= 
   \left( k, -\tilde\omega, 0\right)_{\alpha}
        n
      -\left( 0, 0, k-\tilde\omega\right)_{\alpha}
        m,\\
    e^{(2)}_{\alpha}(\tilde\omega,k) &= 
    -\left( k, -\tilde\omega, 0\right)_{\alpha}
        m
      + \left( 0, 0, \tilde\omega+k\right)_{\alpha}
       n\, ,
\end{aligned}
\end{equation}
where $m,n$ are two two-component spinors satisfying $\gamma^{tx}m=m$ and $\gamma^{tx}n=-n$.
We now consider the $d=3$ parity operator that
takes $(t,x,y)\to (t,x,-y)$. Using the conventions described in appendix A we have
\begin{align}
e^{(1)}_\alpha(\tilde\omega,k)&\to {\cal P}_{\alpha}{}^\beta \gamma^ye^{(1)}_\beta(\tilde\omega,k)=-e^{(1)}_\alpha(\tilde\omega,k)\, ,\nn
e^{(2)}_\alpha(\tilde\omega,k)&\to {\cal P}_{\alpha}{}^\beta \gamma^ye^{(1)}_\beta(\tilde\omega,k)=e^{(2)}_\alpha(\tilde\omega,k)\, ,
\end{align}
where ${\cal P}_\alpha{}^\beta=\diag(1,1,-1)$.
This induces the action $t_{12}\to-t_{12}$ and $t_{21}\to-t_{21}$ at the same value of $(\tilde\omega,k)$ and
hence 
we can conclude that
\begin{align}
t_{12}=t_{21}=0\, .
\end{align}
We next consider the $SO(2)$ rotation that takes $(t,x,y)\to (t,-x,-y)$. A short calculation shows that
\begin{align}
e^{(1)}_\alpha(\tilde\omega,k)&\to -e^{(2)}_\alpha(\tilde\omega,-k)\, ,\nn
e^{(2)}_\alpha(\tilde\omega,k)&\to e^{(1)}_\alpha(\tilde\omega,k)\, ,
\end{align}
and hence
\begin{align}
t_{22}(\tilde\omega,k)=t_{11}(\tilde\omega,-k)\, .
\end{align}
We find that all of the information of the retarded Green's function is contained in 
$t_{11}(\tilde\omega,k)$, if it is formally viewed as a function of {\it both} positive and negative $k$.

Thus our objective is to calculate $t_{11}(\tilde\omega,k)$ and more specifically the spectral function $A(\tilde\omega,k)$
which we define to be the imaginary part:
\begin{align}
A(\tilde\omega,k)\equiv \textrm{Im}\, t_{11}(\tilde\omega,k)\, .
\end{align}
In appendix \ref{sec:spectral-fn} we show that $A(\tilde\omega,k)>0$.

Before concluding this subsection, it is interesting to relate the above discussion to an alternative way of characterising the
four independent components of the general Green's function.
We first introduce the projector onto transverse gamma-traceless parts of vector spinors
\bea
P_{\alpha}{}^{\beta}
   =\delta_{\alpha}{}^{\beta}-\frac{1}{2}\left(
      \tdgamma_{\alpha}-\frac{p_{\alpha}\cancel{p}}{p^2}\right)
   \tdgamma^{\beta}
   -\frac{1}{2p^2}\left(3p_{\alpha}-\tdgamma_{\alpha}\cancel p\right)p^{\beta}\, .
\eea
Then the constraints \eqref{susycond} imply that we can write $G_{{\alpha}{\beta}}(p)$ in
the general form
\begin{equation}\label{cfn}
\begin{aligned}
   G_{\alpha\beta}(p) &= 
   P_{\alpha}{}^{\gamma}\left[t_{\gamma\delta\rho}\tdrho^{\rho}+2(t_1 p^2+t_2\cancel p)\eta_{\gamma\delta}\right]P^{\delta}{}_{\beta}\\
&=      t_{\alpha\beta\gamma}\tdrho^{\gamma}
        + t_1 \left[ \epsilon_{\alpha\beta\gamma}
           p^{\gamma}\slashed{p}
        + (p^2\eta_{\alpha\beta}-p_{\alpha}p_{\beta})
           \right] \\ & \qquad \qquad \qquad
        + t_2 \left[
           (p^2\eta_{\alpha\beta}-p_{\alpha}p_{\beta})
              \slashed{p}
           +p^2\epsilon_{\alpha\beta\gamma}
           p^{\gamma} \right]\, ,
\end{aligned}
\end{equation}
where 
$t_{\alpha\beta\gamma}$ is totally symmetric,
and satisfies
$p^{\alpha}t_{\alpha\beta\gamma}=0$ and
$t^{\alpha}{}_{\alpha\beta}=0$, and hence has two
independent components. In general the four independent quantities $t_{\alpha\beta\gamma}$, $t_1$
and $t_2$ are arbitrary functions of $p^{\alpha}$. In the case of interest, where we have $O(2)$ invariance, we can
use the above results to conclude that
\begin{align}
t_{\alpha\beta\gamma}=t_1=0,\qquad
t_2=\frac{1}{\tilde\omega+k}t_{11}(\tilde\omega,k)\, .
\end{align}
This easily follows by sandwiching \eqref{cfn} between $\bar m,\bar n$ and $m,n$.
If there is full $d=3$ Lorentz invariance
we would have $t_{\alpha\beta\gamma}=t_1=0$ and $t_2=t_2(p^2)$.
Finally, we note that we also have $\bar e^{(i)}_\alpha e^{(j)\beta}\epsilon^{ij}=-4p^2P_{\alpha}{}^\beta$.

\subsection{Phonino}

Placing the $N=2$ SCFT at finite temperature breaks supersymmetry and leads to a fermionic Goldstino mode. 
When $\mu=0$ it was pointed out that this mode appears as a pole in the spectral function for the supercurrent correlator located at $\omega=0,\,k=0$ 
\cite{Lebedev:1989rz,Kovtun:2003vj,Kratzert:2003cr}. 
This mode is important in the hydrodynamical description of the system 
\cite{Lebedev:1989rz,Kovtun:2003vj} and in this context is known as the ``phonino". Recall that
to construct the hydrodynamics one needs to identify a set of variables
whose thermal expectation values distinguish equilibrium states of the theory. These consist of the charge
densities for all conserved currents. In the present setup there are the usual energy and momentum densities
associated with the energy-momentum tensor, there is the R-charge density associated with the 
R-symmetry current, and there is also the supercharge density associated with the supercurrent.
The hydrodynamical equations are obtained by postulating constitutive relations for
the spatial parts of the currents in terms of the charge densities and then imposing
current conservation.
By analysing the constitutive relations when $\mu=0$ it was shown in \cite{Kovtun:2003vj}
that in addition to normal sound, there is also,
for small $k$, a weakly damped propagating fermionic collective excitation. This phonino or supersound is somewhat analogous to second sound in a superfluid.
The phonino was first studied from a holographic point of view at finite $T$ and $\mu=0$ in \cite{Policastro:2008cx}
for SCFTs in $d=4$. In particular, the diffusion constant for the dispersion relation for the phonino pole was calculated.

Here we will be considering finite temperature and $\mu\ne 0$. In view of our comments above, we expect
that switching on $\mu$ will move the phonino pole to $\omega+\mu=0,\,k=0$, and our results will confirm this
expectation. Recall that $\omega=0$ corresponds to the Fermi energy and hence the phonino is certainly not a gapless excitation.
From our numerical results we will also be able to extract the dispersion relation for the phonino pole.

\section{$N=2$ minimal gauged supergravity in $D=4$}\label{n2gs}
The field content of minimal $N=2$ gauged supergravity in $D=4$ \cite{Freedman:1976aw,Fradkin:1976xz}
consists of a metric, a gauge field ${\cal A}$ and a single Dirac
gravitino $\psi_\mu$ (or equivalently two Majorana gravitini). 
The action is given by
\begin{align}\label{overallaction}
S = &\int d^4 x \sqrt{-g}\bigg[R - {\cal F}_{\mu\nu}{\cal F}^{\mu\nu} + \frac{6}{\ell^2} - 2 \bar\psi_\mu\Gamma^{\mu\nu\rho}D_\nu\psi_\rho 
\nn
&
 \qquad\qquad\qquad-2m\bar\psi_\mu\Gamma^{\mu\nu}\psi_\nu- {i} {\cal F}^{\mu\nu} \bar\psi_\rho \Gamma_{\mu}\Gamma^{\rho\sigma}\Gamma_{\nu}\psi_\sigma   +\dots\bigg]\, ,
\end{align}
where ${\cal F}=d{\cal A}$ and we have not explicitly written the four-Fermi terms, which will not be needed. Here 
\be
D\equiv \nabla - i q{\cal A},\qquad\qquad q=-m\equiv \frac{1}{\ell}\, ,
\ee
where $\nabla$ is the Levi-Civita connection. 
Observe the presence of Pauli terms
which couple the gravitino non-minimally to the field strength of the gauge field\footnote{In a bottom-up
context the role of Pauli term couplings for spin-$\tfrac{1}{2}$ fermions have been studied in some detail in
\cite{Edalati:2010ww,Edalati:2010ge} and \cite{Guarrera:2011my}.}. It is also useful to record the 
supersymmetry transformations about a bosonic configuration. The only non-trivial transformation is for the gravitino and is given by
\be\label{susyvariation}
\delta\psi_\mu =  \left(D_\mu - \frac{m}{2} \Gamma_\mu + \frac{i}{4}{\cal F}_{\nu\rho}\Gamma^{\nu\rho}\Gamma_\mu  \right)\varepsilon\,,
\ee
where $\varepsilon$ is an infinitesimal Dirac spinor.

The AdS-RN black-brane solution solves the equations of motion and is given by
\begin{equation}\label{adsrn1}
   d s^2 = - fdt^2 + \frac{dr^2}{f}  + \frac{r^2}{\ell^2} d{\bf x}^2 \,,\qquad\qquad {\cal A}=\phi dt\,,
\end{equation}
with 
\begin{equation}
\begin{aligned}\label{adsrn2}
 f=\frac{r^2}{\ell^2}-\frac{r_+}{r}\left(\frac{r_+^2}{\ell^2}+\ell^2\mu^2\right)+\ell^2\mu^2\frac{r_+^2}{r^2},\qquad\qquad 
  \phi = \mu\ell \left( 1-\frac{r_+}{r} \right) .
\end{aligned}
\end{equation}
The horizon is located at $r=r_+$ and
the temperature, $T=f'(r_+)/4\pi$, is given by 
\begin{align}
T=(3r_+/\ell^2-\ell^2\mu^2/r_+)/4\pi\, .
\end{align}
Thus, when $\mu=\sqrt{3}r_+/\ell^2$ we have $T=0$ and
\begin{align}\label{eq:fextreme}
  f &= \frac{(r-r_+)^2(r^2+2rr_++3r_+^2)}{\ell^2r^2}, \qquad (T=0)\, .
\end{align}
Furthermore, as $r\to r_+$ the $T=0$ black-brane solution approaches $AdS_2\times \mathbb{R}^2$
with the radius of the $AdS_2$ given by $L_{(2)}=\ell/\sqrt{6}$. 
Indeed after introducing a new coordinate via  $r-r_+=L^2_{(2)}/z$ the solution
approaches the exact solution 
\be\label{ads2back}
ds^2=\frac{L^2_{(2)}}{z^2}\left[-dt^2+dz^2\right]+\frac{r_+^2}{\ell^2} d{\bf x}^2,\qquad {\cal A}=\frac{L_{(2)}}{\sqrt{2}z}dt\, .
\ee

We will study the equation of motion of the gravitino (at the linearised level) in the AdS-RN background \eqref{adsrn1},\eqref{adsrn2} and also
in the $AdS_2\times \mathbb{R}^2$ background \eqref{ads2back}.
It will be convenient to
use the local supersymmetry transformations \eqref{susyvariation} to work in the gauge
\be\label{gaugechoice}
D^\mu \psi_\mu = \Gamma^\mu \psi_\mu =0\, ,
\ee
in which the linearised equation for the gravitino is given by 
\bea\label{graveq}
 \left( \slashed{D} -m- \frac{i\lambda}{2}{\cal F}^{\mu\nu}\Gamma_{\mu\nu} \right)\psi_\rho +i\lambda{\cal F}^{\mu\nu}\Gamma_{\mu}\Gamma_\rho \psi_\nu=0\, .
\eea
Here we have introduced $\lambda\equiv 1$ to highlight where the Pauli coupling enter in some places in the sequel. 
Note that \eqref{gaugechoice} does not quite fix the gauge completely. There are residual gauge transformations of the form 
\eqref{susyvariation} where $\varepsilon$ satisfies
\be\label{ressusy}
(\Gamma^\mu D_\mu-2m)\varepsilon=0\, .
\ee

\section{Effective two-dimensional formulation of the gravitino equation}\label{twodform}

In solving the gravitino equation \eqref{graveq} in the AdS-RN background \eqref{adsrn1},\eqref{adsrn2}
and the $AdS_2\times \mathbb{R}^2$ background \eqref{ads2back},
it will be convenient to, effectively, carry out a dimensional reduction on the two spatial directions $x^i$
to two space-time dimensions. This is advantageous since the Lorentz representations of Spin(1,1) 
are all one-dimensional. 

It is convenient to consider the general metric and gauge connection of the form
\begin{equation}
\begin{aligned}
   ds^2 &= d\hat s_2^2 + H^2 dx^i dx^i \,, \\
   d\hat s_2^2 &= \hat g_{mn}dy^m dy^n\,, & 
   {\cal A} &= {\cal A}_m dy^m\, ,
\end{aligned}
\end{equation}
where $H=H(y)$, $\hat g_{mn}=\hat g_{mn}(y)$ and
$\mathcal{A}_m=\mathcal{A}_m(y)$.  
We choose a frame
\begin{equation}
   e^{\hat{m}} = \hat{e}^{\hat{m}} \,, \qquad
   e^{\hat i} = H dx^{\hat i}\, ,
\end{equation}
in which the spin connection has non-zero components
$\omega^{\hat i}{}_{\hat m} = e^{\hat i} \partial_{\hat m}\ln H$ and
   $\omega^{\hat m}{}_{\hat n} = \hat{\omega}^{\hat m}{}_{\hat n}$.
We then use the planar symmetry of the background to expand the frame components of the
gravitino in plane waves:
\begin{equation}\label{psidef}
   \psi_{\hat\mu}(y^m,\mathbf{x}) 
      = \int\frac{d^2k}{(2\pi)^2} H^{-1}\chi_{\hat\mu}(y^m,\mathbf{k})
         e^{i\mathbf{k}\cdot\mathbf{x}}\, ,
\end{equation}
where, for later convenience, we have included the factor of $H^{-1}$ in the
definition of $\chi_{\hat\mu}$. We now focus on the
two-dimensional $\hat m$ components of $\chi_{\hat\mu}$ since the transverse
$\chi_{\hat i}$ can be determined via the conditions
$\Gamma^\mu\psi_\mu=D^\mu\psi_\mu=0$. 

Substituting into the equation of motion for the gravitino \eqref{graveq} we find
\begin{equation}\label{pop}
   \left(\slashed{\hat{D}} + i H^{-1}\slashed{\mathbf{k}} - m - \tfrac{1}{2}i
      \lambda F_{np}\Gamma^{np}\right) \chi_{\hat m}
      + (\partial_{\hat m}\ln H) \Gamma^n\chi_n
      + i\lambda F^{np}\Gamma_n\Gamma_{\hat m}\chi_p = 0\, ,
\end{equation}
where $\slashed{\mathbf{k}}\equiv k_i \Gamma^{\hat i}$.
We would now like to expand $\chi_{\hat m}$ in terms of irreducible
$Spin(1,1)$ representations. To do this we first note that we can
decompose the spinor components under the two-dimensional chirality operator $\Gamma^{(2)}\equiv \Gamma^{\hat {t}\hat{r}}$ 
and also $\Gamma^{(3)}\equiv (1/k)\Gamma^{(2)}\slashed{\mathbf{k}}=\Gamma^{\hat{t}\hat{r}\hat{x}}$. This is possible
since these two operators commute (for more details on our spinor conventions in $D=4$ see appendix~\ref{spconv}).
Let us first decompose under
$\Gamma^{(3)}$ writing
\begin{equation}\label{psidef2}
   \chi_{\hat m} = \chi^\eta_{\hat m} + \chi^\rho_{\hat m}\, ,
\end{equation}
with $\Gamma^{(3)} \chi^\eta_{\hat m}=+\chi^\eta_{\hat m}$ and $\Gamma^{(3)} \chi^\rho_{\hat m}=-\chi^\rho_{\hat m}$.
It will be important to note that $\chi^\eta_{\hat m}$ and $\chi^\rho_{\hat m}$ are not mixed by \eqref{pop}.
Next consider the two-dimensional lightcone frame
\begin{equation}
   d\hat{s}^2 = e^+ e^- \,, \qquad
   e^{\pm} = e^{\hat r}\pm e^{\hat t} \,.
\end{equation}
We can then introduce a basis for $\chi_{\hat m}$ given by 
\def\uth{u^{(3/2)}}
\def\uoh{u^{(1/2)}}
\def\umoh{u^{(-1/2)}}
\def\umth{u^{(-3/2)}}
\def\vth{v^{(3/2)}}
\def\voh{v^{(1/2)}}
\def\vmoh{v^{(-1/2)}}
\def\vmth{v^{(-3/2)}}
\bea\label{chidef}
   \chi^\eta &= &\uth e^+ \otimes \eta^+ + 
   \uoh e^+ \otimes \eta^- + 
   \umoh e^- \otimes \eta^+ + 
   \umth e^- \otimes \eta^-\, , \nn
      \chi^\rho &= &\vth e^+ \otimes \rho^+ + 
   \voh e^+ \otimes \rho^- + 
   \vmoh e^- \otimes \rho^+ + 
   \vmth e^- \otimes \rho^-\, ,
\eea
where $\Gamma^{(2)}\eta^\pm=\pm\eta^\pm$, $\Gamma^{(2)}\rho^\pm=\pm\rho^\pm$
 and hence
the superscript on $u,v$ refers to the helicity of the $Spin(1,1)$ representation.
The equations of motion for $\chi^\eta$ then reduce to %
\begin{equation}\label{thegeneqns}
\begin{aligned}
   {\cal D}_- \uth + K \uoh
      + (\partial_+ \ln H) \umoh&= 0 \,, \\
         {\cal D}_+ \uoh + K' \uth
      + (\partial_+ \ln H) \uoh&= 0 \,, \\
         {\cal D}_- \umoh + \bar K' \umth
      + (\partial_- \ln H) \umoh&= 0 \,, \\
         {\cal D}_+\umth + \bar K \umoh
      + (\partial_- \ln H) \uoh&= 0 \,, \\
\end{aligned}
\end{equation}
where we have introduced the covariant derivatives
\begin{equation}
   {\cal D}_\pm u^{(s)} = (\partial_\pm - iqA_\pm + 2s\Omega_\pm)u^{(s)} \,, 
   \qquad
   \Omega_\pm = \hat{\omega}_{\pm +-} \,, 
\end{equation}
and
\begin{equation}
\begin{aligned}
   K &= \frac{1}{2}\left[m + ikH^{-1} -2 i\lambda F_{0}\right] \,, \\
   K' &= \frac{1}{2}\left[m - ikH^{-1} - 2i\lambda F_{0}\right]\,, 
\end{aligned}   
\end{equation}
where ${\cal F}\equiv F_{0}e^+\wedge e^-$. The equations of motion
for $\chi^\rho$ give rise to equations for $v$ of the same form but with $k\to-k$, or equivalently
$K\leftrightarrow K'$. 

We can make a similar reduction for the residual gauge
transformation. Decomposing the Fourier modes as 
\def\woh{w^{(1/2)}}
\def\wmoh{w^{(-1/2)}}
\def\zoh{z^{(1/2)}}
\def\zmoh{z^{(-1/2)}}
\begin{equation}
   \varepsilon(x^m,\mathbf{x}) 
      = \int\frac{d^2k}{(2\pi)^2} H^{-1}[    \woh\eta^+ + \wmoh \eta^- + \zoh \rho^+ + \zmoh \rho^-     ]
         e^{i\mathbf{k}\cdot\mathbf{x}}\,.
\end{equation}
with $w$ and $z$ functions of $(x^m,{\bf k})$,
one finds that the constraint which the residual supersymmetry transformations must satisfy,
given in
\eqref{ressusy},
can be written
\begin{equation}
\begin{aligned}\label{residual}
   {\cal D}_- \woh + P \wmoh&= 0 \,, \\
  {\cal D}_+ \wmoh + \bar{P} \woh &= 0 \,,
\end{aligned}
\end{equation}
with 
\begin{equation}
   P = m + \frac{ik}{2H}\, .
\end{equation}
The equations for $z^\pm$ are of the same form but with $k\to -k$ or, equivalently, $P\leftrightarrow\bar{P}$.
The supersymmetry variations \eqref{susyvariation} imply that
\begin{equation}\label{resdual}
\begin{aligned}
   \delta \uth &= \left({\cal D}_+ - \partial_+\ln H\right)\woh  \,, \\
   \delta \uoh &= \left({\cal D}_+ - \partial_+\ln H\right)\wmoh
        + \left(\tfrac{1}{2}m + i \lambda F_0\right)\woh \,, \\
   \delta \umoh &= \left({\cal D}_- - \partial_-\ln H\right)\woh 
        + \left(\tfrac{1}{2}m - i \lambda F_0\right)\wmoh \,, \\
   \delta \umth&= \left({\cal D}_- - \partial_-\ln H\right)\wmoh  \,.
\end{aligned}
\end{equation}
with similar equations for $\delta v^{(s)}$. 

In the following we will just focus on the $\eta$ dependent equations, namely 
the equations for $u^{(s)}$ given in \eqref{thegeneqns} and the residual supersymmetry transformations given in \eqref{resdual},
since the corresponding $\rho$ dependent equations involving $v^{(s)}$ 
can be obtained via $k\to -k$.

We now write down the explicit gravitino equations in the AdS-RN black-brane background 
\eqref{adsrn1},\eqref{adsrn2}. 
We will assume a time dependence of the form $e^{-i\omega t}$.
Adopting the obvious orthonormal frame $e^{\hat t}=f^{1/2}dt$, $e^{\hat r}=f^{-1/2}dr$, $e^{\hat i}=(r/\ell)dx^i$
we find that the gravitino equations \eqref{thegeneqns} take the explicit form
 \begin{equation}\label{thegeneqnsG}
\begin{aligned}
  \left[\partial_r+\frac{i(\omega+q\phi)}{f}+\frac{3f'}{4f}\right]\uth + \frac{1}{f^{1/2}}\left[m+\frac{ik\ell}{r}+i\lambda\phi' \right]\uoh
      + \frac{1}{r} \umoh&= 0 \,, \\
        \left[\partial_r-\frac{i(\omega+q\phi)}{f}-\frac{f'}{4f}+\frac{1}{r}\right]\uoh + \frac{1}{f^{1/2}}\left[m-\frac{ik\ell}{r}+i\lambda\phi' \right]\uth
    &= 0 \,, \\
          \left[\partial_r+\frac{i(\omega+q\phi)}{f}-\frac{f'}{4f}+\frac{1}{r}\right]\umoh + \frac{1}{f^{1/2}}\left[m+\frac{ik\ell}{r}-i\lambda\phi' \right]\umth
    &= 0 \,, \\
               \left[\partial_r-\frac{i(\omega+q\phi)}{f}+\frac{3f'}{4f}\right]\umth + \frac{1}{f^{1/2}}\left[m-\frac{ik\ell}{r}-i\lambda\phi' \right]\umoh
      + \frac{1}{r} \uoh&= 0 \,. \\
        \end{aligned}
\end{equation}
We reiterate that $q=-m=\ell^{-1}$ and $\lambda=1$.
Observe that these equations are invariant under $u^{(s)}\to (u^{(-s)})^*$ and we will return to this point
in section \ref{sec:ZeroFreq}.

The residual supersymmetry transformations acting on the $u^{(s)}$ given in \eqref{resdual} take the explicit form
\bea\label{eq:SUSYvar_explicit}
\delta u^{(3/2)}&=& \frac{f^{1/2}}{2} \left[ \partial_r - \frac{i(\omega+q\phi)}{f}  - \frac{f'}{4f} - \frac{1}{r} \right]w^{(1/2)}\,,\nn
\delta u^{(1/2)}&=& \frac{f^{1/2}}{2} \left[ \partial_r - \frac{i(\omega+q\phi)}{f}  + \frac{f'}{4f} - \frac{1}{r} \right]w^{(-1/2)} + \frac{1}{f^{1/2}}\left[ m - i \lambda  \partial_r\phi \right]w^{(1/2)}\,,\nn
\delta u^{(-1/2)}&=& \frac{f^{1/2}}{2} \left[ \partial_r + \frac{i(\omega+q\phi)}{f}  + \frac{f'}{4f} - \frac{1}{r} \right]w^{(1/2)} + \frac{1}{f^{1/2}}\left[ m + i \lambda  \partial_r\phi \right]w^{(-1/2)}\,,\nn
\delta u^{(-3/2)}&=& \frac{f^{1/2}}{2} \left[ \partial_r + \frac{i(\omega+q\phi)}{f}  - \frac{f'}{4f} - \frac{1}{r} \right]w^{(-1/2)}\,,
\eea
where the residual supersymmetry parameters $w^{(\pm1/2)}$ satisfy \reef{residual} which can be written as 
\bea\label{eq:explicit_residual}
\left[\partial_r + \frac{i(\omega + q \phi)}{f} + \frac{f'}{4f}  \right]w^{(1/2)} + \frac{1}{f^{1/2}}\left[2m + \frac{ik\ell}{r}  \right]w^{(-1/2)}&=&0\,,\nn
\left[\partial_r - \frac{i(\omega + q \phi)}{f} + \frac{f'}{4f}  \right]w^{(-1/2)} + \frac{1}{f^{1/2}}\left[2m - \frac{ik\ell}{r}  \right]w^{(1/2)}&=&0\,.
\eea
Note that in these expressions $\omega$ only appears in the combination $(\omega + q \phi)$ which approaches $(\omega + {\ell}^{-1}\mu)$ at the $AdS_4$ boundary. This shows that $\omega=0$ is the Fermi energy.

\section{Asymptotics and the Green's function}\label{asym}
In the remainder of the paper we will solve the gravitino equations \eqref{thegeneqnsG}, taking into account
the residual supersymmetry transformations \eqref{eq:SUSYvar_explicit},\eqref{eq:explicit_residual}. As we are interested
in obtaining the retarded Green's function, we will impose ingoing boundary conditions at the black-brane horizon located
at $r=r_+$. The boundary conditions for $T\ne 0$ and $T=0$ are different and will be discussed separately in later sections.
The behaviour of the solutions at the asymptotic AdS boundary, located at $r\to\infty$, allow us to 
extract the source and the expectation value of the supercurrent in the dual SCFT. The imposition of the ingoing boundary conditions will
mean that the expectation value is fixed by the source, as in \eqref{form}, and this allows us to extract the retarded Green's function. 
In this section we explain this procedure in more detail.

\subsection{Asymptotic expansion}
In analysing\footnote{We found it useful, as a warmup, to first study a more general Rarita-Schwinger equation in arbitrary dimensions propagating in pure AdS space. The analysis is presented in appendix in \eqref{gb}.} the asymptotic behaviour of solutions as $r\to\infty$, it  will be very helpful to return temporarily to the covariant gravitino equations
given in \eqref{gaugechoice},\eqref{graveq}.
Furthermore, it will be helpful
to consider the radial components, $\psi_{\hat r}$, and the $d=3$ components, $\psi_{\hat\alpha}$,
of the bulk gravitino field separately, where hats denote $D=4$ tangent space indices. 
It is also helpful to further separate each of these into positive and negative chirality components
with respect to $\Gamma^{\hat r}$, denoted by superscript $\pm$, respectively, in the following. At the end of this section we will reconnect with the equations for the $\hat t$ and 
$\hat r$ components that we derived in the last section.

We find that as $r\to\infty$ the asymptotic expansion of \eqref{gaugechoice},\eqref{graveq} takes the schematic form
\begin{align}
\psi
&=r^{-1/2}\psi_{-1/2}+r^{-3/2}\psi_{-3/2}+r^{-5/2}\psi_{-5/2}
   +r^{-7/2}\psi_{-7/2}+(r^{-7/2}\log r)\phi_{-7/2}+\dots 
\end{align}
In particular we find that log terms only start to appear at the order given.
In more detail, introducing $\pm$ superscripts corresponding to positive and negative chirality with respect to 
$\Gamma^{\hat{r}}$, respectively, at leading order we find
\begin{align}
\psi_{-1/2,\hat\alpha}^+&=0 \, , \nn
\psi_{-1/2,\hat\alpha}^-&=A_{\hat\alpha} \, , \nn
\psi_{-1/2,\hat r}^+&=0 \, , \nn
\psi_{-1/2,\hat r}^-&=0 \, , 
\end{align}
with $(1+\Gamma^{\hat r})A_{\hat\alpha}=0$ and $\Gamma^{\hat\alpha}A_{\hat\alpha}=0$.
This data is enough to obtain the next order of the expansion
and we get
\begin{align}\label{32exp}
\psi_{-3/2,{\hat{\alpha}}}^+&=-i\ell^2\cancel p
A_{\hat{\alpha}}+\tfrac{1}{2}i\ell^2\Gamma_{\hat{\alpha}}(p\cdot A) \, , \nn
\psi_{-3/2,{\hat{\alpha}}}^-&=0 \, , \nn
\psi_{-3/2,\hat r}^+&=0 \, , \nn
\psi_{-3/2,\hat r}^-&=-\tfrac{1}{2}i\ell^2(p\cdot A)
\end{align}
At the next order, new data appears. We find that 
\begin{align}
\psi_{-5/2,{\hat{\alpha}}}^+ &=\tilde B_{\hat{\alpha}} \, , \nn
\psi_{-5/2,{\hat{\alpha}}}^-&=\tfrac{1}{2}\ell^4\left[p_{\hat{\alpha}}
(p\cdot A)-p^2A_{\hat{\alpha}}\right] \, , \nn 
\psi_{-5/2,\hat r}^+&=-\tfrac{1}{2}\ell^4\cancel p (p\cdot A) \, , \nn 
\psi_{-5/2,\hat r}^-&=-i\ell(g\cdot A) \, , 
\end{align}
where $(1-\Gamma^{\hat r})\tilde B_{{\hat{\alpha}}}=0$ and
\begin{align}
g_{\hat{\alpha}}\equiv (\ell\mu r_+,0,0)\, .
\end{align}
New data also appears at the next order. We find 
\begin{align}
\psi_{-7/2,{\hat{\alpha}}}^+&=\tfrac{1}{4}i\ell^6\left[(p^2\Gamma_{\hat{\alpha}}+2p_{\hat{\alpha}}\cancel
   p)(p\cdot A)-2p^2\cancel pA_{\hat{\alpha}}\right] \, , \nn 
\psi_{-7/2,{\hat{\alpha}}}^-&=\tfrac{1}{3}\ell^3\left[  \tfrac{5}{3}\Gamma_{\hat{\alpha}}\cancel p (g\cdot A)
+\left(\tfrac{3}{2}\cancel g \Gamma_{\hat{\alpha}}-\tfrac{1}{2}\Gamma_{\hat{\alpha}}\cancel g\right)(p\cdot A)  -2\cancel g\cancel p A_{\hat{\alpha}} \right]\nn
&\qquad -\tfrac{1}{6}c A_{\hat{\alpha}}
-\tfrac{1}{9}C^{\hat\beta\hat\gamma}\Gamma_{\hat{\alpha}}\hat\Gamma_{\hat\beta}
A_{\hat\gamma} 
-\tfrac{1}{3}\Gamma_{\hat{\alpha}} B_r
-\tfrac{1}{3}i\ell^2\cancel p\tilde
B_{\hat{\alpha}} \, , \nn 
\psi_{-7/2,\hat r}^+&=B_{\hat r} \, , \nn
\psi_{-7/2,\hat r}^-&=\tfrac{1}{4}i\ell^6p^2 (p\cdot A) \, , 
\end{align}
where $(1-\Gamma^{\hat r})B_{\hat r}=0$ and we have defined 
\begin{align}\label{defofcmat}
C_{{\hat{\alpha}}\hat\beta}&={c}\,\diag(2,1,1), \qquad c=-\tfrac{1}{2}(r_+^3+\ell^4\mu^2r_+)\, .
\end{align}
Note that $C_{\hat{\alpha}}{}^{\hat{\alpha}}=0$.
Finally, for the log terms we get
\begin{align}
\phi_{-7/2,{\hat{\alpha}}}^+&=0 \, , \nn
\phi_{-7/2,{\hat{\alpha}}}^-&=\tfrac{2}{3}\ell^3\Gamma_{\hat{\alpha}}
\cancel p (g\cdot
A)-\tfrac{1}{3}C^{\hat\beta\hat\gamma}\Gamma_{\hat{\alpha}}\Gamma_{\hat\beta}
A_{\hat\gamma} \, , \nn 
\phi_{-7/2,\hat r}^+&=-2\ell^3\cancel p (g\cdot A)+
C^{\hat\beta\hat\gamma}\Gamma_{\hat\beta} A_{\hat\gamma} \, , \nn 
\phi_{-7/2,\hat r}^-&=0 \, . 
\end{align}

At this point we have shown that the 
expansion is determined by the negative chirality, gamma-traceless $A_{\hat{\alpha}}$, the positive chirality $\tilde B_{\hat{\alpha}}$ satisfying various constraints arising from \eqref{gaugechoice}, 
which we will not write down explicitly, 
as well as the positive chirality spinor $B_{\hat r}$. To proceed it is now very helpful 
to carry out a similar expansion of the residual gauge transformations. In particular, this will reveal that
$B_{\hat r}$ can be gauged away.

In more detail we find that at first order
\begin{align}
\varepsilon_{1/2}^+=0,\qquad
\varepsilon_{1/2}^-=-\frac{i}{\ell}\varepsilon_{1/2}\,,
\end{align}
with $(1+\Gamma^{\hat r})\varepsilon_{1/2}=0$. This data allows us to continue to three more orders and we find
\begin{align}
\varepsilon_{-1/2}^+=-\tfrac{1}{3}\ell\cancel p\varepsilon_{1/2}, \qquad
\varepsilon_{-1/2}^-=0\,,
\end{align}
and
\begin{align}
\varepsilon_{-3/2}^+=-\tfrac{1}{2}\cancel g \varepsilon_{1/2},\qquad
\varepsilon_{-3/2}^-=\tfrac{1}{6}i\ell^3p^2\varepsilon_{1/2}\,,
\end{align}
and
\begin{align}
\varepsilon_{-5/2}^+=\tfrac{1}{6}\ell^5p^2\cancel p \varepsilon_{1/2},\qquad
\varepsilon_{-5/2}^-=-i\left(-\frac{\ell^2}{6}\cancel p\cancel g
   -\frac{\ell^2}{9}\cancel g \cancel p  +\frac{c}{6\ell}
   \right)\varepsilon_{1/2} \,.
\end{align}
At the next order we find a new gauge parameter that has positive chirality. The only information that we require
is that it can be used to gauge away $B_r$, so we shall not write down the next order of the expansion explicitly.

We can now use these results to determine the gauge transformations of
$A_{\hat{\alpha}}$ and $\tilde B_{\hat{\alpha}}$. For
$A_{\hat{\alpha}}$ we find that  
$\delta A_{\hat{\alpha}}=(p_{\hat{\alpha}}-\frac{1}{3}\Gamma_{\hat{\alpha}} \cancel{p})\varepsilon_{1/2}$. It turns out that
$\tilde B_{\hat{\alpha}}$ is not the physical data to focus on. 
The asymptotic expansion of the gravitino equations can be specified by the data $A_{\hat{\alpha}}$ and $\tilde B_{\hat{\alpha}}$ or alternatively $A_{\hat{\alpha}}$ combined with a linear combination of $\tilde B_{\hat{\alpha}}$ and 
$A_{\hat{\alpha}}$. We will fix this ambiguity, {\it uniquely}, by
demanding that the latter data is gamma traceless, conserved and
gauge-invariant. In terms of the symmetry algebra of the
asymptotic AdS space, this decomposes into independent, irreducible
representations. In more detail we define 
\begin{align}
B_{\hat{\alpha}}=\tilde B_{\hat{\alpha}}+\tfrac{1}{2}i\ell\cancel g A_{\hat{\alpha}}+M_{{\hat{\alpha}}}{}^{\hat\beta}A_{\hat\beta}\,,
\end{align}
where
\begin{align}
M_{{\hat{\alpha}}\beta}=\bar P_{\hat{\alpha}}{}^{\hat\gamma} T_{\hat\gamma\hat\sigma}\bar P^{\hat\sigma}{}_{\hat\beta},\qquad 
\bar
P_{{\hat{\alpha}}}{}^{\hat\beta}=\delta^{\hat\beta}_{\hat{\alpha}}-\tfrac{1}{3}\Gamma_{\hat{\alpha}}\Gamma^{\hat\beta}\,, 
\end{align}
and
\bea
T_{{\hat{\alpha}}{\hat\beta}}&=&\frac{3i\ell}{p^2}g_{({\hat{\alpha}}}p_{{\hat\beta})}\cancel p
-\frac{3i\ell}{4p^2}p_{\hat{\alpha}} p_{\hat\beta} \cancel g-\frac{3i\ell}{4p^4}(g\cdot p) p_{\hat{\alpha}} p_{\hat\beta}\cancel p\nn
&&-\frac{3i}{2\ell^2 p^2}C_{\hat\gamma({\hat{\alpha}}}\Gamma^{\hat\gamma} p_{{\hat\beta})}
+\frac{9i}{8\ell^2 p^4}C_{\hat\gamma\hat\sigma}\Gamma^{\hat\gamma} p^{\hat\sigma} p_{\hat{\alpha}} p_{\hat\beta}\,.
\eea
We then find that $B_{\hat{\alpha}}$ satisfies
\begin{align}\label{bfconds}
(1-\Gamma^{\hat r})B_{\hat\alpha}=0,\qquad
\Gamma^{\hat \alpha} B_{\hat \alpha}=0,\qquad p^{\hat\alpha} B_{\hat\alpha}=0,\qquad \delta B_{\hat \alpha}=0\, .
\end{align}
as advertised.
It is convenient to also record here the constraints satisfied by the $A_{\hat\alpha}$ data and their gauge transformations
\begin{align}\label{afconds}
(1+\Gamma^{\hat r})A_{\hat\alpha}=0,\qquad \Gamma^{\hat \alpha}
A_{\hat \alpha} =0,\qquad \delta A_{\hat\alpha}=\left( 
p_{\hat{\alpha}}-\tfrac{1}{3}\Gamma_{\hat \alpha} \cancel
p\right)\varepsilon_{1/2}\, . 
\end{align}

\subsection{Extracting the Green's function}\label{sec:GreenFun}
The asymptotic expansion is specified by $A_{\hat\alpha}$ and $B_{\hat\alpha}$ satisfying \eqref{bfconds},\eqref{afconds}
with, in particular,
\bea
\psi_{\hat\alpha}=
r^{\Delta-d}A_{\hat\alpha}+\dots+ r^{-\Delta}B_{\hat\alpha}+\dots 
\eea
where $\Delta=5/2$ and $d=3$. Using the expression for the $D=4$ gamma-matrices in terms of the $d=3$ gamma matrices presented
in appendix A, we can write
\begin{align}
A_{\hat\alpha}\equiv \ell^{1/2}(0,a_\alpha),\qquad B_{\hat\alpha}\equiv \ell^{9/2}(b_\alpha,0),\qquad 
\varepsilon_{1/2}\equiv \ell^{1/2}(0,\epsilon),
\end{align}
and the factors of $\ell$ have been added to
give the canonical $d=3$ dual field theory dimensions (we deduce from \eqref{overallaction} that $\psi(r,p)$ has mass dimension $-5/2$ and recall the comments after \eqref{gdefij}). 
The conditions \eqref{bfconds},\eqref{afconds} can then be written in terms of
the $d=3$ spinors $a_\alpha$, $b_\alpha$ and $\epsilon$ as follows:
\begin{align}\label{condsonanadb}
\gamma^\alpha a_\alpha =0,\qquad \delta
a_\alpha=\left(\delta_\alpha^\beta-\tfrac{1}{3}\gamma_\alpha
\gamma^\beta\right)p_\beta\epsilon\nn 
\gamma^\alpha b_\alpha=0,\qquad p^\alpha b_\alpha=0,\qquad \delta b_\alpha=0\, .
\end{align}

Since the supercurrent is an operator in the dual SCFT with scaling dimension $\Delta=5/2$, 
the source should be fixed by the $r^{-1/2}$ 
expansion data and the expectation value by the $r^{-5/2}$ expansion data.
After comparing \eqref{condsonanadb} with \eqref{susycond}, \eqref{sceconds}, the above expansion allows
us to identify $a_\alpha$ and $b_\alpha$ as the source and the (gauge-invariant) 
expectation value of the supercurrent in the dual $d=3$ field
theory, respectively. Furthermore, \eqref{form} allows us to write
\begin{align}\label{baexp}
b_\alpha=iG_{\alpha\beta} a^\beta
=it_{ij}e^{(i)}_\alpha\bar e^{(j)}_\beta a^\beta\,,
\end{align}
where $e^{(i)}_{\alpha}$, $i=1,2$, are the basis of 3d vector spinors
that we introduced in section \ref{sec:corrfn}. 
Recall that in section \ref{sec:corrfn} we argued that the $O(2)$ symmetry of the system can be used to show that $t_{12}=t_{21}=0$ and that
$t_{22}(\omega,k)=t_{11}(\omega,-k)$. Using these results we can obtain the convenient formula
\begin{align}\label{otherversiont11}
t_{11}=-\frac{i\bar e^{(2)}_\alpha b^\alpha}{2p^2\bar e^{(1)}_\beta a^\beta}\,.
\end{align}

We can actually derive these results directly. In doing so we will obtain another formula for $t_{11}$ that will make contact with
the gravitino equations that we derived in section 4 that involved the components $\psi_{\hat t}$ and $\psi_{\hat r}$.
Recall that we decomposed the gravitino into $\eta$ and $\rho$ sectors (see \eqref{psidef}, \eqref{psidef2} and \eqref{chidef}) and that the resulting
equations in the two sectors do not mix (in the $\eta$ sector the equations for $u^{(s)}$ are given explicitly in \eqref{thegeneqnsG}
and in the $\rho$ sector the equations for the $v^{(s)}$ can be obtained from \eqref{thegeneqnsG}
after taking $k\to -k$). To proceed we use the constraints \eqref{condsonanadb} to write
\begin{align}
b_{\hat x} &=-\frac{\tilde \omega}{k}b_{\hat t} \, , &
b_{\hat y} &=\left(\tdrho^{\hat{t}\hat{y}}-\frac{\tilde\omega}{k}\tdrho^{\hat{x}\hat{y}}\right) b_{\hat t}\,,\nn
a_{\hat x} &=-\frac{\tilde \omega}{k}a_{\hat t}+\frac{1}{k}p\cdot a 
\, ,&
a_{\hat y} &=\left(\tdrho^{\hat{t}\hat{y}}-\frac{\tilde\omega}{k}\tdrho^{\hat{x}\hat{y}}\right) a_{\hat t}+\tdrho^{\hat{x}\hat{y}}\frac{1}{k}p\cdot a\,.
\end{align}
We can also use \eqref{32exp}, which says that asymptotically we have
\be\label{eq:arcoeff}
p\cdot a=\frac{2i}{\ell^2}a^{(1)}_{\hat r}\,,
\ee
the superscript is a reminder that it appears at one lower order in the expansion in $1/r$.
Then taking the $\hat t$ component of \eqref{baexp} we are led to
\bea
b_{\hat t}=-2[t_{11} n -t_{21} m]\bar n\left( p^2a_{\hat t}-\frac{i}{\ell^2}(k-2\tilde \omega)a^{(1)}_{\hat r}\right)
 \nn
 -2[t_{22} m- t_{12} n]\bar m\left(p^2a_{\hat t}+\frac{i}{\ell^2}(k+2\tilde \omega)a^{(1)}_{\hat r}\right)\,.
 \eea
Now, after imposing ingoing boundary conditions at the black-brane horizon (which we discuss in subsequent sections) we will
have two solutions in the $\eta$ sector and two solutions
in the $\rho$ sector which allow us to extract the four components $t_{ij}$. 
Consider first the $\eta$ sector solutions. Since $A_{\hat t}$ and 
$A^{(1)}_{\hat r}\equiv \frac{\ell^2}{2i}p\cdot A$ both
have negative chirality with respect to $\Gamma^r$, they must be proportional to $(\eta^++\eta^-)=(0,m)^T$ in the basis we are using.
Similarly, since $B_{\hat t}$ has positive chirality with respect to $\Gamma^r$, it must be proportional to $(\eta^+-\eta^-)=(0,n)^T$.
Thus, by considering the $\eta$ sector 
solutions, and recalling $\bar m n=-\bar n m=1$, we deduce that $t_{21}=0$ and that 
\be\label{eq:CorrelatorOne}
t_{11}=\frac{\bar m b_t}{2\bar n(- p^2 a_t+\tfrac{i}{\ell^2}(k-2\tilde\omega)a^{(1)}_r)}\,.
\ee
By a similar chain of reasoning in the $\rho$ sector we deduce that $t_{12}=0$ and that 
\be\label{eq:CorrelatorTwo}
t_{22}=\frac{\bar n b_t}{2\bar m( p^2 a_t+\tfrac{i}{\ell^2}(k+2\tilde\omega)a^{(1)}_r)}\,.
\ee
That $t_{22}(\omega,k)=t_{11}(\omega,-k)$ follows from the fact that the equations for the $v^{(s)}$ in the $\rho$
sector are exactly the same as those for the $u^{(s)}$ in the $\eta$ sector after taking $k\to -k$ and also that
we have $(\rho^++\rho^-)=(0,n)^T$ and $(\rho^+-\rho^-)=-(m,0)^T$ (note the minus sign).

To summarise, we can obtain $t_{11}$, which contains all of the independent information in the retarded Green's function, by solving
the $\eta$ sector gravitino equations for the $u^{(s)}$ given in \eqref{thegeneqnsG}, with ingoing boundary conditions, extracting
the asymptotic behaviour at the $AdS$ boundary and then substitute into \eqref{eq:CorrelatorOne}. It is worth emphasising the following point.
The equations \eqref{thegeneqnsG} immediately give the $\hat t$ and $\hat r$ components of the gravitino and hence
$a_{\hat t}$ and $a_{\hat r}$. However, to obtain the correct $b_{\hat t}$, one also needs to consider the  $\hat x$ and $\hat y$ components and ensure
that the conditions \eqref{bfconds} are satisfied.

\section{The Green's function for $T\ne 0$}\label{nonzeroT}
Having understood the boundary asymptotics of the $N=2$ gravitino, which is independent of whether the AdS-RN solution is 
extremal ($T=0$) or non-extremal ($T>0$) we must now address the near-horizon behaviour, which shows a dichotomy between the two cases due to the different singularity structure of the equations.
In both cases, in order to determine the retarded correlation function we want to specify ingoing boundary conditions. In this section we consider $T>0$, analysing the near-horizon behaviour and
then presenting the results of our numerical integrations for the Green's function $t_{11}$, focussing on the spectral function $A=$Im $t_{11}$.
We will consider $T=0$ in subsequent sections.

 \subsection{Near-horizon series}
At temperature $T\neq 0$ the function $f$ appearing in the AdS-RN black-brane metric \eqref{adsrn1} near the horizon has a simple zero:
\be
f(r) = 4 \pi T (r-r_+)+\dots 
\ee
Thus, near the horizon the tortoise coordinate is given by
\be
r_* = \frac{\log (r-r_+)}{4\pi T} \,.
\ee
Furthermore, the simple zero in $f$ means that the equations \reef{thegeneqnsG} have a {\it regular} singular point at $r=r_+$ and we can apply a version of Frobenius' method to the system of four coupled ODEs. This is done
most easily 
in the first-order formalism, where the independent solutions are
determined as the leading-order eigenvalues and eigenvectors of the
matrix multiplying the non-derivative terms in the coupled
equations. Concretely, we write the system of differential equations
\reef{thegeneqnsG} near the horizon as 
\be
\frac{d\mathbf{u}(\omega,k,r)}{dr} - \frac{1}{r-r_+}\mathsf{A}(\omega,k,r)\mathbf{u}(\omega,k,r) = 0\,,
\ee
where the function $\mathbf{u}\equiv(u^{(3/2)},u^{(1/2)},u^{(-1/2)},u^{(-3/2)})$ and $\mathsf{A}(\omega,k,r)$ is a $4\times 4$ matrix, one can show that we have a series solution in ascending powers of $(r-r_+)^{1/2}$, with
\be
\mathsf{A}=\sum_{n=0}^\infty \mathsf{A}_{n}(\omega,k) (r-r_+)^{n/2}\,,\qquad \mathbf{u} = \sum_{n=0}^\infty \mathbf{u}_{n}(\omega,k)(r-r_+)^{n/2}\,.
\ee
The leading-order result for the matrix $\mathsf{A}(\omega,k)$
reads
\be
\mathsf{A}_{0} = {\diag} \left(-\frac{3}{4} - \frac{i\omega}{4\pi T}\,,\, \frac{1}{4} + \frac{i\omega}{4\pi T} \,,\, \frac{1}{4} - \frac{i\omega}{4\pi T}\,, \, -\frac{3}{4} + \frac{i\omega}{4\pi T} \right)\,,
\ee
Being diagonal, we immediately obtain the indicial roots and the associated eigenvectors and hence
the four different solutions at the horizon.

Since ingoing solutions should behave like $e^{-i\omega(t+ r_*)}$ near the horizon, we deduce that we should impose the horizon boundary conditions
\bea\label{eq:horizonIN}
u^{(1/2)}\Bigr|_{r=r_+} = u^{(-3/2)}\Bigr|_{r=r_+} =0\quad\Leftrightarrow \quad u^{(1/2)}_0 = u^{(-3/2)}_0=0\,.
\eea
The two independent ingoing solutions then have leading-order behaviour
\bea\label{eq:finiteTnearhorizon}
u^{(s)} &=& (r-r_+)^{-\frac{s}{2}-\frac{i\frak{w}}{2}} \left( u^{(s)}_0  + {\cal O}\left((r-r_+)^{1/2}\right) \right)\,,\qquad s = \tfrac{3}{2},-\tfrac{1}{2}
\eea
where 
we have introduced the often-employed dimensionless measure of frequency at finite temperature:
\begin{align}
\frak{w}\equiv \frac{\omega}{2\pi T}\, ,
\end{align}
and the $u^{(s)}_0$
are  two arbitrary complex coefficients. Note that the near-horizon expansion is a series in ascending powers of $(r-r_+)^{1/2}$ and that these two free parameters determine the entire series expansion of $\mathbf{u}$ 
(all four
components are non-vanishing)
to arbitrarily high order in principle\footnote{And to very high order in practise, using computer algebra. In our numerical procedures (see appendix C) we expand to order $(r-r_+)^{15/2}$ to set the boundary conditions near the horizon.}. 
Gauge fixing restricts this to a single physical parameter in this sector as we shall see now. In particular, this parameter arising in the $\eta$ sector
of the gravitino and the corresponding parameter in the $\rho$ sector correspond to the two physical polarisation states
of the gravitino.

\subsection{Near-horizon SUSY}
We now consider the residual supersymmetry transformations \eqref{eq:SUSYvar_explicit},\reef{eq:explicit_residual}. Writing
${\bf w}=(w^{(1/2)},w^{(-1/2)})$ the equations \reef{eq:explicit_residual} have an expansion
\be
\frac{d\mathbf{\bf w}(\omega,k,r)}{dr} - \frac{1}{r-r_+}\mathsf{B}(\omega,k,r)\mathbf{\bf w}(\omega,k,r)  = 0\,,
\ee
where, as before
\be
\mathsf{B}=\sum_{n=0}^\infty \mathsf{B}_{n}(\omega,k) (r-r_+)^{n/2}\,,\qquad \mathbf{w} = \sum_{n=0}^\infty \mathbf{w}_{n}(\omega,k)(r-r_+)^{n/2}\,.
\ee
This time find the leading-order matrix
\be
\mathsf{B}_{0} = {\diag} \left( -\frac{1}{4} - \frac{i\omega}{4\pi T}\,,\,- \frac{1}{4} + \frac{i\omega}{4\pi T} \right)\,,
\ee
and thus, the ingoing gauge transformation satisfies the horizon boundary condition
\be
w^{(-1/2)}\Bigr|_{r=r_+} = 0\quad\Leftrightarrow \quad w^{(-1/2)}_0 =0\,.
\ee
We can then develop the most general ingoing gauge transformation parameters
as a series with leading form
\bea
w^{(1/2)} &=& (r-r_+)^{-\frac{1}{4}-\frac{i\frak{w}}{2}}\left( w_{0}^{(1/2)} + {\cal O}\left(  (r-r_+)^{1/2} \right)  \right)\,,
\eea
with the arbitrary complex number $w^{(1/2)}_0$ 
determining the entire series for $\mathbf{w}(\omega,k,r)$.
Plugging this into the SUSY variations \reef{eq:SUSYvar_explicit} finally gives the desired expressions for the leading ingoing pieces
\be
\delta u^{(s)} =(r-r_+)^{-\frac{s}{2}-\frac{i\frak{w}}{2}}\left( f^{(s)}(\frak{w}) + \cdots \right),\qquad s = \tfrac{3}{2}\,,-\tfrac{1}{2}
\ee
where
\bea
f^{(3/2)}(\frak{w}) &=& -w^{(1/2)}_0\sqrt{\pi T} \left(\tfrac{1}{2} + i \frak{w}  \right)\nn
f^{(-1/2)}(\frak{w}) &=& w^{(1/2)}_0\frac{1}{4r_+^2\sqrt{\pi T}}\frac{2 k^2 \ell^2-2 k \left(\ell^2 \mu +i  r_+\right)+\ell^2 \mu^2+4 i  r_+ (\mu +\omega )+r_+^2\ell^{-2}}{\frac{1}{2} - i \frak{w}}\,.\nn
\eea
From this calculation we can see that the two 
parameters defining the near-horizon ingoing solutions transform under the residual gauge transformations as
\bea
\delta u^{(s)}_0 = f^{(s)}(\frak{w})\,,
\eea
which can be used to set one of the $u^{(s)}$ 
to zero. Alternatively one can focus on a single gauge-invariant combination of initial conditions 
\be\label{eq:gaugeinvariantcombo}
u_0^{(-1/2)} - \frac{f^{(-1/2)}(\frak{w})}{f^{(3/2)}(\frak{w})}u_0^{(3/2)}\,.
\ee
In practise, for our numerics, we always choose a specific gauge, but have checked extensively that the resulting expressions extracted for the correlator are manifestly independent of gauge. This is not only reassuring but also gives a strong check on our numerical algorithms.

\subsection{Numerical results at $T>0$}
 \begin{figure}[t!]
\begin{center}
 a)\, \includegraphics[width=0.45\textwidth]{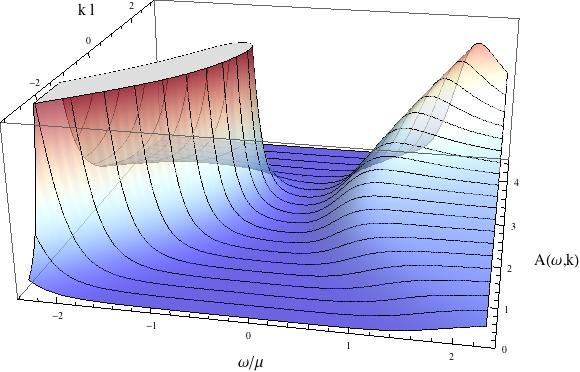} b)\, \includegraphics[width=0.45\textwidth]{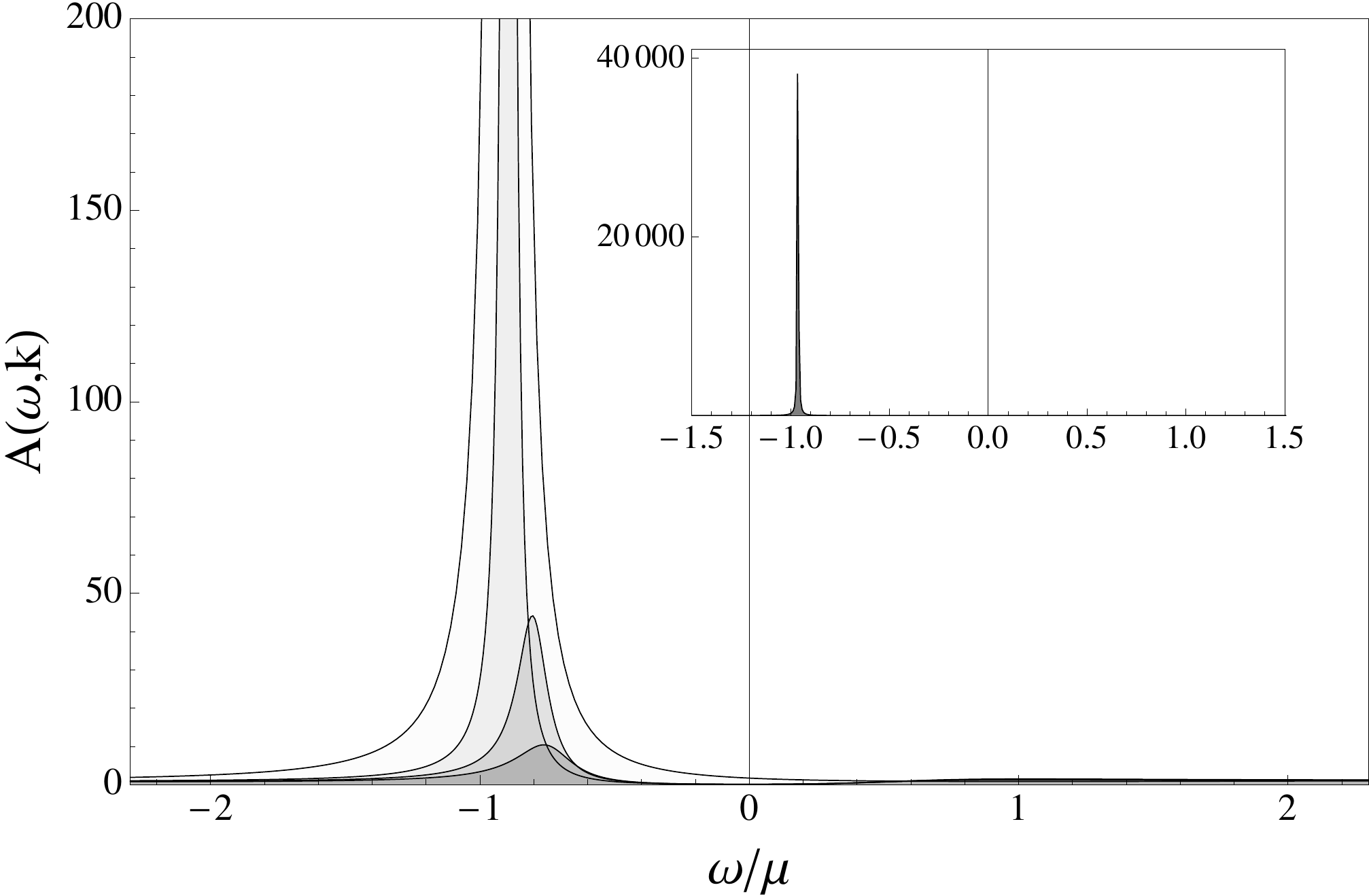}
\caption{\it Representation of spectral density $A(\omega,k)$ as a function of frequency and momentum at $T/\mu=0.04$. Panel a) shows a surface plot in the $(\omega,k)$ plane with $\omega$-slices taken in intervals of $\Delta(k\ell)=0.5$. Panel b) shows an alternative representation where a selection of such $\omega$ slices for positive $k\ell \in (0.1,1.1)$ is shown superimposed on one plot, with larger values of $kl$ in darker shades of grey. Furthermore the scale of the peak at $k\ell=0.1$ is shown separately in an inset. The gap between the two regions of high spectral density is clearly visible in the first panel (for more information see Fig. \ref{fig:ARPES1}).
\label{fig:ARPES0}}
\end{center}\end{figure}
\begin{figure}[t!]
\begin{center}
 a)\, \includegraphics[width=0.45\textwidth]{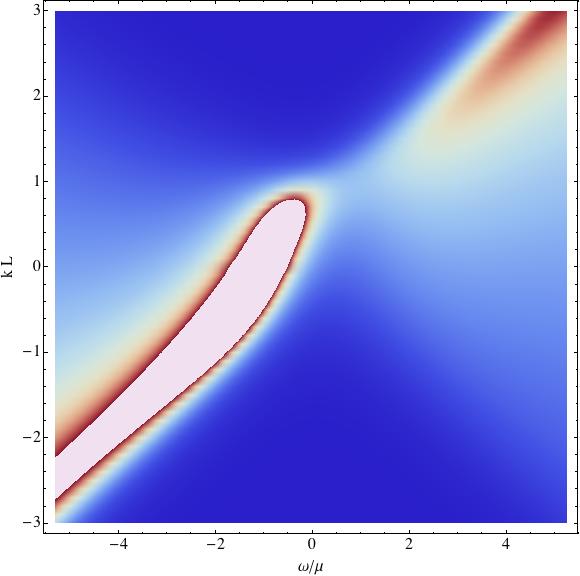}\hskip1em
 b)\,\includegraphics[width=0.45\textwidth]{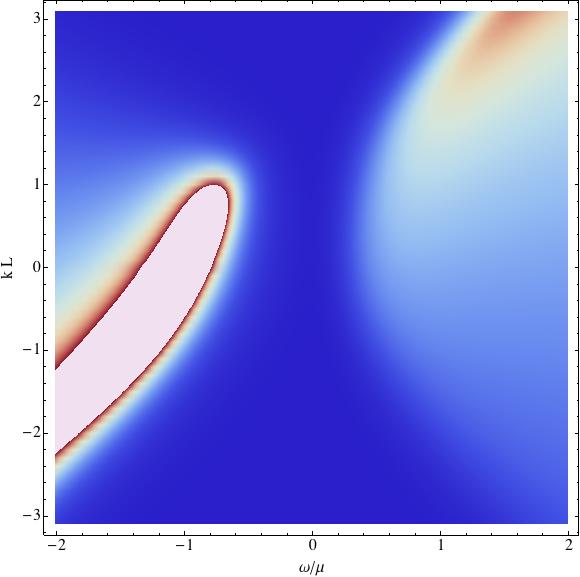} 
\vskip2em
 c) \,\includegraphics[width=0.5\textwidth]{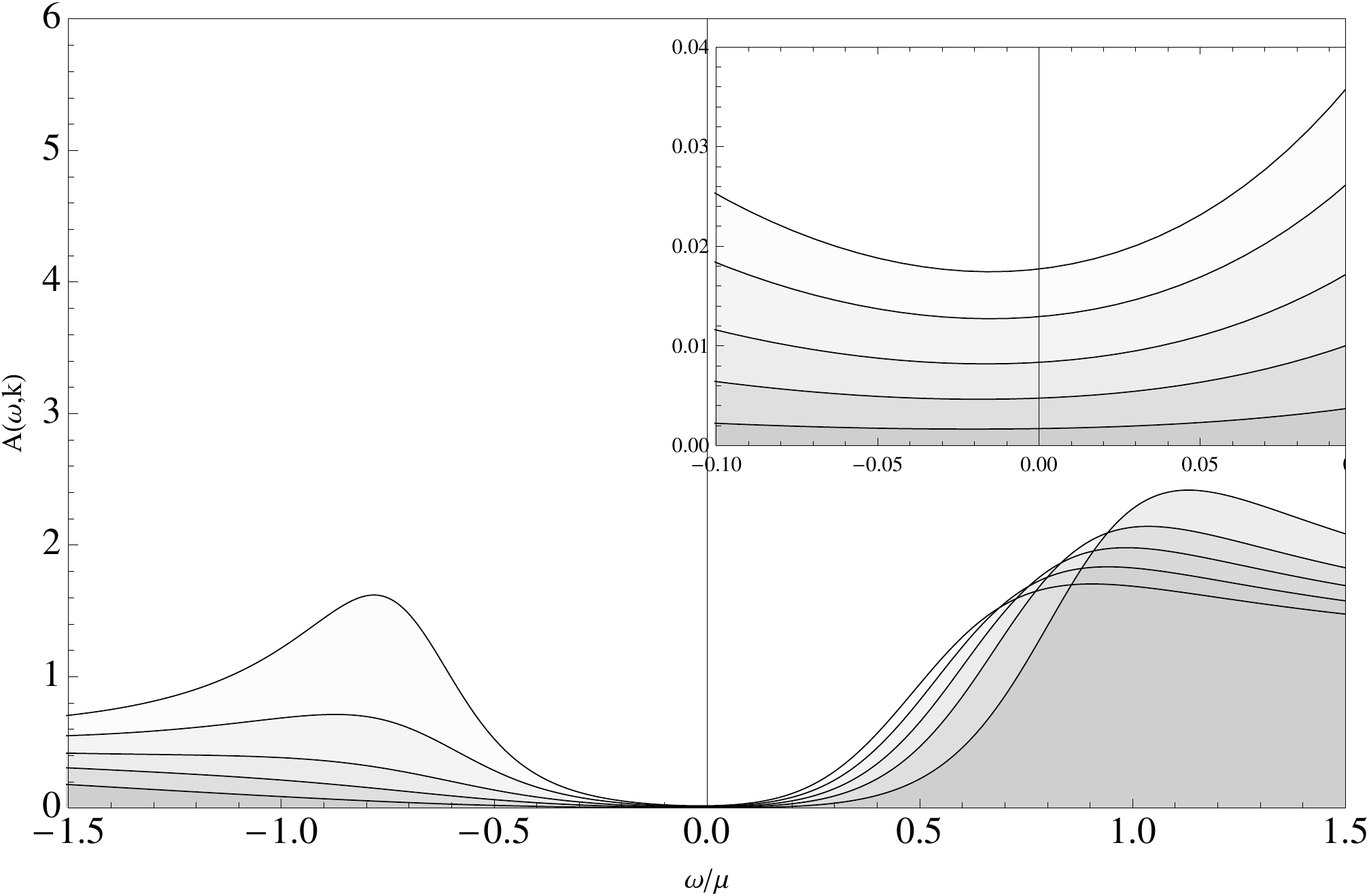}
\caption{\it Spectral density $A(\omega,k) $ for high and low temperatures
(in units of the chemical potential): $T/\mu = 0.37$ for panel a) and $T/\mu = 0.04$ for panels b) \& c).
In panels a) and b) values of low spectral density are in shades of blue, whereas regions of high density are in shades of red and the largest peak is cut-off and the cut-out region shown in light pink. At finite temperature the most prominent feature is the large spectral density region due to the phonino, or supersound pole, in the left bottom part of the $(\omega\,,\,k)$ plane. There is a second region of larger spectral density in the positive-frequency half plane and as the temperature is lowered a region of suppressed spectral weight of size ${\cal O}(\mu)$ opens up between these two regions,
that is approximately power-law. However, as illustrated in c), for momenta $k\ell \in (1.2,2.1)$, thermal excitations cause $A(\omega\ll \mu,k)$ to be nonzero at any $T\neq 0$ (see also zoomed region around origin in inset). However, $A(\omega,k)$ can be very small near the origin.\label{fig:ARPES1}}
\end{center}\end{figure}

Using the numerical procedure outlined in appendix \ref{app:numerics} we can now solve the gravitino equations
in the $\eta$ sector and then extract $t_{11}(\omega,k)$ from \eqref{eq:CorrelatorOne}. We are especially interested
in the spectral function $A(\omega,k)={\rm Im}\,t_{11}(\omega,k)$ which encodes 
information about the density of states at finite temperature. A surface plot of the spectral density is shown in Fig. \ref{fig:ARPES0} together with another representation of the type used in the companion paper \cite{Gauntlett:2011mf}. A collection of other
ARPES-like\footnote{ARPES, or angle resolved photon emission spectroscopy, is an experimental technique in which the charged (surface) excitations of a material are measured and represented as a density in the $(\omega,k)$ plane.}plots showing the spectral density $A(\omega,k)$ as a function of $\omega$ and $k$ are shown in Fig. \ref{fig:ARPES1}. Regions of high spectral density are displayed in `warmer' colours. 

The most striking feature on the $(\omega,k)$ plane is the region of high spectral weight around $k=0, \omega=-\mu$. This long-wavelength mode
is associated with the {\it phonino}, a collective fermionic excitation of any supersymmetric medium at finite temperature and density, that we discussed in section 2.2. The pole is at $k=0$ but frequency $\omega=-\mu$ due to the presence of a finite chemical potential, which arises as a weak gauging of the $R$-symmetry of the boundary theory.

There is a second region of high spectral weight, albeit suppressed with respect to the phonino
excitation, located at positive $k$. For all $k$ there is a depletion of the density of states near the origin, with $A(\omega,k)$ getting closer to zero around $\omega=0$ as the temperature is lowered. 
We will see in later sections that at strictly zero temperature, the spectral weight actually vanishes at the origin and that there is a power-law soft gap in the spectrum.
\subsubsection{Supersound propagation and diffusion}
Returning once more to the phonino, that is to small $k$ and small $\tilde\omega$, we assume that the retarded propagator takes the form dictated by the presence of a simple pole
located at $\tilde\omega=k=0$:
\be
t_{11}(\tilde\omega,k)\sim \frac{-Z(k,\mu,T)}{\tilde\omega - \Omega (k,\mu,T)+ i \Gamma(k,\mu,T)}
\ee
with $Z,\Omega,\Gamma$ real and $Z$ and $\Gamma$ positive at $k=0$.
A straightforward analysis of $A(\tilde\omega,k) = {\rm Im}\, t_{11}(\tilde\omega,k)$ shows that the 
spectral peak at $\tilde\omega = \Omega$ has height $Z/\Gamma$ and that
the full width at half maximum is $\Delta\tilde\omega = 2 \Gamma$.

Given the above analysis, our numerical results allow us to fit a dispersion relation, encoding the location of the maximum of the peak as well as its broadening (its full width at half maximum), of the form
\be\label{drone}
\tilde{\frak{w}} = v_s {\frak{q}} - i \Sigma \left( T, \mu \right){\frak{q}}^2+ \cdots
\ee
where $\frak{q}=k/2\pi T$ 
and to the given accuracy 
we 
find\footnote{Note that there are two sources of error here. Firstly numerical, but secondly also the error due to the fact that the dispersion itself is only true in the sense of a power series for small values of $\tilde{\frak{w}}$ and $\frak{q}$.
} the propagation speed
\be\label{drtwo}
v_s \sim 0.50\,,
\ee
and we shall return to the diffusion coefficient presently, after giving an analytic argument for the constant speed of propagation $v_s$.
Indeed our fitted value for $v_s$ is consistent with the value $v_s=1/2$ that is expected for conformally invariant
field theories \cite{Kovtun:2003vj}. In fact we can obtain some further insight as follows.
The normalisable phonino mode should have $A_\alpha=0$ in the asymptotic expansion of the gravitino.
In an expansion in small values of $\tilde\omega$ and $k$ we find that at leading order there is a solution
to the gravitino equations with
$B_{\hat{\alpha}}= C_{\hat\alpha\hat\beta}\Gamma^{\hat{\beta}}\epsilon$, where $\epsilon$ is a constant spinor and 
the diagonal matrix $C$ was given in \eqref{defofcmat}. This can be viewed as a large gauge transformation and 
satisfies $\Gamma^{\hat{\alpha}}B_{\hat{\alpha}}=0$. The condition
$p^{\hat{\alpha}}B_{\hat{\alpha}}=0$ then becomes the Dirac equation $(2\tilde\omega\Gamma^{\hat t}+k\Gamma^{\hat x})\epsilon=0$
with phase velocity $v_s=1/2$. 

Now returning to the diffusion coefficient, we note that in the absence of a chemical potential, simple dimensional analysis would dictate that $\Sigma = {\rm const}$. One then defines the diffusion constant $D_s= 2 \pi T \Sigma$, governing the broadening of the peak in $\tilde \omega,k$ space. Since in the present context we have a non-zero chemical potential, dimensional analysis is no longer sufficient to ensure that $D_s/(2\pi T)$ is a constant, but rather it allows for it to be a non-trivial function of the dimensionless ratio of temperature and chemical potential, $\frac{D_s(T/\mu)}{2\pi T}$.

\begin{figure}[t!]
\begin{center}
 \includegraphics[width=0.7\textwidth]{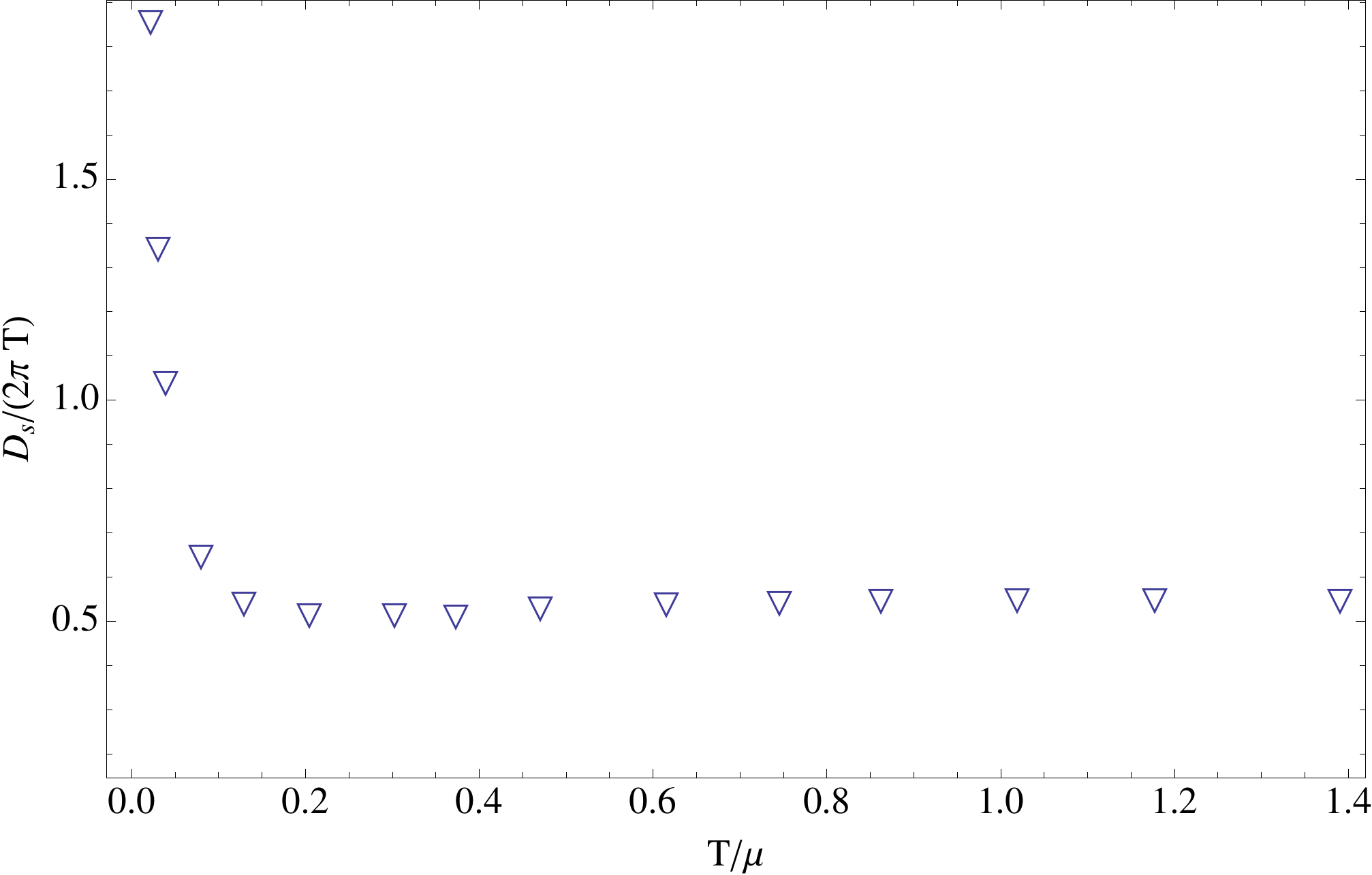}
\caption{\it Supersound diffusion rate. For high values of $T/\mu$ the diffusion coefficient approaches a constant, 
which is expected since in this limit it should approach the $\mu=0$ form and by dimensional analysis must be a constant.\label{fig:diffcoeff}}
\end{center}\end{figure}

Our extensive numerical data allows us to determine $D_s$ for a number of values of $T/\mu$ and we show the results in Fig \ref{fig:diffcoeff}. Note that it approaches a constant in the appropriate limit $T\gg \mu$, where the zero chemical potential result should apply. It would be interesting to carry out a more exhaustive analysis, which is however, beyond the scope of the present work. Finally, it is worthwhile to point out that the same dimensional analysis argument would in principle allow for the propagation speed $v_s$ to be a non-trivial function of $T/\mu$, but the additional requirement of conformal symmetry, which is at heart of the brief analytical argument we gave above, ensures that it is actually a constant, in complete agreement with our numerical results.

We should compare the results of this section to the work of \cite{Policastro:2008cx} which used an analytical expansion at small frequency and momentum, 
but zero chemical potential, to extract the analogous information in five bulk dimensions where it was found that $v_s=1/3$ and 
$\Sigma = 4\sqrt{2}/9$.  Note, that due to the absence of finite $\mu$ the diffusion coefficient is a constant, as expected from dimensional analysis.

\section{Extremal $T=0$ case}\label{zeroT}
 \begin{figure}[h!]
\begin{center}
 a)\, \includegraphics[width=0.45\textwidth]{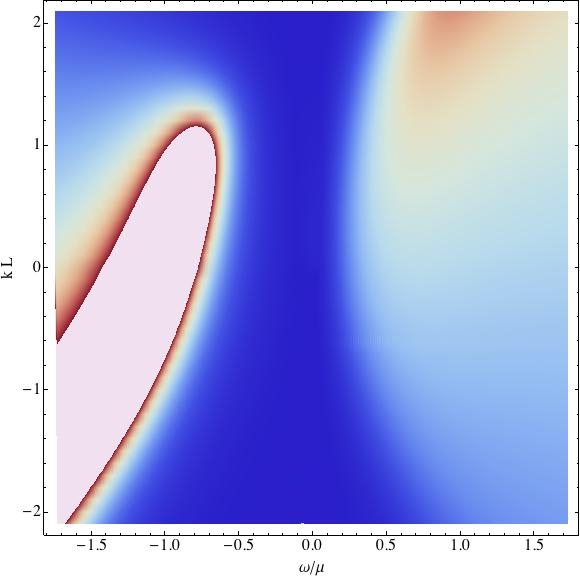}\\
\vskip2em
 b)\,\includegraphics[width=0.45\textwidth]{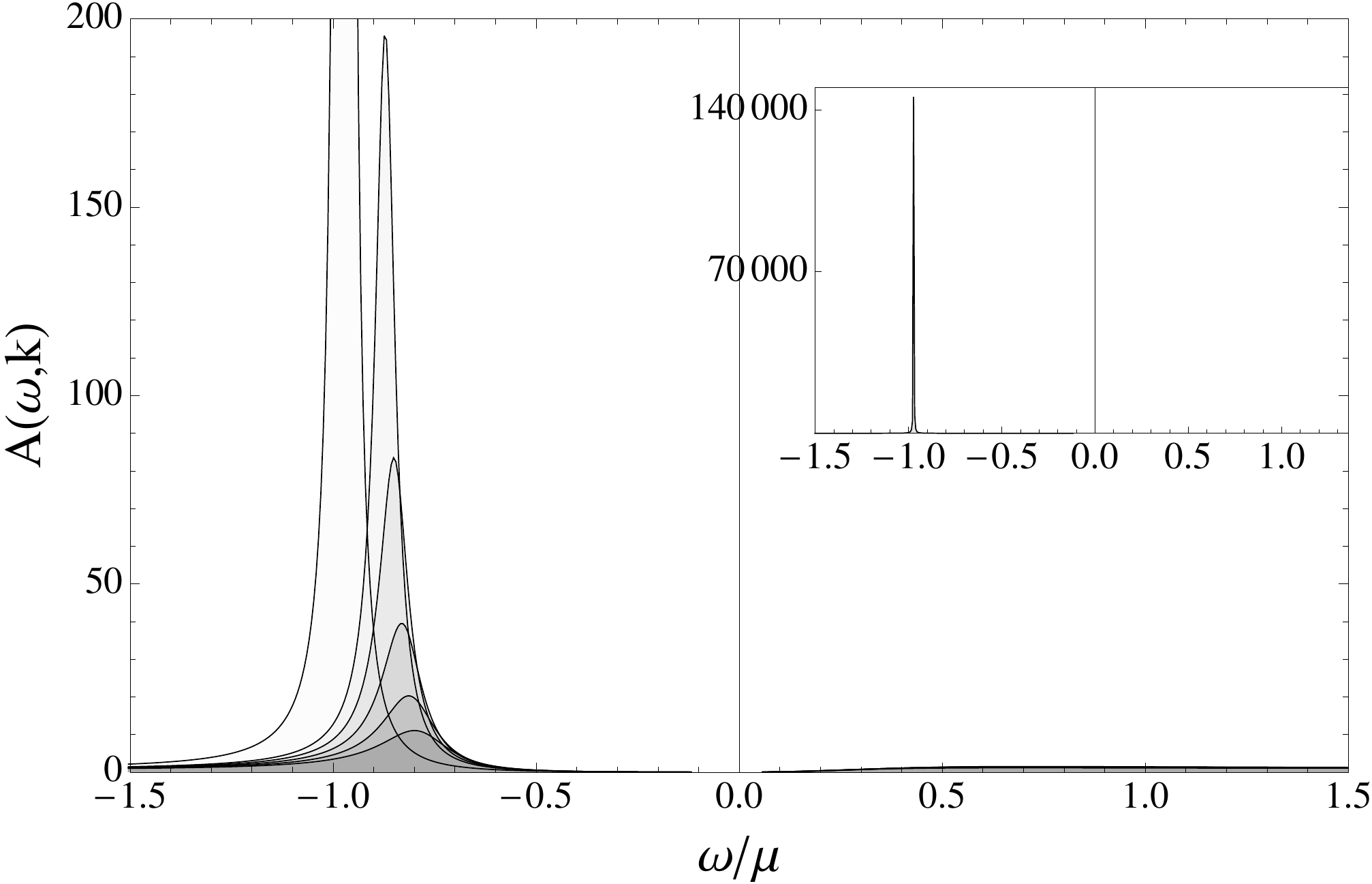}\hskip1em c)\,\includegraphics[width=0.45\textwidth]{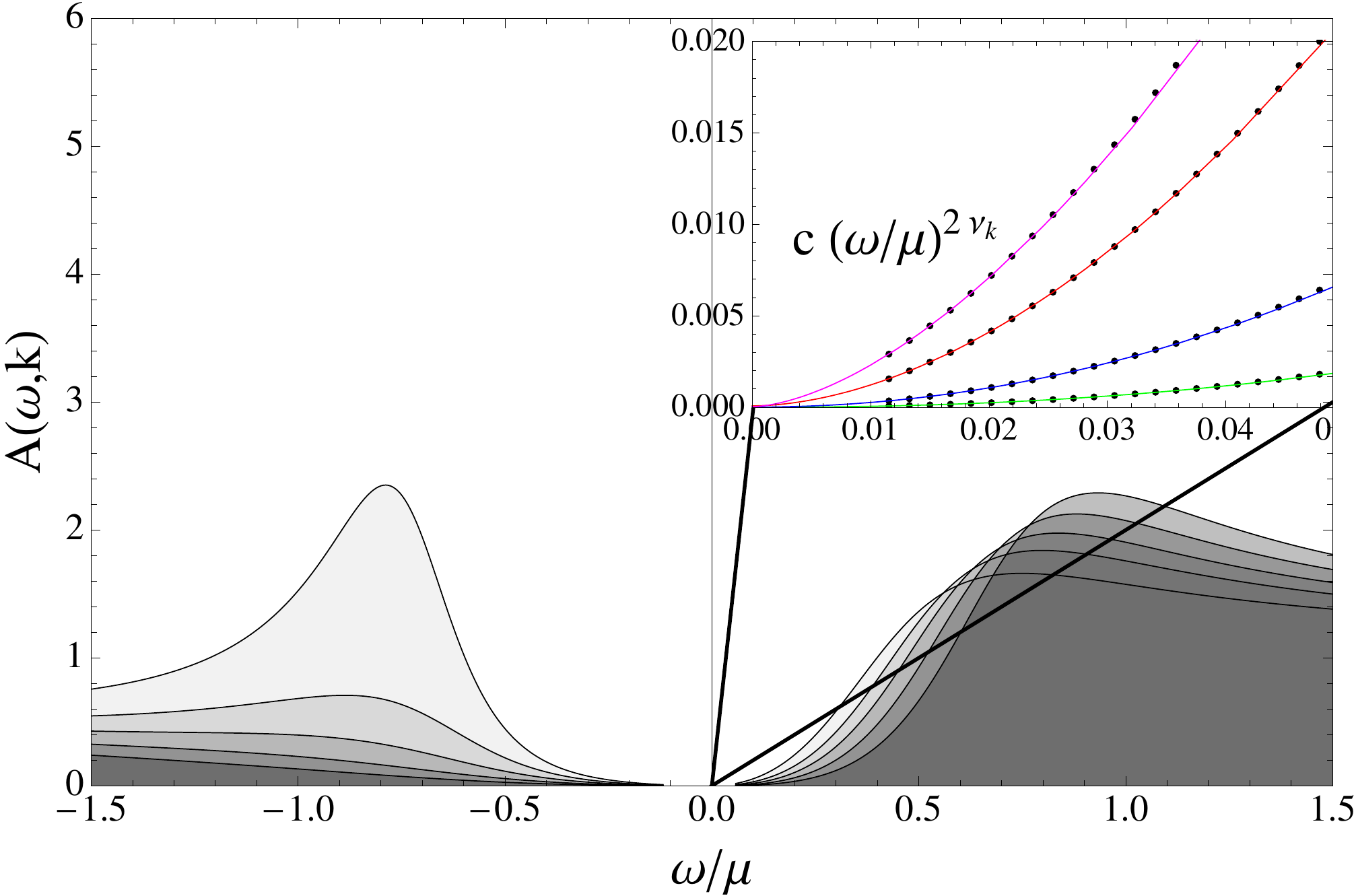}
\caption{\it a) Spectral density $A(\omega,k) $ at $T=0$. As before, the largest peak associated with the phonino pole, is cut-off and the cut-out region shown in pink to distinguish it from the white areas of the plots. Values of low spectral density are in shades of blue, whereas regions of high density are in shades of red. b) The spectral density at discrete values of $k\ell \in (0.1,1.1),$ with increasing values of momentum in darker shades of grey. c) For higher values of $k\ell\in(1.2,2.1)$, as is already apparent from panel a), the bump at positive frequency becomes the dominant feature of the spectrum. The inset highlights the emergence of a power-law gap at zero frequency by fitting the numerics close to the origin to the analytic result $\propto\omega^{2\nu_k}$, where as shown in section \ref{sec:LeadingOrder}, $\nu_k=\sqrt{\frac{7}{12} + \frac{k^2}{2\mu^2}    }$.
\label{fig:ARPES2}}
\end{center}\end{figure}
\subsection{Near horizon series}
At zero temperature the horizon at $r=r_+$ is an irregular\footnote{At zero frequency this irregular behaviour is absent and we can develop an ordinary Frobenius series. We can sidestep this subtlety for now, but will return to it in section \ref{sec:ZeroFreq} where it will allow us a valuable analytical insight into the spectral function at zero frequency.} singular point of the gravitino equations, arising from the confluence of the two regular singular points at $r_\pm$. However, we are still able to develop a systematic expansion by factoring out the leading essentially singular behaviour. The remaining equations then, again, allow a Frobenius series expansion, as we describe now. The extremal limit of the function $f$ appearing in the metric \reef{eq:fextreme} has a double zero at the horizon
\be
f \sim L_{(2)}^2 (r-r_+)^2 +\cdots
\ee
which gives rise to a confluent singularity in the gravitino equation. The tortoise coordinate near the horizon is 
given by
\be
r_* = -\frac{L_{(2)}^2}{r-r_+}\,.
\ee
After defining the quantity 
\be
\Delta(\omega) =\frac{q\ell}{2\sqrt{3}} + \frac{4 L_{(2)}^2\omega}{3 r_+}  =\frac{1}{2\sqrt{3}}+\frac{2}{3\sqrt{3}}\frac{\omega}{\mu}\,,
\ee
we find that there are four independent solutions, specified by
the four complex coefficients $u^{(s)}_0$, with
the following leading behaviour
\bea\label{eq:ZeroTLeading}
u^{(s)} = e^{\frac{i L_{(2)}^2\omega}{r-r_+}} (r-r_+)^{-s-i\Delta(\omega)} \left(u_0^{(s)} + {\cal O}(r-r_+) \right),\qquad s = \tfrac{3}{2},-\tfrac{1}{2}\nn
u^{(s)} = e^{-\frac{i L_{(2)}^2\omega}{r-r_+}} (r-r_+)^{s+i\Delta(\omega)} \left(u^{(s)}_{0} + {\cal O}(r-r_+)  \right),\qquad s= \tfrac{1}{2},-\tfrac{3}{2}\,.
\eea
Note that the leading essentially singular behaviour is factored out. Also
note that, in contrast to the finite-temperature solution, the series solutions proceed in ascending powers of $(r-r_+)$ (rather than its square root as we saw
for $T>0$).
Since the first two solutions behave in terms of the tortoise coordinate as $e^{-i\omega r_*}$, we identify them as ingoing at the horizon, whereas the latter two are outgoing. Hence, also at zero temperature we impose the boundary conditions 
$u^{(1/2)}_0=u^{(-3/2)}_0=0$, as in \reef{eq:horizonIN}, leaving two solutions.

The parameter of the residual gauge transformations can be similarly factored, allowing us two write down two independent solutions, specified by the complex parameters $w^{(\pm1/2)}_0$, with leading behaviour
\bea
w^{(1/2)}  &=& e^{\frac{i L_{(2)}^2\omega}{r-r_+}} (r-r_+)^{-\frac{1}{2} - i \Delta(\omega)} (w^{(1/2)}_0 + {\cal O}(r-r_+))\,, \nn
w^{(-1/2)}  &=& e^{\frac{-i L_{(2)}^2\omega}{r-r_+}} (r-r_+)^{-\frac{1}{2} + i \Delta(\omega)} (w^{(-1/2)}_0 + {\cal O}(r-r_+))\,.
\eea
We thus impose $w^{(-1/2)}_0=0$ as the ingoing horizon boundary condition.
Substituting the resulting expansion into the gauge variation equations \reef{resdual} gives us 
\be
\delta u^{(s)} = e^{-i\omega r_*}g^{(s)}(\omega) w_0^{(1/2)}(r-r_+)^{-s - i\Delta(\omega)}\left( 1 + {\cal O}(r-r_+) \right)\,,
\ee
for
$s = \tfrac{1}{2},-\tfrac{3}{2}\,,$
where the $g^{(s)}(\omega)$ are helicity-dependent constants, the precise form of which will not be needed. 
Again, we see that by virtue of the residual gauge transformation we can 
set either of $u^{(3/2)}_0=0$ or $u^{(-1/2)}_0=0$.  As at finite temperature there is a gauge-invariant combination that follows from the action of the gauge transformation on the horizon parameters, but since its exact form, the exact analog of \reef{eq:gaugeinvariantcombo}, is not needed and not very illuminating we do not present it here.
Thus, as expected, gauge fixing restricts us to a single physical parameter in the $\eta$ sector
and, similarly, a single parameter in the $\rho$ sector and these correspond to the two physical polarisation states
of the gravitino.

\subsection{Numerical results at $T=0$}

Using the numerical procedure outlined in appendix \ref{app:numerics} we can now solve the gravitino equations and extract $t_{11}(\omega,k)$ using \eqref{eq:CorrelatorOne}. Recall, once more, that knowing $t_{11}(\omega,k)$ for positive and negative values of the momentum also gives us $t_{22}(\omega,k)$.
A plot of the spectral density $A(\omega,k)$=Im $t_{11}(\omega,k)$ is shown in Fig. \ref{fig:ARPES2}.

We find a structure that is very similar to the finite temperature results, as one might have expected by continuity. The phonino pole is still responsible for the dominant feature on the $(\omega,k)$ plane, although of course its interpretation in terms of hydrodynamics is rather subtle at zero temperature. A discussion of this issue would go beyond the scope of this paper. Again, there is a second region of high spectral weight at larger positive values of momentum, separated from the phonino peak by a power-law gap. Interestingly, at zero temperature, the spectral density actually vanishes at $\omega=0$ exhibiting a characteristic power-law $\propto\omega^{2\nu_k}$, with $\nu_k=\sqrt{\frac{7}{12} + \frac{k^2}{2\mu^2}}$, for small values of frequency, as shown clearly in Fig. \ref{fig:ARPES2}c. As the exact expression for $\nu_k$ suggests, we can in fact derive this power law analytically as a low-frequency expansion of the retarded correlator. This expansion makes heavy use of the exact solution of the gravitino equations in the near-horizon geometry of the extremal limit, $AdS_2\times \mathbb{R}^2$, of the AdSRN black-brane, which we will turn to now.

\section{Analytic solution in the $AdS_2\times \mathbb{R}^2$ background}\label{ads2appr}
In this section we pause to solve the gravitino equations exactly in the $AdS_2\times \mathbb{R}^2$ background.
In the next section we will use these results to analyse the non-analytic behaviour of the spectral function
in the extremal AdS-RN black-brane background for small $\omega$.

The $AdS_2\times \mathbb{R}^2$ background is given by
\be
ds^2=\frac{L^2_{(2)}}{z^2}\left[-dt^2+dz^2\right]+\frac{r_+^2}{\ell^2} dx^i dx^i,\qquad {\cal A}=\frac{L_{(2)}}{\sqrt{2}z}dt\,,
\ee
and we note that the boundary of the $AdS_2$ is at $z=0$ and the Poincar\'e horizon is at $z=\infty$, 
where $z = L_{(2)}^2/(r-r_+)$.
In the frame 
\be
e^\pm=\frac{L_{(2)}}{z}\left[-dz\pm dt\right]\,,
\ee
we find $\Omega_\pm=\mp 1/(4L_{(2)})$, ${\cal A}_\pm=\pm1/(2\sqrt{2})$, $f=-1/(2\sqrt{2}L_{(2)})$, and
hence the gravitino equations (in the $\eta$ sector) \eqref{thegeneqns} read
\bea\label{uads2eqs}
A_{-\frac{3}{2}}^- \uth -K^+_{\mu_k}\uoh &=&0\,, \nn
%\nn
A_{\frac{1}{2}}^+ \uoh -K^+_{\bar\mu_k}\uth &=&0\,,\nn
%\nn
A_{\frac{1}{2}}^- \umoh -K^-_{\mu_k}\umth &=&0\,, \nn
%\nn
A_{-\frac{3}{2}}^+\umth -K^-_{\bar\mu_k}\umoh&=&0\,,
\eea
where we have defined the differential operators
\be
A_n^\pm :=z\partial_z + n \pm i\left[  \omega z   + \frac{qL_{(2)}}{\sqrt{2}}   \right]\,
\ee
and also
\be
K_{\mu_k}^\pm = L_{(2)}\mu_k \pm  \frac{i\lambda}{\sqrt{2} },\qquad \qquad \mu_k= m + \frac{ik\ell}{r_+}\,.
\ee

Notice that in \eqref{uads2eqs} the Pauli coupling essentially contributes a shift of the momentum (this was also observed in 
bottom-up models for spin 1/2 fermions in \cite{Edalati:2010ww,Edalati:2010ge}).
Similarly, the residual supersymmetry transformations \eqref{resdual} read
\bea\label{susytranrs}
\delta\uth&=&-\frac{1}{2L_{(2)}}A^+_{\frac{1}{2}}\woh\,,\nn
\delta\uoh&=&-\frac{1}{2L_{(2)}}\left[A^+_{-\frac{1}{2}}\wmoh-L_{(2)}\left(m-\frac{i}{\sqrt{2} L_{(2)}}\right)\woh\right]\,,\nn
\delta\umoh&=&-\frac{1}{2L_{(2)}}\left[A^-_{-\frac{1}{2}}\woh-L_{(2)}\left(m+\frac{i}{\sqrt{2} L_{(2)}}\right)\wmoh\right]\,,\nn
\delta\umth&=&-\frac{1}{2L_{(2)}}A^-_{\frac{1}{2}}\wmoh\,,
\eea
where,
from \eqref{residual}, the residual gauge transformation parameters satisfy
\bea\label{diraceqrs}
A^-_{-\frac{1}{2}}\woh-L_{(2)}\left(2m+\frac{ik\ell}{r_+}\right)\wmoh&=&0\,,\nn
A^+_{-\frac{1}{2}}\wmoh-L_{(2)}\left(2m-\frac{ik\ell}{r_+}\right)\woh&=&0\,.
\eea

From \eqref{uads2eqs} we can derive the following second-order equations for the component functions 
\bea\label{secordeqs}
\left(A^+_{\frac{1}{2}} A^-_{-\frac{3}{2}} - K_{\mu_k}^+ K_{\bar\mu_k}^+\right) \uth &=& 0\,,\nn
\left(A^-_{-\frac{3}{2}}A^+_{\frac{1}{2}}  - K_{\mu_k}^+ K_{\bar\mu_k}^+\right) \uoh &=&0\,,\nn
\left(A^+_{-\frac{3}{2}} A^-_{\frac{1}{2}} - K_{\mu_k}^- K_{\bar\mu_k}^- \right)\umoh&=& 0\,,\nn
\left(A^-_{\frac{1}{2}}A^+_{-\frac{3}{2}}  - K_{\mu_k}^- K_{\bar\mu_k}^-\right) \umth &=&0\,.
\eea
These equations can be solved exactly. Before doing so, let us first analyse the asymptotic behaviour
near the $AdS_2$ boundary $z\rightarrow 0$:
\bea\label{expansionnh}
\uth,\uoh &\sim& A z^{\frac{1}{2} - \nu_+} \left( 1 + {\cal O}(z)\right) + B z^{\frac{1}{2} + \nu_+} \left( 1 + {\cal O}(z)  \right)\,,\nn
\umoh,\umth &\sim&A z^{\frac{1}{2} - \nu_-} \left( 1 + {\cal O}(z)\right) + B z^{\frac{1}{2} + \nu_-} \left( 1 + {\cal O}(z)  \right)\,,
\eea
where
$\nu_\pm = \left[1 + L_{(2)}^2 \left( |\mu_k|^2 - \frac{1}{2}q^2
   \right) \pm i\sqrt{2}L_{(2)}(q+m\lambda) -
   \frac{1}{2}\lambda^2\right]^{1/2}$. 
It is interesting to observe that the exponents $\nu_\pm$ always have imaginary parts, unless $q=-m\lambda$, which is precisely the case for the supersymmetric $N=2$ theory we are interested in. Indeed in
this case the two exponents in fact coincide: $\nu_\pm=\nu_k$ with
\be
\nu_k = \sqrt{\frac{1}{2} + L_{(2)}^2 \left( |\mu_k|^2 - \frac{q^2}{2} \right)    }=
\sqrt{\frac{7}{12} + \frac{k^2}{2\mu^2}    }\,.
\ee
The expansion \eqref{expansionnh} implies that the scaling dimension of the operators, labelled by the spatial momentum $k$, 
in the dual one-dimensional CFT, is given by $1/2+\nu_k$. 

We now see that there is no possibility for the log-periodic 
behaviour that was seen in the bottom-up models for spin-$\frac{1}{2}$in \cite{Liu:2009dm}, 
since this feature only arises when the IR conformal dimensions are not real. 
The reality of $\nu_k$ is due to the charge to mass ratio in minimal $N=2$ supergravity and this can be viewed as being 
uniquely fixed by the structure of the 
embedding in string/M-theory. 
From a physical point of view the same ratio also means that pair production of charged Fermions is not energetically favourable
in the constant electric field in the $AdS_2$ region \cite{Faulkner:2009wj}. This suggests
that the system does not exhibit an instability towards forming an electron star \cite{Hartnoll:2009ns,Hartnoll:2010gu,Cubrovic:2011xm}.
Note that in \eqref{expansionnh} we have labeled the
source behaviour with the letter `$A$' and the expectation value with `$B$'. 

We can in fact write down the general solutions of the equations \eqref{secordeqs} the asymptotics of which
we just explored. The second-order equations all follow the template
\be
\left[{z^2}\partial^2_z  +   \frac{1 - 4 \nu_k^2}{4}  -2i\alpha(s)\omega z + (w z)^2 \right]u^{(s)}=0\,,
\ee
where
\bea
\alpha(s)= s + \frac{iqL_{(2)}}{\sqrt{2}}\,.
\eea
This is Whittaker's equation and there are two linearly independent solutions which, for each $u^{(s)}$, 
we can group into
\be\label{eq:Whittaker}
a_{\rm IN} W_{\alpha,\nu_k}\left( -2i\omega z \right)+ a_{\rm OUT} W_{-\alpha,\nu_k}\left(2i\omega z \right)\,,
\ee
where $W_{\alpha,\nu_k}$ is Whittaker's confluent hypergeometric function.
The first term in \reef{eq:Whittaker} corresponds to a purely ingoing solution, whereas the second term gives the purely 
outgoing one\footnote{Note that in the extreme AdS-RN black-brane, near the horizon at $r\sim r_+$ the tortoise coordinate 
is given by $r_* \sim -\frac{L_{(2)}^2}{r-r+} \sim -z$. Thus the ingoing  boundary condition is
$e^{-i\omega (t +r_*)} \sim e^{-i\omega(t - z)}$.}. This can be seen from the asymptotic behaviour for large $x$
\be
W_{\alpha,\nu_k}(i x) \sim e^{-\frac{ix}{2}}x^{\alpha}(1 + \cdots)\,.
\ee
It is now easy to write down the ingoing and outgoing solutions. As we
are most interested in the retarded Green's function we will only show the ingoing solutions, for which 
$u^{(s)} = a^{(s)}W_{\alpha(s),\nu_k}\left( -2i\omega z \right)$.
The first-order equations \eqref{uads2eqs} will now relate the constant coefficients $a^{(s)}$, so that only two free complex parameters remain. 
Using the identities
\bea\label{whitrec}
t \frac{d}{dt} W_{\alpha\,,\nu_k}(t) &=& \left(\tfrac{1}{2}t - {\alpha}  \right)W_{\alpha,\nu_k}(t) -  W_{\alpha + 1,\nu_k}(t)\,,\nn
&=&-\left(\tfrac{1}{2}t-\alpha\right) W_{\alpha,\nu_k}(t) -
\left[\nu_k^2-\left(\alpha-\tfrac{1}{2}\right)^2\right]W_{\alpha - 1,\nu_k}(t)\,, 
\eea
we find
$a^{(1/2)} = - a^{(3/2)} K_{\bar \mu_k}^+$, $a^{-(3/2)} = - a^{(-1/2)}  K_{\bar\mu_k}^-$, and so we can now write
\bea\label{finalinm}
\uth&=&a^{(3/2)}W_{\alpha(3/2),\nu_k}\left(-2i\omega z\right)\,,\nn
\uoh&=&-a^{(3/2)}K^+_{\bar\mu_k}W_{\alpha(1/2),\nu_k}\left(-2i\omega z\right)\,, \nn
\umoh&=&a^{(-1/2)}
W_{\alpha(-1/2),\nu_k}\left(-2i\omega z\right)\,,\nn
\umth&=&-a^{(-1/2)}K^-_{\bar\mu_k}W_{\alpha(-3/2),\nu_k}\left(-2i\omega z\right)\,.
\eea
Analogous expressions hold in the $\rho$-sector. After substitution in
\eqref{chidef} and then in \eqref{psidef} we obtain the ingoing
behaviour of the gravitino in the near-horizon $AdS_2\times
\mathbb{R}^2$ region. 

Finally we consider the residual supersymmetry transformations. We find that we can
again solve \eqref{diraceqrs} using ingoing Whittaker-type functions 
\bea
\woh&=&2d_1 W_{\alpha(1/2),\nu_k}\left(-2i\omega z\right)\,,\nn
\wmoh&=&2d_2 W_{\alpha(-1/2),\nu_k}\left(-2i\omega z\right)\,,
\eea
where $c_i$ are constants satisfying
\be
d_2=L_{(2)}\left(\frac{ik\ell}{r_+}-2m\right)d_1\,.
\ee
After substituting into \eqref{susytranrs}, and also using the supersymmetry values for $q,m,\lambda$, we obtain
\bea\label{susytranrs2}
\delta\uth&=&\frac{d_1}{L_{(2)}}W_{\alpha(3/2),\nu_k}\left(-2i\omega z\right)\,,\nn
\delta\uoh&=&-\frac{d_1}{L_{(2)}}K^+_{\bar\mu_k}W_{\alpha(1/2),\nu_k}\left(-2i\omega z\right)\,, \nn
\delta\umoh&=&-\frac{d_2}{L_{(2)}}K^-_{\mu_k}
W_{\alpha(-1/2),\nu_k}\left(-2i\omega z\right)\,,\nn
\delta\umth&=&\frac{d_2}{L_{(2)}}K^-_{\mu_k}K^-_{\bar\mu_k}W_{\alpha(-3/2),\nu_k}\left(-2i\omega z\right)\,.
\eea
Comparing with \eqref{finalinm} we see that the residual supersymmetry transformations can be used
to set either $\uth=\uoh=0$ or alternatively $\umoh=\umth=0$.

By expanding the solutions \eqref{finalinm} near the boundary of the $AdS_2\times\mathbb{R}^2$ region we can extract the
retarded Green's function of the dual one-dimensional SCFT. Specifically, introducing the constant vector-spinors
\bea\label{vecspin}
s_1 &=& e^+\otimes \eta^+ - \frac{K_{\bar\mu}^+}{\frac{1}{2} - \alpha(3/2) + \nu_k}e^+\otimes \eta^- \,,  \nn
s_2 &=& e^+\otimes \eta^+ - \frac{K_{\bar\mu}^+}{\frac{1}{2} - \alpha(3/2) - \nu_k}e^+\otimes \eta^-\,,  \nn
s_3 &=& e^-\otimes \eta^+ - \frac{K_{\bar\mu}^-}{\frac{1}{2} - \alpha(-1/2) + \nu_k}e^-\otimes \eta^- \,,  \nn
s_4 &=& e^-\otimes \eta^+ - \frac{K_{\bar\mu}^-}{\frac{1}{2} - \alpha(-1/2) - \nu_k}e^-\otimes \eta^- \,,
\eea
we find that after substituting \eqref{finalinm} into \eqref{chidef}, in the limit that $z\to 0$, we have
\bea\label{exactsolsh}
\chi^\eta&\sim& \bar a^{(3/2)} \left[ \left(\frac{z}{L_{(2)}}\right)^{\frac{1}{2}-\nu_k}s_{1} + {\cal G}_R(\omega,\alpha(3/2),\nu_k) \left(\frac{z}{L_{(2)}}\right)^{\frac{1}{2} + \nu_k} s_{2}\right]\nn
&+& \bar a^{(-1/2)}\left[ \left(\frac{z}{L_{(2)}}\right)^{\frac{1}{2}-\nu_k}s_3 + {\cal G}_R(\omega,\alpha(-1/2),\nu_k)   \left(\frac{z}{L_{(2)}}\right)^{\frac{1}{2}+\nu_k}s_4\right]\,,
\eea
and the bars on $a^{(3/2)}$, $a^{(-1/2)}$ indicate that we have absorbed some overall factors.
In these expressions
\bea
{\cal G}_R(\omega,\alpha(s),\nu_k) &=& e^{-i\pi\nu_k}\frac{\Gamma(-2\nu_k)}{\Gamma(2\nu_k)} \frac{\Gamma(\frac{1}{2} - \alpha(s) +\nu_k)}{\Gamma(\frac{1}{2} - \alpha(s) - \nu_k)} \left( 2\omega L_{(2)} \right)^{2\nu_k}\,,
\eea
define, up to normalisation, the retarded Green functions of the emergent IR CFT for each helicity state, that is each one-dimensional representation of Spin$(1,1)$ labeled by $s$. We show more details on this computation, as well as its cousin with advanced boundary conditions,
in appendix \ref{app:asymptotics}.
Note that the residual supersymmetry allows us to choose either the gauge $\bar a^{(3/2)}=0$ or $\bar a^{(-1/2)} = 0 $ if we so wish.

\section{Matched asymptotic expansion}\label{matching}

In this section we derive the leading-order result of a
small-frequency expansion of the retarded correlation function at zero 
temperature and finite chemical potential. This leads to an explicit
expression for the scaling properties of the spectral function at small
frequency. The analysis is very similar to the one first used for
spinors in \cite{Faulkner:2009wj}, but we face additional
complications stemming from gauge invariance as well as the
multi-component nature of the gravitino. Our treatment follows closely
that of \cite{Faulkner:2009wj}, but our emphasis differs in some
respects.

We start by discussing carefully the near-horizon limit in which the
analytic solution in $AdS_2\times\bbR^2$ is applicable, and then how
this can be matched to the leading small-frequency solution at large
$r$.  

\subsection{Scaling limits and the small $\omega$ expansion}

When $T=0$ the only physical parameter in the bulk solution is the chemical
potential, given by $\mu\ell=\sqrt{3}r_+/\ell$. Naively the solutions in
the $AdS_2\times\bbR^2$ background given in section~\ref{ads2appr} are
applicable close to the horizon, as measured relative to the chemical
potential, that is for $(r-r_+)/r_+\ll 1$. This is 
the regime in which the higher-order terms in $f$ and $\phi$
in~\eqref{eq:fextreme} and~\eqref{adsrn2} can be dropped. However,
this limit requires some care since the term $\omega/f$ in
\eqref{thegeneqnsG} diverges faster at small $(r-r_+)$ than the other
terms. To get the precise $AdS_2\times\bbR^2$
equations~\eqref{uads2eqs} one must also take a small-frequency limit
where $\omega/\mu\ll 1$.

To see this explicitly, let us define the new coordinates
\be
\label{eq:zeta-def}
\zeta = \frac{\epsilon\, L_{(2)}^2}{r-r_+}\,,\qquad \tau = \epsilon \,t
\ee
so that taking the limit $\epsilon\rightarrow 0$, keeping $\zeta$ and
$\tau$ fixed, of the metric and gauge field \reef{adsrn1} gives the
exact $AdS_2\times \mathbb{R}^2$ near-horizon geometry. In other words 
\be
ds^2 = \frac{L_{(2)}^2}{\zeta^2} \left(  -d\tau^2 + d\zeta^2\right) + \frac{r_+}{\ell^2} dx^i dx^i + {\cal O}(\epsilon) \,,\qquad {\cal  A} = \frac{\sqrt{3}L_{(2)}}{2 \zeta}d\tau + {\cal O}(\epsilon)\,.
\ee
We now perform an expansion of the gravitino equations \reef{thegeneqnsG} in the parameter $\epsilon$ to find that (at zero $T$) we get the set of equations
\bea\label{eq:inner}
\frac{i\omega\zeta}{\epsilon}u^{(3/2)}+\left[\zeta\partial_\zeta - \frac{3}{2} - \frac{iqL_{(2)}}{\sqrt{2}}  \right]u^{(3/2)} - K_{\mu_k}^+ u^{(1/2)} + {\cal O}(\epsilon)&=&0\,,\nn
\frac{i\omega\zeta}{\epsilon}u^{(1/2)} +\left[\zeta\partial_\zeta +\frac{1}{2} + \frac{iqL_{(2)}}{\sqrt{2}}  \right]u^{(3/2)} - K_{\bar\mu_k}^+ u^{(3/2)} + {\cal O}(\epsilon)&=&0\,,\nn
\frac{i\omega\zeta}{\epsilon}u^{(-1/2)}+\left[\zeta\partial_\zeta +\frac{1}{2} - \frac{iqL_{(2)}}{\sqrt{2}}  \right] u^{(-1/2)} - K_{\mu_k}^- u^{(-3/2)} + {\cal O}(\epsilon)&=&0\,,\nn
\frac{i\omega\zeta}{\epsilon}u^{(-3/2)}+\left[\zeta\partial_\zeta -\frac{3}{2} + \frac{iqL_{(2)}}{\sqrt{2}}  \right] u^{(-3/2)} - K_{\bar\mu_k}^- u^{(-1/2)} + {\cal O}(\epsilon)&=&0\,.
\eea
The frequency $\omega$ appearing here is the Fourier conjugate to the
time coordinate $t$, but in order to obtain a regular near-horizon
limit, we were forced to scale $\epsilon t = \tau$ and thus the
relevant frequency for the near-horizon region is instead $\omega^\tau
= \omega/\epsilon$. With this identification the leading-order
behaviour is actually ${\cal O}(\epsilon^0)$. The equations at leading
order then precisely reproduce the $AdS_2\times \mathbb{R}^2$ problem
\reef{uads2eqs} we solved above, now written in terms of $\zeta$ and
$\tau$ (or rather its Fourier conjugate $\omega^\tau$). 

Hence to get the well-defined set of equations for finite
$\omega^\tau$ and $\zeta$, we want to take the limit 
\be
\omega\ell\rightarrow 0\,\qquad\epsilon\rightarrow 0\,,\qquad {\rm with}\qquad\,\omega^\tau, \zeta \quad{
\rm finite}\,.
\ee
Recall that the corrections to the metric function $f$ and potential
$\phi$ are small provided $(r-r_+)/r_+\ll 1$. Given the coordinate 
redefinition~\eqref{eq:zeta-def}, the limit is thus a good approximation
provided 
\begin{equation}
   0 < \frac{r-r_+}{\ell} < \epsilon , \qquad
   \omega\ell < \epsilon , 
\end{equation}
with the size of $\epsilon\ll\mu\ell$, set by the chemical potential. 

In the language of matched asymptotic expansions~\cite{Liu:2009dm},
this $AdS_2\times\bbR^2$ limit describes the \emph{inner} region for a
perturbative expansion of the differential equations in small
$\omega$. It is subtle because we are considering a range of $r$ where
the $\omega/f$ term is diverging faster than the other terms in
\eqref{thegeneqnsG}, and so cannot simply be dropped even though
$\omega$ is small. The \emph{outer} region is considerably
simpler. It is given by the range of $r$ where the $\omega/f$ term
can be safely ignored relative to the other terms. Since $1/f$
diverges as $\ell/(r-r_+)$ relative to the other terms, the outer
region is defined by  
\begin{equation}
   \epsilon' < \frac{r-r_+}{\ell} < \infty\,,
\end{equation}
with $\epsilon'\gg\omega\ell$. In each region we can expand the
solution as a power series in $\omega\ell$
\begin{equation}
\begin{aligned}
   \psi_I &= \psi_I^{(0)}(r) + (\omega\ell) \psi_I^{(1)}(r) + \cdots , \\
   \psi_O &= \psi_O^{(0)}(r) + (\omega\ell) \psi_O^{(1)}(r) + \cdots ,
\end{aligned}
\end{equation}
where the subscripts ``I'' and ``O'' denote inner and outer regions. 

In this case the matched expansion is very straightforward.  If we
take $\omega\ell$ small enough we can simply make the two regions
overlap and match the solutions. Specifically, by taking a
double-scaling limit  
\begin{equation}
   \omega\ell \ll \epsilon' < \epsilon \ll \mu\ell , 
\end{equation}
both inner and outer solutions will be valid in the region
$\epsilon'<(r-r_+)/\ell<\epsilon$. 
Let us now turn to the resulting leading-order matched solution.

\subsection{Leading-order solution}
\label{sec:LeadingOrder}

By comparing \reef{eq:inner} with \reef{uads2eqs} and \reef{finalinm}
it is clear that the leading-order (in $\omega\ell$) ingoing solution in
the inner region takes the form 
\be\label{eq:innerleading}
u^{(s)}_I\left(z;\omega,k\right) \propto 
W_{\alpha(s),\nu_k}\left( -2i\omega z\right)\,,
\ee
with the exact coefficients of the individual components as given in
\reef{finalinm} and where we are using the coordinate
$z=L_{(2)}^2/(r-r_+)=\ell^2/6(r-r_+)$. In the matching region we have 
\begin{equation}
   \epsilon' < \ell/6z < \epsilon \, , \qquad 
   \omega\ell \ll\epsilon' \, , 
\end{equation}
and hence $\omega z\ll 1$. Thus we can use the asymptotic, near-boundary
expansions \reef{vecspin} from the $AdS_2$ analysis of section
\ref{ads2appr}. 
Explicitly we have 
\begin{equation}
\label{eq:inner-asymp}
\begin{aligned}
   \chi^\eta_I &\sim \bar a^{(3/2)} \left[ 
       \left(\frac{L_{(2)}^2}{r-r_+}\right)^{\frac{1}{2}-\nu_k} s_1 
       + {\cal G}_R(\omega,\alpha(\tfrac{3}{2}),\nu_k) 
       \left(\frac{L_{(2)}^2}{r-r_+}\right)^{\frac{1}{2}+\nu_k} s_2
       \right] \\ & \qquad
    + \bar a^{(-1/2)} \left[ 
       \left(\frac{L_{(2)}^2}{r-r_+}\right)^{\frac{1}{2}-\nu_k} s_3 
       + {\cal G}_R(\omega,\alpha(-\tfrac{1}{2}),\nu_k) 
       \left(\frac{L_{(2)}^2}{r-r_+}\right)^{\frac{1}{2}+\nu_k} s_4
       \right] \\ & \qquad 
    + \mathcal{O}(\omega\ell) \, . 
\end{aligned}
\end{equation}
Note that ${\cal G}_R(\omega,\alpha(s),\nu_k)$ are non-analytic in
$\omega$. In addition, we could have fixed the gauge, e.g. by setting
$\bar{a}^{(3/2)}=0$, but we prefer, for the time being, not to do so.   

The leading-order outer solution is obtained by solving, at least
formally, Eqs. \reef{thegeneqnsG} {\it at} $\omega=0$. As was already
observed in \cite{Liu:2009dm} for spin $1/2$ fermions, at zero temperature and strictly zero
frequency the essentially singular nature of the horizon is absent
(the problematic terms are proportional to $\omega$) and we can
develop an ordinary Frobenius series. The leading terms in this
expansion give the approximate outer solution in the overlap
region. Prior to gauge fixing there are four independent solutions and
these can be fixed by their asymptotic behaviour near the horizon. For this purpose we
can define $\hat{\chi}_{(i)}$ with $i=1,2,3,4$ such that for
$\epsilon'<(r-r_+)/\ell<\epsilon\ll\mu\ell$ we have 
\bea\label{behavnh}
\hat{\chi}_{(1),(3)}(r,k) \sim \left( \frac{L_{(2)}^2}{r-r_+}
\right)^{\frac{1}{2} - \nu_k}s_{1,3},\qquad 
\hat{\chi}_{(2),(4)}(r,k) \sim \left( \frac{L_{(2)}^2}{r-r_+}
\right)^{\frac{1}{2} + \nu_k}s_{2,4}\,. 
\eea
As expected, we see that this behaviour is identical to the
small-argument asymptotics of the inner region. Matching to the inner
solution fixes the coefficient of each $\hat{\chi}_{(i)}$ component and
we have 
\begin{equation}
\label{leadingorder}
\begin{aligned}
   \chi^\eta_O(r,\omega,k) 
      &= \bar{a}^{(3/2)}\Big[ \hat{\chi}_{(1)}(r,k)
          + {\cal G}_R(\omega,\alpha(\tfrac{3}{2}),\nu_k)
              \hat{\chi}_{(2)}(r,k) \Big] 
      \\ & \qquad 
      + \bar{a}^{(-1/2)} \Big[ \hat{\chi}_{(3)}(r,k)
          + {\cal G}_R(\omega,\alpha(-\tfrac{1}{2}),\nu_k)
          \hat{\chi}_{(4)}(r,k) \Big]
      + \mathcal{O}(\omega\ell) . 
\end{aligned}
\end{equation}
More generally, recalling the asymptotic expansion of the Whittaker functions
\eqref{whittexp}, we can extract the
non-analytic $\omega$-dependence of the outer solution by writing
\begin{equation}\label{eq:FullMatched}
   \chi^\eta_O(r,\omega,k) 
      = \chi^\eta_{(1)}(r,\omega,k)
          + {\cal G}(\omega,\nu_k) \chi^\eta_{(2)}(r,\omega,k) \, . 
\end{equation}
where $\chi^\eta_{(i)}$ are {\it analytic} in $\omega$, with
\begin{equation}
\label{eq:outer-pair}
\begin{aligned}
   \chi^\eta_{(1)}(r,\omega,k)
      &= \bar{a}^{(3/2)}\hat{\chi}_{(1)}(r,k)
          + \bar{a}^{(-1/2)} \hat{\chi}_{(3)}(r,k) +{\cal O}(\omega \ell), \\
   \chi^\eta_{(2)}(r,\omega,k)
      &= \bar{a}^{(3/2)}\hat{\chi}_{(2)}(r,k)
          + \bar{a}^{(-1/2)} 
             \frac{\big(\nu_k-\frac{i}{2\sqrt{3}}\big)
                  \big(\nu_k-\frac{i}{2\sqrt{3}}-1\big)}
                {\big(\nu_k+\frac{i}{2\sqrt{3}}\big)
                  \big(\nu_k+\frac{i}{2\sqrt{3}}+1\big)}
                \hat{\chi}_{(4)}(r,k) +{\cal O}(\omega \ell) \, , 
\end{aligned}
\end{equation}
and we have defined
\begin{equation}\label{eq:GR-def}
\begin{aligned}
   {\cal G}(\omega,\nu_k) 
      &\equiv {\cal G}_R(\omega,\alpha(\tfrac{3}{2}),\nu_k) , \\
      &= e^{-i\pi\nu_k}\frac{\Gamma(-2\nu_k)}{\Gamma(2\nu_k)}
          \frac{\Gamma(-1-\frac{i}{2\sqrt 3}+\nu_k)}
              {\Gamma(-1-\frac{i}{2\sqrt 3} - \nu_k)} 
              \left( 2\omega L_{(2)} \right)^{2\nu_k}\,.
\end{aligned}
\end{equation}
Note that our choice in factoring out ${\cal G}(\omega,\nu_k)$ is
simply one possible option corresponding to ${\cal
  G}_R(\omega,\alpha(s),\nu_k)$ with $s=3/2$. Any other choice of
$s$, related to this by the familiar recursion identities of the Euler 
gamma function, would be equally valid. Furthermore all choices lead
to the same non-analytic $\omega$ dependence 
$\propto\left(\omega L_{(2)}\right)^{2\nu}$ of the end result.

We now have all the ingredients to write down the answer for
$t_{11}(\omega,k)$. To leading order in the expansion we have
\begin{equation}
\label{eq:MatchedProp}
   t_{11}(\omega,k)
      = - \frac{i\bar{e}^{(2)}_\alpha \left( b_{(1)}^{\alpha,0} 
              + {\cal G}(\omega,\nu_k) b_{(2)}^{\alpha,0}
              +\mathcal{O}(\omega\ell) \right)}
           {2p^2\bar{e}^{(1)}_\beta \left( a_{(1)}^{\beta,0}
              + {\cal G}(\omega,\nu_k) a_{(2)}^{\beta,0}
              +\mathcal{O}(\omega\ell) \right)} \, , 
\end{equation}
and of course the analogous expression for $t_{22}(\omega,k)$. 
In
these expressions the quantities $b_{(i)}^{\alpha,n}$ and $a_{(i)}^{\beta,n}$
are the various near-boundary data (c.f.~\eqref{otherversiont11})
of the asymptotic expansion of the two
solutions~\eqref{eq:outer-pair} at $n$-th order in the $\omega\ell$ perturbative expansion. We have no analytic expressions for
these coefficients as functions of $k$, but they can be determined
numerically from the outer region equations.

\subsection{Scaling behaviour of spectral function at zero frequency}
\label{sec:ZeroFreq}

The immediate utility of the exact expressions obtained in the last
subsection is that we can use them to give a simple
analytic argument for the scaling properties of the spectral function
near the origin.
We shall establish the scaling property by combining the action of a
discrete symmetry with the specific form of the horizon boundary
condition at zero frequency and zero temperature.

The discrete symmetry of interest is the three-dimensional
inversion $(t,x,y)\to(-t,-x,-y)$. This can be viewed as the
combination of time-reversal symmetry $T$ and the rotation
$R:(t,x,y)\to(t,-x,-y)$ already discussed in section
\ref{sec:corrfn}. Here we will focus on the action of these symmetries
on the bulk gravitino field.

The action of time-reversal is
\begin{equation}
   T:\qquad \psi_\mu(t,\mathbf{x},r) \mapsto
      \psi^{(T)}_\mu(t,\mathbf{x},r)
      = -{\cal B}_T\,{\cal P}_\mu{}^\nu \psi_\nu(-t,\mathbf{x},r)^*\,,
\end{equation}
where ${\cal P}=diag(1,-1,-1,-1)$ is the parity inversion matrix in
4d and ${\cal B}_T$ induces the transformations $\Gamma^{\hat t}
\mapsto {\cal B}^{-1}_T\Gamma^{\hat t}{\cal B}_T=-\Gamma^{\hat t}$
while leaving all other gamma matrices untouched. The matrix
$\mathcal{B}_T$ is uniquely defined up to an arbitrary phase, which
here we fix so that ${\cal B}_T = \Gamma^{\hat r\hat x\hat y}$. The complex
conjugation is included in line with the usual action for solutions of
complex wave equations which include a coupling to a $U(1)$ field. It
ensures that the bulk gravitino equation is time-reversal covariant
given the usual map of the gauge potential
$\mathcal{A}_\mu(t,\mathbf{x},r)\mapsto
\mathcal{P}_\mu{}^\nu\mathcal{A}_\nu(-t,\mathbf{x},r)$ under $T$. From the
boundary field theory perspective, where $\psi_\mu$ is related to the
expectation value of $S_\alpha$, it is a consequence of the anti-linearity of
the time-reversal operator. 

Recall that the black-hole background is static and purely electrically charged
(only $\mathcal{A}_{\hat t}$
non-zero) and hence is invariant under time-reversal . Hence $T$ should be
a symmetry of the bulk gravitino equations of motion. To see this,
note that  the corresponding action on
$\chi_{\hat{\mu}}(t,r,\mathbf{k})$ defined in \eqref{psidef} is  
\begin{equation}
   T : \begin{pmatrix} 
          \chi_{\hat{t}}(t,r,\mathbf{k}) \\
          \chi_{\hat{r}}(t,r,\mathbf{k})
       \end{pmatrix}
       \mapsto
       \begin{pmatrix} 
          -\Gamma^{\hat r\hat x\hat y}
             \chi_{\hat{t}}(-t,r,-\mathbf{k})^* \\
          \Gamma^{\hat r\hat x\hat y}
             \chi_{\hat{r}}(-t,r,-\mathbf{k})^* 
       \end{pmatrix} \, .
\end{equation}
Expanding in the basis~\eqref{chidef}, this leads to
\begin{equation}
\label{Tuv}
   T : \begin{cases} 
      u^{(s)}(t,r,\mathbf{k}) \mapsto  v^{(-s)}(-t,r,-\mathbf{k})^* 
      \\ 
      v^{(s)}(t,r,\mathbf{k}) \mapsto - u^{(-s)}(-t,r,-\mathbf{k})^* 
   \end{cases}\, . 
\end{equation}
Given time-dependence of the form $e^{-i\omega t}$, the frequency
$\omega$ is unchanged under this map, and we see that \eqref{Tuv} is
manifestly a symmetry of the bulk equations \eqref{thegeneqnsG}.   

The action of the rotation $R$ on the gravitino can similarly be
written as 
\be
R:\qquad \psi_\mu(t,\mathbf{x},r) \mapsto
\psi^{(R)}_\mu(t,\mathbf{x},r) = \Gamma^{\hat x\hat y} 
   {\cal R}_\mu{}^\nu \psi_{\nu}(t,-\mathbf{x},r)\,.
\ee
Here ${\cal R}$ is the rotation matrix 
${\cal R}=diag(1,-1,-1,1)$ and the matrix $\Gamma^{\hat x\hat y}$ effects the appropriate
rotation on the spinor indices. Acting on the components we have 
\begin{equation}
\label{Ruv}
   R : \begin{cases}
      u^{(s)}(t,r,\mathbf{k}) \mapsto -v^{(s)}(t,r,-\mathbf{k}) 
      \\
      v^{(s)}(t,r,\mathbf{k}) \mapsto  u^{(s)}(t,r,-\mathbf{k}) 
      \end{cases} \, . 
\end{equation}
Combining the two symmetries we see that the action of
three-dimensional inversion
\begin{equation}
\label{eq:TRdef}
   TR:\qquad \psi_\mu(t,\mathbf{x},r) 
      \mapsto \psi^{(TR)}_\mu(t,\mathbf{x},r) 
      =- \Gamma^{\hat r} {\cal I}_\mu{}^\nu 
         \psi_{\nu}(-t,-\mathbf{x},r)^* \,          ,
\end{equation}
where $\mathcal{I}=diag(-1,-1,-1,1)$, induces the transformation
\begin{equation}
\label{TRuv}
   TR : \begin{cases} 
      u^{(s)}(t,r,\mathbf{k}) \mapsto u^{(-s)}(-t,r,\mathbf{k})^* 
      \\ 
      v^{(s)}(t,r,\mathbf{k}) \mapsto v^{(-s)}(-t,r,\mathbf{k})^* 
   \end{cases}\, . 
\end{equation}
This is the $u^{(s)}\to (u^{(-s)})^*$ symmetry of the equations
\eqref{thegeneqnsG} that we mentioned earlier.

Although TR is a symmetry on the bulk equations we note that this
operation induces a symmetry of the boundary correlator $t_{ij}$ only
if the boundary conditions also respect the symmetry. The action on
the three-dimensional boundary theory can be derived from the
embedding of the $d=3$ Clifford algebra in the $d=4$ one via the
decomposition \reef{eq:threeDfourD} and the decomposition of the
asymptotic expansion variables $A_{\hat\alpha}\equiv(0,a_\alpha)$,
$B_{\hat\alpha}\equiv(b_\alpha,0)$ described in
section \ref{sec:GreenFun}. Given~\eqref{eq:TRdef} we find 
\begin{equation}\label{eq:TRaction}
   TR : \begin{cases}
           a_\alpha(p) \mapsto -a_\alpha^*(p) \\
           b_\alpha(p) \mapsto  b_\alpha^*(p) 
        \end{cases} \, . 
\end{equation}
Hence, given the definition \eqref{baexp} and since the basis
vector-spinors $e^{(i)}_\alpha$ are real, we have the action on the
correlation function 
\be
   t_{11}(\omega,k)\mapsto t_{11}(\omega,k)^*\, ,
\ee
and similarly for $t_{22}$. 

Let us now examine the action of the TR transformation on the
horizon boundary condition. We will argue that at zero temperature, as
the frequency goes to zero, it becomes a symmetry. This limit is 
subtle because of the irregular singular point at the horizon. However
here we can use our small $\omega$ analysis of the previous section. We saw
that, for small $\omega$, ingoing boundary conditions imply that
solution in the $AdS_2\times\bbR^2$ region can be written in terms of
Whittaker functions as in \eqref{eq:innerleading}. In the overlap
region, the corresponding near boundary expansion is given in
\eqref{eq:inner-asymp}. In this region it is then straightforward to
take the zero $\omega$ limit since
$\mathcal{G}_R(\omega,\alpha(s),\nu_k)\sim (\omega
L_{(2)})^{2\nu_k}\to 0$ for all $s$. In the limit we thus have 
\be\label{eq:horizonBoundaryCondition}
\chi^\eta(r) = (r-r_+)^{-\frac{1}{2} + \nu_k}\big( 
   u_{0}^{(3/2)} s_1 + u_0^{(-1/2)} s_3 \big) + \dots\,,
\ee
where the constant vector-spinors $s_1$ and $s_3$ are defined in
\reef{vecspin} and have the form 
\begin{equation}
\begin{aligned}
   s_1 &= e^+\otimes \eta^+ - c_1 e^+ \otimes \eta^- \, , &&&&&
   s_3 &= e^-\otimes \eta^+ - c_2 e^- \otimes \eta^- \, ,
\end{aligned}
\end{equation}
where $c_1 c_2^*=1$. Note that for the strictly $\omega=0$ equations of
motion, the essentially singular nature of the horizon is absent and
there are four independent solutions~\eqref{behavnh}. What the
argument above implies is that at zero frequency we should retain only
solutions $\hat{\eta}_{(1)}$ and $\hat{\eta}_{(3)}$ that are regular at the
horizon (see also \cite{Gauntlett:2011mf,Belliard:2011qq}).
(Note that
this is not the case for the zero-frequency limit of the $T>0$
solutions discussed in section \ref{nonzeroT}.) The gauge
parameter similarly has two types of solutions $\varepsilon \sim
(r-r_+)^{-\frac{1}{2} \pm \nu_k}$ and, by a similar argument, we discard the
lower sign, keeping the regular solution. Feeding the expansion through
\reef{resdual} to obtain the gauge variations of the gravitino we find
that we can gauge away either of $ u_{0}^{(3/2)}$ or $u_{0}^{(-1/2)}$
completely.   

Given the action \eqref{TRuv} we have, under the inversion map, 
\begin{equation}
\begin{aligned}
   s_1^{(TR)} &= e^-\otimes \eta^- - c^*_1 e^- \otimes \eta^+
      = -\frac{1}{c_2} s_3 \, , \\
   s_3^{(TR)} &= e^+\otimes \eta^- - c^*_2 e^+ \otimes \eta^+ 
      = -\frac{1}{c_1} s_1 \, ,
\end{aligned}
\end{equation}
Hence, in the overlap region, 
\begin{equation}
\label{eq:horizonBoundaryCondition2}
   \chi^{\eta(TR)}(r) = - (r-r_+)^{-\frac{1}{2} + \nu_k}\left[ 
       \frac{u_0^{(-1/2)\,*}}{c_1} s_1 
       + \frac{u_0^{(3/2)\,*}}{c_2} s_3 \right] + \dots\,,
\end{equation}
which is the same form as the original solution. Thus the TR symmetry
respects the ingoing boundary conditions in the zero $\omega$
limit. This fact relies crucially on the conjugation property of
$c_{1,2}$. More precisely one can show that the TR-transformed
solution is, up to linear rescalings, gauge equivalent to the original
solution \reef{eq:horizonBoundaryCondition}. Since the
Rarita-Schwinger equation is a linear equation, equivalence up to
linear rescalings is sufficient to ensure equality of the correlation
functions at infinity. The numerator and denominator of
\reef{eq:CorrelatorOne} and \reef{eq:CorrelatorTwo} scale
homogeneously with such linear factors and so they cancel in the final
result. 

We have thus shown that the TR action is a symmetry of the
system of equations {\it and} the horizon boundary conditions at zero
temperature, in the limit $\omega=0$. This implies the condition
\be
t_{11}(0,k) = t_{11}(0,k)^*
\ee
on the retarded propagator of the boundary theory. An equivalent
argument, involving the $\rho$ sector equations can be made for
$t_{22}(0,k)$. Thus, at zero frequency,
the retarded Green's function
is purely real and hence the spectral function $A(0,k) =
{\rm Im} \, t_{11}(0,k)$ vanishes.

In fact we can make a more general statement about the reality of certain expansion coefficients in the retarded Green's function \reef{eq:MatchedProp}. Since the boundary condition at higher orders in perturbation theory is fully determined by the boundary condition at order zero (consider the expansion of \reef{finalinm}, or alternatively \reef{eq:FullMatched}, in powers of $\omega\ell$; also  see \cite{Faulkner:2009wj}), we deduce that if the latter is TR invariant, so is the former. Note now that the arguments above also show that the quantities $a_{\hat t,(1)}^n$ and $b_{\hat t,(1)}^n$ at $n$-th order in perturbation theory of the outer region, obeying the boundary conditions specified in \reef{behavnh} are mapped under TR into
\be
a^n_{\hat t,(1)}(k) = -\left(a^n_{\hat t,(1)}(k)\right)^* \,,\qquad b^n_{\hat t,(1)}(k)= \left( b^n_{\hat t,(1)}(k)\right)^*\,.
\ee
An exactly analogous argument can be made for the latter boundary condition in \reef{behavnh} involving $s_2\,, s_4$ implying that 
\be
a^n_{\hat t,(2)}(k) = -\left(a^n_{\hat t,(2)}(k)\right)^* \,,\qquad b^n_{\hat t,(2)}(k) = \left( b^n_{\hat t,(2)}(k)\right)^*\,.
\ee

If we now combine this fact with
the result \reef{eq:MatchedProp}, we can see the numerically observed
power-law scaling for low frequencies analytically. 
Expanding \reef{eq:MatchedProp} for small $\omega$ we find to
leading order
\be
t_{11} (\omega,k)
   = t_{11}(0,k) \left( 1 + \sum_{n=0}^\infty C_n(k)(\omega\ell)^n {\cal G}(\omega,\nu_k) + D_n(k)(\omega\ell)^n\right)\, ,
\ee
where the $C_n(k)$ and $D_n(k)$ are $k$-dependent constants depending on the 
$\omega$-independent coeficients $a_{(i)},b_{(i)}$.  To be explicit, expanding\footnote{For this computation it is convenenient to make the gauge choice $p\cdot a=0$. Note that we should use the relation \reef{eq:arcoeff} to find the TR action on quantities like $a_{\hat r,(1)}$. Note also that all quantities including the $e_\alpha^i$ and $p^2$ have to be expanded in powers of $\omega\ell$.} the numerator of \reef{eq:MatchedProp} in powers of $\omega\ell$ gives the first two coefficients
\bea
C_1(k) &=& \frac{\bar m b_{\hat t,(2)}^0}{\bar m b_{\hat t,(1)}^0} - \frac{\bar n a_{\hat t,(2)}^0}{\bar n a_{\hat t,(1)}^0}\nn
D_1(k) &=&  \frac{\bar m b_{\hat t,(1)}^1}{\bar m b_{\hat t,(1)}^0}  - \frac{\bar n a_{\hat t,(1)}^1}{\bar n a_{\hat t,(1)}^0} + \frac{2\mu}{\ell p_0^2}\,,
\eea
where $p_0^2 = p^2|_{\omega=0}$ and the action of TR \reef{eq:TRaction} induces the simple map
\be
C_1(k) \rightarrow C_1(k)^*\,,\qquad D_1(k) \rightarrow D_1(k)^*\,.
\ee
Since we saw above that at $\omega=0$ the boundary conditions are TR invariant, we can conclude that these coefficients are real. In fact, a little more thought shows that all higher coefficients $D_n(k)$ and $C_n(k)$ are all mapped to their complex conjugates under the action of TR. Hence we conclude that they also are real.
Finally, putting these arguments together with $t_{11}(0,k) \in \mathbb{R}$, we can extract the scaling relation at leading order
\be
{\rm Im} (t_{11}(\omega,k)) \propto \omega^{2\nu_k}\, ,
\ee
valid for the spectral function near the origin, exactly as we saw in our
numerical results. 
\section{Quasi-normal modes and their physics}\label{leaver}
In this section we establish the location in the $\omega,k$ plane of the quasi-normal modes (QNMs) of the gravitino in the AdS-RN black-brane geometry. Recall (e.g. \cite{Kovtun:2005ev}) 
that the QNMs should have ingoing boundary conditions at the future horizon
and normalisable boundary conditions at the AdS boundary. The latter condition implies that the QNMs are associated with poles
of the Green's function when considered as functions of complex $\omega$. Furthermore, the QNMs with 
$\omega$ close to the real axis are associated
with prominent features of the Green's function when considered as functions of real $\omega$. Determining the QNM spectrum
is thus a non-trivial check on the structure of the supercurrent two-point function we studied above. 
The results of this section, to our knowledge, are the first determination of the QNM spectrum of the gravitino in an AdS black-brane background.
Our approach is numerical and makes use of Leaver's method as first developed in \cite{Leaver:1990zz} and more recently applied to AdS/CMT\footnote{We thank the authors of \cite{Denef:2009yy,Denef:2009kn} for sharing technical details of their work with us.} in \cite{Denef:2009yy,Denef:2009kn}. For simplicity we will just deal with $T\ne 0$. Our results here are an independent confirmation of the pole structure in Figures \ref{fig:ARPES0}-\ref{fig:ARPES2}.

\subsection{Quasinormal modes for $T>0$}
\begin{figure}[h!]
\begin{center}
 a)\, \includegraphics[width=0.45\textwidth]{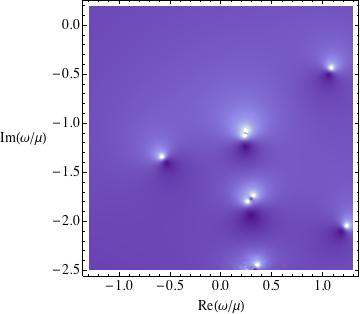}\hskip1em
 b)\,\includegraphics[width=0.45\textwidth]{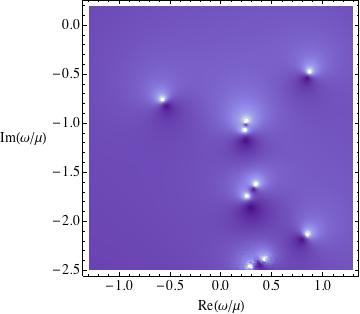}\\
 c) \includegraphics[width=0.45\textwidth]{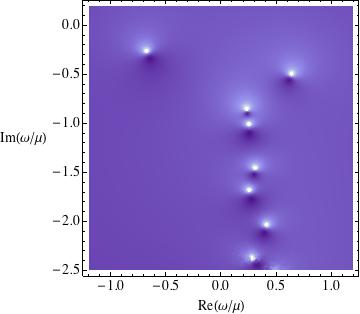}\hskip1em
 d)\,\includegraphics[width=0.45\textwidth]{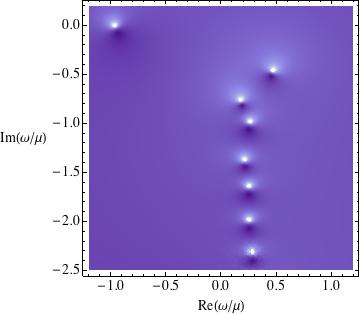}\\
\caption{\it Here we show the highest -lying quasinormal frequencies in the complex $\omega$ plane. Background parameters are chosen so that temperature $T/\mu=0.08$. We set $\ell k=2.1$ in panel a), $\ell k=1.7$ in panel b),
$\ell k=1.1$ in panel c) and  $\ell k=0.1$ in panel d). One can clearly see the pole moving to $\tilde\omega=0$ as $k\rightarrow 0$, corresponding to the supersymmetric sound pole at finite T. There is a line of poles just off the imaginary axis, which becomes a branch cut in the zero-temperature limit. We note that the Leaver method does not give any information about the strength of the pole. The figures show density plots of $|{\rm det}M'(\omega)/{\rm det}M(\omega)|$, where the matrix $M(\omega)$ is defined in \reef{eq:Mmatrix}.
\label{fig:QNMPlotI}}
\end{center}\end{figure}
We now turn to describing the appropriate ingoing boundary conditions at the horizon.
At finite temperature the horizon at $r=r_+$ is a regular singular point of the gravitino equations and we can 
use the expansion studied in section \ref{nonzeroT} above.
In our implementation of the Leaver method we shall specifically {\it not} choose a gauge at the horizon (having fixed the gauge near the boundary), which is the opposite approach to that employed in the calculation of the spectral density above, where we fixed the gauge at the horizon, but not near the boundary. These are merely matters of convenience, of course since the physical quantities we extract, such as the location of the QNMs, are gauge-invariant objects.
\begin{figure}[h!]
\begin{center}
 a)\, \includegraphics[width=0.45\textwidth]{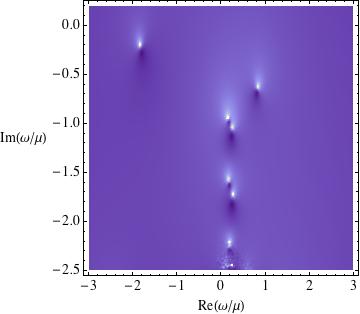}\hskip1em
 b)\,\includegraphics[width=0.45\textwidth]{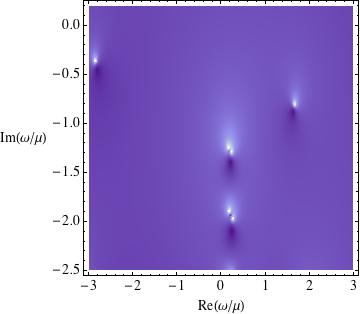}\\
\caption{\it Here we show the highest-lying quasinormal frequencies in the complex $\omega$ plane for $k\ell<0$. Background parameters are chosen so that temperature $T/\mu=0.08$. We set $\ell k=-1.5$ in panel a) and $\ell k=-2.8$ in panel b). We see that the phonino pole bounces off the real axis at $k\ell=0$ and moves back down into the negative ${\rm Im}\,\omega$ plane with ${\rm Re}\,\omega<-\mu$. This corresponds to the spread of the high spectral density region associated with supersound to negative $\omega$ at negative $k\ell$ in Fig. \ref{fig:ARPES1}.
\label{fig:QNMPlotII}}
\end{center}\end{figure}

\def\z{Z}
We now have assembled all the information necessary to set up a version of Leaver's method, adapted to our system. In this approach one factors out the desired leading behaviour at the horizon as well as infinity and expands the solutions around the midpoint of the interval between horizon and infinity. For this purpose it is very useful to change coordinates to $\z=\ell^2/r$ and we shall do so in this section. In addition, we find it convenient for the numerics to set $\ell = r_+ = 1$, 
using the scaling symmetries of the equations,
so that the temperature is entirely set by the chemical potential $\mu$. Thus, in detail, we expand our vector spinor 
(in the $\eta$ sector) to some arbitrary, but fixed order $N$ as
\bea
u^{(3/2)} &=& (\z-1)^{-\frac{3}{4} - \frac{i\frak{w}}{2}}\z^{3/2} \sum_{n=0}^N a^{(3/2)}_n (\z-\tfrac{1}{2})^n\,,\nn
u^{(1/2)} &=& (\z-1)^{-\frac{1}{4} - \frac{i\frak{w}}{2}}\z^{3/2} \sum_{n=0}^N a^{(1/2)}_n (\z-\tfrac{1}{2})^n\,,\nn
u^{(-1/2)} &=& (\z-1)^{\frac{1}{4} - \frac{i\frak{w}}{2}}\z^{3/2} \sum_{n=0}^N a^{(-1/2)}_n (\z-\tfrac{1}{2})^n\,,\nn
u^{(-3/2)} &=& (\z-1)^{\frac{3}{4} - \frac{i\frak{w}}{2}}\z^{3/2} \sum_{n=0}^N a^{(-3/2)}_n (\z-\tfrac{1}{2})^n\,,\,.
\eea
We then substitute this ansatz into the gravitino equations of motion and collect the resulting linear equations into a matrix equation of the form
\be\label{eq:Mmatrix}
\frac{d u^{(s)}_n}{d\z} - \sum_{s'=-3/2}^{3/2}\sum_{m=0}^NM^{s s'}_{nm} \left(\frak{w},\frak{q},T  \right)u^{(s')}_m=0\,,
\ee
for complex frequencies $\frak{w}$ where, as before, $\frak{w}\equiv \omega/2\pi T$ and
$\frak{q}\equiv{k}/2\pi T$. Solutions of this equation are forced to be trivial, unless the condition
\be
\det\left[ M^{s s'}_{nm} \left(\frak{w}^*,\frak{q},T  \right)\right]=0\,,
\ee
is met for some special complex frequency $\frak{w}^*$. The determinant is taken over all indices including the helicity index $s$. Solutions to this equation are the quasinormal frequencies of the background, under spin-3/2 perturbations. In practise we extract the matrix $M_{nm}^{ss'}$ using computer algebra and proceed to numerically solve the determinant condition.

We present our results for a finite temperature  $T/\mu\sim0.08$ in Fig. \ref{fig:QNMPlotI}. The behaviour of the highest QNMs at other finite temperatures is qualitatively very similar and we do not present explicit plots here. We see that there are two highest-lying poles, one in the negative $\omega$ plane and one in the positive $\omega$ plane. The former corresponds to the phonino pole and approaches the real axis as $k\rightarrow 0$ and $\tilde\omega \rightarrow 0$, while the latter corresponds to the second region of high spectral weight identified in Fig. \ref{fig:ARPES1}. We see clearly how, as $k$ is increased, these two poles cross over so that for lower momenta the phonino pole dominates, while for low $k$ it is the latter pole that resides closest to the real axis, consistent with it dominating the spectrum. We also see another set of poles going down the complex $\omega$ plane near the axis (but just off it). It should be noted that the representation of the poles as computed in the Leaver method can be slightly misleading as the strength of the poles on the plot is not related to its residue, as it would be, for instance, if one computed the full spectral function in the complex $\omega$ plane. The poles going down the imaginary axis are in fact much weaker than the two leading ones we discuss above and correspondingly do not give an appreciable imprint on the physical spectral function of the supercurrent. A similar point was made, for the case of a bosonic field in \cite{Brattan:2010pq}. We also studied the QNM structure at $T=0$, and found results that are consistent with the corresponding spectral density presented in Fig. \ref{fig:ARPES2}.

\section{Final comments}\label{conclude}
In this paper and its companion \cite{Gauntlett:2011mf} we have carried out a detailed analysis of
the spectral function of the supercurrent for the general class of $d=3\,, N=2$ SCFTs which have $D=10$ or $D=11$ supergravity duals. This was achieved by using the consistent KK truncation results of
\cite{Gauntlett:2007ma} and then solving the gravitino equations of $N=2$ $D=4$ minimal gauged supergravity
in the AdS-RN black-brane background.
A major feature of the spectral function 
is the presence of the phonino pole at $k=0$ and $\omega+\mu =0$. We calculated
the dispersion relation for this pole deducing a speed consistent with the hydrodynamical results of
\cite{Kovtun:2003vj} and also the diffusion constant. It would be interesting to go beyond our numerical results
for the dispersion relation and obtain its value analytically. It would also be interesting to extend the analysis of \cite{Kovtun:2003vj} to finite chemical potential.
A second major feature of the spectral function is the depletion of spectral weight around $\omega =0$, which is
accentuated for low temperatures. Indeed at zero temperature we showed that the spectral function vanishes
at $\omega =0$ giving rise to a power-law soft gap that is controlled by the locally quantum critical point
dual to the $AdS_2\times\mathbb{R}^2$ solution. In the latter background we obtained an analytic expression
for the Green's function by solving the gravitino equations exactly in terms of Whittaker functions.    

It is noteworthy that there is no Fermi surface in the supercurrent correlator at (bulk) tree level, a result that has been
recently confirmed in \cite{Belliard:2011qq}.
While it is possible that this indicates that there is no Fermi surface in these top-down models, we think it is
more likely to be a consequence of the particular correlator being considered and in particular
a Fermi surface will be seen in other Fermion correlators and/or at higher loop level. We hope to analyse a concrete
top down example in the future to develop a better understanding of this issue.
One possibility is to study the top-down models given in \cite{Gauntlett:2009zw} and furthermore 
examine the behaviour of the supercurrent
and other fermion correlators in the superfluid phase that appears at low temperatures 
\cite{Gauntlett:2009dn,Gauntlett:2009bh}. We saw in section 8 that the conformal dimensions of the operators in the one-dimensional conformal field theory in the far IR, dual to the $AdS_2$ region, are all real. This means that the ``log-oscillatory" behaviour seen in the bottom-up models of \cite{Liu:2009dm,Faulkner:2009wj} is absent. In the bottom-up models this behaviour was correlated with the presence of a Fermi surface and one might wonder if there is some deeper connection between these two phenomena. 
It would be worthwhile extending our calculations to the alternative quantisation of the gravitino field to check which of these features persist. Although this just entails changing the $AdS_4$ boundary conditions, some care will be required since  the alternative quantisation breaks the boundary supersymmetry and hence the gravitino will have additional propagating modes in the bulk.

\subsection*{Acknowledgements}
We would like to thank Joe Bhaseen, Alex Buchel, Aleksey Cherman, Diego Hofman, Andrew Green, Sean Hartnoll, Joao Laia, John McGreevy, Rob Myers, Subir Sachdev,
Koenraad Schalm, David Tong, David Vegh, Claude Warnick and Jan Zaanen for helpful discussions and correspondence. JPG is supported by an EPSRC Senior Fellowship and a Royal Society Wolfson Award. JS is supported by EPSRC and Trinity College Cambridge and thanks the GGI, Firenze and also, along with JPG, the Aspen Center for Physics for hospitality during this work.
\appendix

\section{Spinor conventions}\label{spconv}
Throughout we have chosen a mostly minus signature convention.
For the $d=3$ gamma-matrices $\gamma^\alpha$ we will use the explicit real basis
\be
\tdrho^{ t}=i\sigma_2,\qquad \tdrho^{ x}=\sigma_1,\qquad \tdrho^{ y}=\sigma_3\,,
\ee
with $\tdrho^{ t x y}=+1$. For a $d=3$ spinor $s$ we define $\bar s= s^\dagger \gamma^t$.
Parity in $d=3$ will be taken to act on the coordinates via $(t,x,y)\to (t,x,-y)$ and acting on 
spinors via
\begin{align}\label{parspin3}
s(t,x,y)\to \gamma^ys(t,x,-y)\,.
\end{align}
With this definition observe that the bi-linear $\bar s \gamma^\alpha s$ transforms as a vector. However, the bi-linear
$\bar s s$ transforms as a pseudo-scalar (a scalar bi-linear can only be constructed from a spinor transforming
as in \eqref{parspin3} with another spinor that is chosen to transform under parity with an extra overall minus sign).
We will also define the basis of $d=3$ spinors
\begin{align}
m=   \left( \begin{array}{c}
1  \\
0  
 \end{array} \right), \qquad
n= \left( \begin{array}{c}
0  \\
1  
 \end{array} \right)\,,
\end{align}
 which satisfy $\gamma^{tx}m=m$, $\gamma^{tx}n=-n$
and $\bar m n=-\bar nm=1$.

For the $D=4$ gamma-matrices, with tangent space indices, we will choose the real representation
\be\label{eq:threeDfourD}
\Gamma^{\hat\alpha}=\tdrho^{\alpha}\otimes\sigma_1,\qquad \Gamma^{\hat r}=1\otimes \sigma_3.
\ee
In the four-dimensional space spanned by coordinates $(t,r,x,y)$ we have projectors made from the following chirality operators
\be
\Gamma^{(2)} \equiv \Gamma^{ \hat t}\Gamma^{\hat r}\,,\qquad \Gamma^{(3)}\equiv
\frac{1}{|\mathbf{k}|}\Gamma^{(2)} \slashed{\mathbf{k}}=\Gamma^{\hat t\hat r\hat x}\,.
\ee
Since these commute, $\left[   \Gamma^{(2)}\,,\,\Gamma^{(3)}  \right] = 0$,
we may label the components of four-dimensional spinors (and vector spinors) in terms of simultaneous eigenvalues under them. Let us denote their simultaneous eigenspinors as
\bea
\Gamma^{(3)}\eta^\pm = \eta^\pm\,\qquad \Gamma^{(3)}\rho^\pm = -\rho^\pm\,,\qquad \Gamma^{(2)}\eta^\pm = \pm \eta^\pm\,,\qquad \Gamma^{(2)} \rho^\pm = \pm \rho^\pm\,.
\eea
We find it convenient to work in a basis with
\begin{align}
\eta^+=   \left( \begin{array}{c}
n  \\
m  
 \end{array} \right), \quad
 \eta^-=   \left( \begin{array}{c}
-n  \\
m  
 \end{array} \right), \qquad
 \rho^+=   \left( \begin{array}{c}
-m \\
n 
 \end{array} \right), \quad
\rho^-= \left( \begin{array}{c}
m  \\
n 
 \end{array} \right)\,.
\end{align}
We also have
\bea
\Gamma^{\hat t}\eta^\pm = \pm \eta^\mp\,,\qquad\Gamma^{\hat t}\rho^\pm = \pm \rho^\mp\,,\nn
\Gamma^{\hat r}\eta^\pm = - \eta^\mp\,,\qquad \Gamma^{\hat r}\rho^\pm =  - \rho^\mp\,,\nn
\Gamma^{\hat x}\eta^\pm = \pm \eta^\pm\,,\qquad \Gamma^{\hat x}\rho^\pm = \mp \rho^\pm\,,\nn
\Gamma^{\hat y}\eta^\pm = \mp \rho^\pm\,,\qquad \Gamma^{\hat y}\rho^\pm = \mp \eta^\pm\,.
\eea

\section{Positivity of the Spectral function}
\label{sec:spectral-fn}

\newcommand{\comm}[2]{\big[ #1,#2 \big]}
\newcommand{\anti}[2]{\big\{ #1,#2 \big\}}
By definition the retarded correlation function in position space has the form 
\begin{equation}
   \tilde{G}_{\alpha\beta}(x-y) 
      %= \left< S_\alpha(x) \bar{S}_\beta(y) \right>  
      =i\theta(x^0-y^0)\left< \anti{S_\alpha(x)}{\bar{S}_\beta(y)} \right> .
\end{equation}
If we define
\begin{equation}
   \tilde{\eta}_{\alpha\beta}''(x-y) 
      = \tfrac{1}{2}\left< \anti{S_\alpha(x)}{\bar{S}_\beta(y)} \right> , 
\end{equation}
then, after a Fourier transform, the spectral function is given by 
\begin{equation}
   A_{\alpha\beta}(\omega,\mathbf{k})\equiv 
      \im G_{\alpha\beta}(\omega,\mathbf{k})
      = \eta''_{\alpha\beta}(\omega,\mathbf{k}) ,
\end{equation}
where $\eta''_{\alpha\beta}(\omega,\mathbf{k})$ is the Fourier
transform of $\tilde{\eta}_{\alpha\beta}''(x-y)$.

Formally we can write a complete set of states of the Hilbert space as
\begin{equation}
   \id = \int d\mu(q)d\mu(\lambda) \ket{q,\lambda}\bra{q,\lambda}\, ,
\end{equation}
where $q=(E_{\mathbf{q}},\mathbf{q})$, with
$E_{\mathbf{q}}=\sqrt{\mathbf{q}^2+m(\lambda)^2}$, labels the 
total three-momentum and $\lambda$ is a formal label for the other
degrees of freedom, while $d\mu(q)=d^2q/2E_{\mathbf{q}}(2\pi)^2$
and $d\mu(\lambda)$ are the appropriate measures. Taking the thermal
expectation value at finite chemical potential $\mu$, we have, by inserting
a complete set of states and tracing,  
\begin{equation}
\label{eq:KLexp}
\begin{aligned}
   \left< S_\alpha(x)\bar{S}_\beta(y)\right> 
      &= \int d\mu(q)d\mu(\lambda)d\mu(q')d\mu(\lambda')
           e^{-\beta E + \beta\mu N}
           \\ & \qquad \qquad 
           \bra{q,\lambda}S_\alpha(x)\ket{q',\lambda'}
           \bra{q',\lambda'}\bar{S}_\beta(y)\ket{q,\lambda} \\
      &= \int d\mu(q)d\mu(\lambda)d\mu(q')d\mu(\lambda')
           e^{-\beta E + \beta\mu N}
           e^{-i(q-q')(x-y)}
           \\ & \qquad \qquad 
           \bra{q,\lambda}S_\alpha(0)\ket{q',\lambda'}
           \bra{q',\lambda'}\bar{S}_\beta(0)\ket{q,\lambda} , 
\end{aligned}
\end{equation}
where $E=q^0$ is the energy of the state $\ket{q,\lambda}$ and $N$ is the
$R$-charge. We define the matrix element that appears
in~\eqref{eq:KLexp} 
\begin{equation}
   M_\alpha(q,q',\lambda,\lambda')
      = \bra{q,\lambda}S_\alpha(0)\ket{q',\lambda'} .
\end{equation}
The gamma-tracelessness and weakly gauged conservation of the
supercurrent $S_\alpha$ implies that 
\begin{equation}
   \gamma^\alpha M_\alpha = 0 , \qquad
   p^\alpha M_\alpha = 0 , 
\end{equation}
where $p=(\tilde{\omega},\mathbf{k})
=(q^0-q^{\prime 0}+\mu,\mathbf{q}-\mathbf{q}')$. Thus we can expand
$M_\alpha$ in terms of the basis~\eqref{explicitbb}
\begin{equation}
   M_\alpha(q,q',\lambda,\lambda')
      = Z_i(q,q',\lambda,\lambda') e^{(i)}_\alpha . 
\end{equation}
Changing variables to the three-momentum $q-q'$ so that the measure can be
written as 
\begin{equation}
   d\mu(q)d\mu(q') d\mu(\lambda)d\mu(\lambda')
      = d\tilde{\mu}(q,q',\lambda,\lambda') \frac{d^3(q-q')}{(2\pi)^3}\, ,
\end{equation}
we have 
\begin{equation}
   \tilde{\eta}''_{\alpha\beta}(x-y)
      = \int \frac{d^3(q-q')}{(2\pi)^3}
           A_{ij}(p) e^{(i)}_\alpha \bar{e}^{(j)}_\beta 
           e^{-i(q-q')(x-y)} \, , 
\end{equation}
where
\begin{equation}\label{anspop}
\begin{aligned}
   A_{ij}(p) 
      &= \int d\tilde{\mu}(q,q',\lambda,\lambda')
         e^{-\beta E + \beta\mu N} 
         \\ & \qquad 
         \times \left[
            Z_i(q,q',\lambda,\lambda')Z^*_j(q,q',\lambda,\lambda')
            + Z_i(q',q,\lambda',\lambda)Z^*_j(q',q,\lambda',\lambda)
            \right] \, .  
\end{aligned}
\end{equation}
In the notation of section~\ref{sec:corrfn} we have $A_{ij}(p)=\im t_{ij}(p)$. 
In particular we deduce that
\begin{equation}
   A(p) \equiv \im t_{11}(p)= A_{11}(p) > 0 ,
\end{equation}
because the integrand in \eqref{anspop} is positive definite.

\section{The Rarita-Schwinger equation in $AdS_{d+1}$}\label{gb}

In this appendix we consider the Rarita-Schwinger equation for an {\it uncharged} gravitino in $d+1$ spacetime dimensions.
We found this to be a useful exercise in order to elucidate some of the analogous calculations for the charged
gravitino in $3+1$ spacetime dimensions. The material presented here may also have other applications.

The Rarita-Schwinger equation for an {\it uncharged} gravitino in $d+1$ spacetime dimensions is given by
\be\label{eq:RS}
\Gamma^{\mu\nu\rho}\nabla_\nu\psi_\rho - \frac{\hat m}{\ell}\Gamma^{\mu\nu}\psi_\nu = 0\,,
\ee
where $\nabla$ is the Levi-Civita connection.
At this stage the parameter $\ell>0$ is arbitrary, but is inserted for convenience, since
shortly we will restrict to the case of $AdS_{d+1}$ with radius taken to be $\ell$.  For definiteness we will take $\hat m> 0$.
The case that is most relevant 
to the bulk of this paper is $d=3$ and $\hat m=-\ell m =+1$.

Following \cite{Grassi:2000dm} we write
\be
\Gamma^{\alpha\beta\gamma} = \Gamma^\alpha\Gamma^\beta\Gamma^\gamma - \eta^{\alpha\beta}\Gamma^\gamma - \eta^{\beta \gamma}\Gamma^\alpha + \eta^{\alpha\gamma}\Gamma^\beta\,,
\ee
whence upon contraction, first with $\nabla_\mu$, then with $\Gamma_\mu$, we obtain
\bea\label{eq:gravitinocontract}
0 &=&  \cancel{\nabla} \cancel{\nabla} (\Gamma\cdot\psi) - \nabla^2 (\Gamma\cdot\psi) - \cancel{\nabla}(\nabla\cdot\psi)  + \nabla_\mu (\cancel{\nabla} \psi^\mu)- \frac{\hat m}{\ell}\left( \cancel{\nabla}\Gamma\cdot\psi - \nabla\cdot \psi  \right)\nn 0&=& (d-1) \left[ \cancel{\nabla} (\Gamma\cdot \psi) -  \nabla \cdot \psi     \right] - \frac{\hat m d}{\ell}\Gamma \cdot \psi\,.
\eea
We can use various curvature identities to remove the double derivative terms from the first expression. 
Since $\psi^\mu$ is in the spin-3/2 representation we have
\be
\left[ \nabla_\mu, \nabla_\nu  \right] \psi^\rho = R_{\mu\nu}{}^{\rho\sigma}\psi_\sigma + \tfrac{1}{4}R_{\mu\nu}{}^{\sigma\delta}\Gamma_{\sigma\delta}\psi^\rho\,,
\ee
from which we can deduce
\bea
\cancel{\nabla}\cancel{\nabla}(\Gamma\cdot\psi) &=& \nabla^2 (\Gamma\cdot \psi) -\tfrac{1}{4}R (\Gamma\cdot\psi)\nn
\nabla_\mu (\cancel{\nabla}\psi^\mu) &=& \cancel{\nabla}(\nabla\cdot \psi) + \tfrac{1}{2}\Gamma^\mu R_{\mu\nu}\psi_\nu\,.
\eea
As advertised, these identities allow us to eliminate all two-derivative terms from the first equation in \reef{eq:gravitinocontract}. 

The resulting expressions are particularly simple in $AdS_{d+1}$ spacetime, for which $R_{\mu\nu}=-(d/\ell^2)g_{\mu\nu}$.
Indeed after a bit of algebra we conclude
\be
0 = \left(\hat m^2 - \frac{(d-1)^2}{4}  \right)(\Gamma\cdot \psi)\,.
\ee
Thus, we have to distinguish two cases. If the gravitino is physically massless, i.e. if
\be
\hat m = \frac{(d-1)}{2}
\ee
this equation is satisfied identically. If not, we must set
\be\label{eq:constraint}
\Gamma\cdot \psi = 0\,, \qquad\qquad \nabla\cdot\psi = 0
\ee
where the second constraint follows from the second equation in  \reef{eq:gravitinocontract} once $\Gamma\cdot\psi=0$ is imposed. 
If the gravitino is physically massless, as in the supersymmetric case we consider in the bulk of the paper, 
we can in fact use the local supersymmetry, to impose these same two constraints, taking note, however, that they do not fix the gauge freedom completely.  More specifically, the supersymmetry transformations are (setting all other
bosonic fields in the supergravity to zero) 
\begin{equation}\label{zone}
   \delta\psi_{\hat\mu} 
      = \left(D_{\hat\mu} + \frac{1}{2\ell}\Gamma_{\hat\mu}\right)\varepsilon \,,
\end{equation}
and the residual supersymmetry transformations read
\begin{equation}\label{ztwo}
   \left( \slashed{D} + \frac{\hat{m} + 1}{\ell} \right) \varepsilon = 0 \,, 
\end{equation}

The equations Eq. \reef{eq:constraint} project out the spin-$1/2$ parts of the Rarita-Schwinger equation and we thus see that solving \reef{eq:RS} is equivalent to solving
\be
\left( \cancel{\nabla} + \frac{\hat m}{\ell} \right)\psi_\mu = 0\,,\qquad\Gamma\cdot \psi = 0\,,\qquad \nabla\cdot \psi = 0\,.
\ee

\subsection{Asymptotic analysis of Rarita-Schwinger in $AdS_{d+1}$}
\label{app:asymptoticsRS}

Let us write down the asymptotic form of these equations at the boundary of $AdS_{d+1}$. This allows us
to deduce the leading asymptotic behaviour. We write the $AdS_{d+1}$ metric in Poincar\'e coordinates
\begin{equation}
   \dd s^2 = \frac{r^2}{\ell^2}\dd \tilde{s}^2_d 
       + \frac{\ell^2}{r^2} \dd r^2\,,
\end{equation}
where $\dd \tilde{s}^2_d $ is the flat metric of $\mathbb{R}^{1,d-1}$.
The equations can be succinctly written as 
\be
\Gamma^{\hat r} \left(r\partial_r + \tfrac{1}{2}d\right)\psi_{\hat \mu} 
+ \frac{\ell^2}{r}\slashed{\tilde{D}}\psi_{\hat \mu} -
m\ell\psi_{\hat\mu} = - \Gamma_{\hat \mu}\psi_{\hat r}\,, 
\ee
where $\tilde{D}_\alpha$ is the connection on $\dd\tilde{s}_d^2$. 
If we have a power series of the form $\psi_{\hat{\mu}}=\sum_p
C^{(p)}_{\hat{\mu}} r^{-p}$ the equations imply
\begin{equation}
\label{eq:iterate}
\begin{aligned}
   \left[ \left(p-\tfrac{1}{2}d\right)\Gamma^{\hat{r}} 
      - \hat{m}\right]C^{(p)}_{\hat{\alpha}} 
      &= \ell^2 \slashed{\tilde{D}} C^{(p-1)}_{\hat{\alpha}} 
       + \Gamma_{\hat{\alpha}} C^{(p)}_r ,  \\
   \left[ \left(p-\tfrac{1}{2}d-1\right)\Gamma^{\hat{r}} 
      - \hat{m}\right]C^{(p)}_{\hat{r}} 
      &= \ell^2 \slashed{\tilde{D}} C^{(p-1)}_{\hat{r}} .
\end{aligned}
\end{equation}
We now have to distinguish two cases, the first of which is more relevant for this paper.

\subsubsection*{$\hat{m}\not\in \mathbb{Z}+\frac{1}{2}$}
Then we can develop an asymptotic series the leading behaviour of which is
\begin{equation}
\begin{aligned}
   \psi_{\hat\alpha} &= 
      r^{\Delta-d} A_{\hat\alpha}^{(0)}
          \left(1 + {\cal O}(r^{-1}) \right)  
      + r^{-\Delta} B_{\hat\alpha}^{(0)}
          \left(1 + {\cal O}(r^{-1}) \right)\,,\\
   \psi_{\hat r} &= 
      r^{\Delta-d-1} A_{\hat r}^{(1)}
          \left(1 + {\cal O}(r^{-1}) \right)  
      + r^{-\Delta-1} B_{\hat r}^{(1)}
          \left(1 + {\cal O}(r^{-1}) \right) \,,
\end{aligned}
\end{equation}
where $\Delta=\frac{1}{2}d+\hat{m}$ and
\begin{equation}
\begin{aligned}
   (\dblone + \Gamma^{\hat r})A_{\hat\alpha}^{(0)} &=0 \,, &
   (\dblone - \Gamma^{\hat r})B_{\hat\alpha}^{(0)}=0 \,, \\
   (\dblone + \Gamma^{\hat r})A_{\hat r}^{(1)} &=0 \,, &
   (\dblone - \Gamma^{\hat r})B_{\hat r}^{(1)}=0 \,.
\end{aligned}
\end{equation}
Note from~\eqref{eq:iterate} that successive terms have opposite
$\Gamma^{\hat r}$ chirality. Thus the $B^{(p)}$ series does not mix
with the $A^{(p)}$ series unless $\hat{m}\in\bbZ+\frac{1}{2}$. 

\subsubsection*{$\hat{m}\in \mathbb{Z}+\frac{1}{2}$}
In this case $B^{(0)}$ and $A^{(2\hat{m})}$ have the same
chirality. To get the full solution we need to include a 
logarithmic term so the asymptotic solution takes the form
\begin{equation}
\begin{aligned}\label{logsexp}
   \psi_{\hat\alpha} &= 
      r^{-\Delta+2\hat{m}} A_{\hat\alpha}^{(0)}
          \left(1 + {\cal O}(r^{-1}) \right)  
      + r^{-\Delta}\ln r \tilde{A}_{\hat\alpha}^{(0)}
          \left(1 + {\cal O}(r^{-1}) \right)\,, \\
   \psi_{\hat r} &= 
      r^{-\Delta+2\hat{m}-1} A_{\hat r}^{(1)}
          \left(1 + {\cal O}(r^{-1}) \right)  
      + r^{-\Delta-1}\ln r \tilde{A}_{\hat r}^{(1)}
          \left(1 + {\cal O}(r^{-1}) \right) \,,
\end{aligned}
\end{equation}
where 
\begin{equation}
\begin{aligned}
   (\dblone + \Gamma^{\hat r})A_{\hat\alpha}^{(0)} &=0 \,, &
   (\dblone - \Gamma^{\hat r})\tilde{A}_{\hat\alpha}^{(0)}=0 \,, \\
   (\dblone + \Gamma^{\hat r})A_{\hat r}^{(1)} &=0 \,, &
   (\dblone - \Gamma^{\hat r})\tilde{A}_{\hat r}^{(1)}=0 \,.
\end{aligned}
\end{equation}
\medskip

Next we consider the conditions $\Gamma\cdot\psi=0$ and
$D\cdot\psi=0$, focusing on the $\hat{m}\notin\bbZ+\frac{1}{2}$ case. To
leading order the gamma-traceless condition is simply
\be
\Gamma^{\hat \alpha}A_{\hat \alpha}^{(0)}=\Gamma^{\hat \alpha}B_{\hat
  \alpha}^{(0)}=0 \,,
  \ee
that is gamma-tracelessness on the boundary. The $D\cdot\psi=0$
condition reads, after using the $\Gamma\cdot\psi=0$ condition,
\begin{equation}
   \left( r\partial_r + d + \tfrac{1}{2}\right) \psi_{\hat r}
      + \frac{\ell^2}{r}\tilde{D}\cdot\psi = 0 \,.
\end{equation}
To leading order this implies
\begin{equation}\label{prop}
\begin{aligned}
   \left( \tfrac{1}{2}(d-1) + \hat{m}  \right) A^{(1)}_{\hat r} 
      &= - \ell^2 \tilde{D}\cdot A^{(0)} \,, \\
   \left( \tfrac{1}{2}(d-1) - \hat{m}  \right) B^{(1)}_{\hat r} 
      &= - \ell^2 \tilde{D}\cdot B^{(0)} \,.
\end{aligned}   
\end{equation}
Thus provided $\hat{m}\neq\frac{1}{2}(d-1)$ we see that
(gamma-traceless) $A^{(0)}_{\hat \alpha}$ and $B^{(0)}_{\hat \alpha}$ are the
independent degrees of freedom. 

The gravitino is dual to an operator with dimension $\Delta$ in the dual CFT. The
$A_{\hat\alpha}^{(0)}$ and $B_{\hat\alpha}^{(0)}$ coefficients specify the deformations of the CFT by sources and the expectation values of the dual operator, respectively. For the $\hat{m}\in \mathbb{Z}+\frac{1}{2}$ case the coefficient specifying the expectation value behaviour is just the $2\hat{m}$-th
term of the (non-log) series in \eqref{logsexp} i.e. we can identify
$B_{\hat \alpha}^{(0)} = A^{(2\hat{m})}_{\hat \alpha}$.

Finally we consider the supersymmetric case where
$\hat{m}=\frac{1}{2}(d-1)$, again focusing on the $\hat{m}\notin\bbZ+\frac{1}{2}$ case. 
First note that the asymptotic expansion now
reads 
\begin{equation}
\begin{aligned}
   \psi_{\hat\alpha} &= 
      r^{-1/2} A_{\hat\alpha}^{(0)}
          \left(1 + {\cal O}(r^{-1}) \right)  
      + r^{-d+1/2} B_{\hat\alpha}^{(0)}
          \left(1 + {\cal O}(r^{-1}) \right)\,, \\
   \psi_{\hat r} &= 
      r^{-3/2} A_{\hat r}^{(1)}
          \left(1 + {\cal O}(r^{-1}) \right)  
      + r^{-d-1/2} B_{\hat r}^{(1)}
          \left(1 + {\cal O}(r^{-1}) \right) \,.
\end{aligned}
\end{equation}
The gamma-traceless condition again implies
\be
\Gamma^{\hat \alpha}A_{\hat \alpha}^{(0)}=\Gamma^{\hat \alpha}B_{\hat
  \alpha}^{(0)}=0 \,.
\ee
The $D\cdot\psi=0$ condition implies 
\begin{equation}
   \tilde{D}\cdot B^{(0)} = 0 \,,
\end{equation}
and $B^{(1)}_{\dot r}$ is now undetermined. To elucidate what is going on we need to consider the residual supersymmetry
transformations. The residual supersymmetry
parameter $\varepsilon$ satisfies \eqref{ztwo}
which becomes
\begin{equation}
   \Gamma^{\hat r}\left( r\partial_r + \tfrac{1}{2}d \right)\varepsilon
       + \tfrac{1}{2}(d+1)\varepsilon + \frac{\ell^2}{r}\slashed{\tilde{D}}\varepsilon = 0 \,.
\end{equation}
The asymptotic solution is then
\begin{equation}
   \varepsilon = 
      r^{1/2} a^{(-1)}
          \left(1 + {\cal O}(r^{-1}) \right)  
      + r^{-d-1/2} b^{(1)}
          \left(1 + {\cal O}(r^{-1}) \right) \,, 
\end{equation}
where
\begin{equation}
\begin{aligned}
   (\dblone + \Gamma^{\hat r})a^{(-1)} &=0 \,, &
   (\dblone - \Gamma^{\hat r})b^{(1)}=0 \,. 
\end{aligned}
\end{equation}
We also have 
\begin{equation}
   a^{(0)} = - \frac{\ell^2}{d} \slashed{\tilde{D}}a^{(-1)} \,.
\end{equation}
The supersymmetry variation of $\psi_{\hat\mu}$ in \eqref{zone} is given by 
\begin{equation}
\begin{aligned}
   \delta \psi_{\hat\alpha} &=
       \frac{1}{2\ell}\Gamma_{\hat\alpha}
          \left(\dblone+\Gamma^{\hat r}\right)\varepsilon
       + \frac{\ell}{r}\tilde{D}_{\hat\alpha}\varepsilon \,, \\
   \delta \psi_{\hat r} &= \frac{r}{\ell}\partial_r\varepsilon
       + \frac{1}{2\ell}\Gamma_{\hat r}\varepsilon \,.
\end{aligned}
\end{equation}
We find the leading variations are (note that the potential $\delta
A^{(-1)}_{\hat{r}}$, $\delta A^{(-1)}_{\hat{\alpha}}$ and $\delta
A^{(0)}_{\hat{r}}$ terms all vanish)  
\begin{equation}
\begin{aligned}\label{propjoe}
   \delta A^{(0)}_{\hat\alpha} 
     &= \ell \left( \tilde{D}_{\hat\alpha} 
         - \frac{1}{d}\Gamma_{\hat\alpha}\slashed{\tilde{D}}
          \right) a^{(-1)} \,, \\
   \delta B^{(1)}_{\hat r} 
     &= - \ell^{-1} db^{(1)} \,. 
\end{aligned}
\end{equation}
We find that $A^{(0)}_{\hat\alpha}$ is gauge-dependent (though remains
gamma-traceless) as is $B^{(1)}_{\hat r}$. We can always then choose a gauge
where $\tilde{D}\cdot A^{(0)}=0$ and $B^{(1)}_{\hat r}=0$. 
The final result matches with our expectations for the supersymmetric case. 
The coefficient giving the expectation value, $B^{(0)}_{\hat{\alpha}}$, is gauge-invariant. On the other
hand the coefficient
fixing the source, $A^{(0)}_{\hat{\alpha}}$, is gauge-dependent,
corresponding to a weak gauging of the supersymmetry in the dual SCFT.

\section{Comments on our numerical approach}\label{app:numerics}

Here we describe briefly some aspects of our numerics, which are designed so as to cope with the highly oscillatory behaviour of the equations, especially at low temperatures (including $T=0$).

The general idea is to integrate equations \eqref{thegeneqnsG} from the horizon at $r=r_+$ out to infinity (or, in practise, to a very large value of the radial variable) and then to extract the boundary quantities $a_{\hat\alpha}$ and $b_{\hat\alpha}$ of Eq. (\ref{condsonanadb}) by fitting the resulting data to the asymptotic expansion of section \ref{asym}. However, especially for small $T/\mu$, the near-horizon behaviour of the gravitino components is highly oscillatory, aggravated by a different singularity structure (see Eqs. \reef{eq:finiteTnearhorizon} and \reef{eq:ZeroTLeading}) at the horizon for different helicity components and thus different rates of oscillations and divergence. This means that the equations are often stiff in the numerical sense and as a result once one has integrated a long way out towards infinity the numerical error has accumulated to an extent that the analytical expansions of section  \ref{asym} are no longer a good fit. Since we need to extract the leading order coefficient at ${\cal O}(r^{-3/2})$ as well as the expectation value at ${\cal O}(r^{-7/2})$, down by {\it two} orders in the expansion, as well as the intermediate orders (in order to verify gauge invariance) a more reliable method is needed that works to very high accuracy.

We circumvent the problems outlined above by side-stepping direct integration in favour of a shooting algorithm. We now briefly summarise our procedure, which works at $T=0$ and $T>0$. We present the procedure for both $\eta$ and $\rho$ sectors at the same time, but in practise, since they completely decouple in the equations \reef{thegeneqnsG} it is easier to just focus on one at a time. Whenever we do numerical calculations we use scaling symmetries to
set $\ell=1$ and $r_+=1$ so that the effective temperature of the background is fixed by the sole remaining dimensionful parameter $\mu$. We also assume that $\omega\neq 0$ for the steps below.
\begin{enumerate}
\item Use computer algebra to develop the near-horizon expansion to a very high order. Both at $T=0$ and $T>0$ we choose ingoing boundary conditions
\be
\psi \sim e^{-i\omega r_*  - i \omega t}\,,
\ee
using the appropriate tortoise variable $r_*$ for each case.
We explicitly implement the gauge transformations on the series so that we can generate from a given choice all other solutions which are in the same gauge orbit. One check on our numerics is, then, that all gauge-equivalent near-horizon boundary conditions give the same physical correlation functions at infinity.
There are four complex numbers specifying the general near-horizon solution, which we can interpret as specifying initial conditions for the two physical helicity states as well as the gauge parameters in the $\eta$ and $\rho$ sectors.
\item Use computer algebra to develop the near-boundary expansion of the gravitino to a very high order. This expansion is specified by eight complex numbers, comprising the independent degrees of freedom in the vector spinors $a_{\hat\alpha}$ and $b_{\hat\alpha}$,  four in the $\eta$ sector and four in the $\rho$ sector.
\item Integrate in from infinity and out from the horizon and match the eight independent components of the gravitino $u^{(s)}, v^{(s)}\,, s  = -\frac{3}{2},-\frac{1}{2},\frac{1}{2},\frac{3}{2}$ at the midpoint, which is of course arbitrary, but can be chosen at a numerically convenient location. We find it numerically more favourable to use a variable $Z=-\ell^2/r$ to do the actual integrations, whence the horizon is at $Z=1$ and asymptotic infinity is at $Z=0$, so that a convenient and obvious choice for the midpoint is $Z_{\rm mid}=0.5$.
\item We then use a shooting algorithm to match the solution, $z_{\rm mid}$, which is equivalent to determining the data at infinity as a function of the near-horizon data as well as the single parameter $\mu$ determining the effective temperature of the background. At every iteration step of the shooting method we must numerically invert an $8\times 8$ matrix, which can be easily and stably done numerically.
\item Run the above steps for a fine grid of values of  $\omega$ (if desired complex) and $k$ and store, e.g. as binary files. We can subsequently use the stored collective data to assemble the various plots and fits shown in the bulk of this paper and the companion paper \cite{Gauntlett:2011mf}. The data for the correlators presented in these works is based on ${\cal O}(500,000)$ data points, each obtained by the shooting algorithm above, which converges after a few steps to an accuracy of ${\cal O}(10^{-10})$. This gave a total running time on a top-range commercial dual-core processor machine of approximately 1 week (in practise several machines with several cores were used in parallel).
\end{enumerate}

\section{Asymptotics of Whittaker Functions}\label{app:asymptotics}
For small values of $\omega z$ we have
\begin{align}\label{whittexp}
&W_{\alpha(s),\nu_k}\left(-2 i \omega z  \right)
\nn
 &  \sim\left(-2i\omega  z\right)^{\frac{1}{2} + \nu_k}\frac{\Gamma(-2\nu_k)}{\Gamma\left(\frac{1}{2} - \alpha(s) - \nu_k\right)} \left( 1 +        \mathcal{O}(\omega z)               \right) 
\nn
&\qquad+ \left(-2i\omega z  \right)^{\frac{1}{2} - \nu_k}\frac{\Gamma(2\nu_k)}{\Gamma\left(\frac{1}{2} - \alpha(s) + \nu_k\right)} \left( 1 +  \mathcal{O}(\omega z)  \right) \nn
& = (-2i\omega L_{(2)})^{1/2-\nu_k}\frac{\Gamma(2\nu_k)}{\Gamma\left(  \frac{1}{2} - \alpha(s) + \nu_k \right)}\left[\left(\frac{z}{L_{(2)}}\right)^{\frac{1}{2}- \nu_k} \left( 1 +  \mathcal{O}(\omega z) \right) \right.\nn
&\left. \qquad+ \left(\frac{z}{L_{(2)}}\right)^{\frac{1}{2} + \nu_k} {\cal G}_R(\alpha(s),\nu_k) \left(  1 +  \mathcal{O}(\omega z)  \right)\right]\,,
\end{align}
where,
in writing the expression in the equality in the fifth row, we could hide the leading Euler gamma functions by choosing a convenient overall normalisation, as we do in the bulk of the paper. The ratio of gamma functions ${\cal G}_R(\alpha(s),\nu_k)$ is the retarded correlation function of the one-dimensional $Spin(1,1)$ representation labeled by $s$:
\be
{\cal G}_R(\alpha(s),\nu_k) = e^{-i\pi \nu_k}\frac{\Gamma(-2\nu_k)  \Gamma\left(\frac{1}{2} - \alpha(s) + \nu_k   \right)}{\Gamma(2\nu_k) \Gamma\left(\frac{1}{2} - \alpha(s) - \nu_k   \right)}(2\omega L_{(2)})^{2\nu_k}\,.
\ee
In a similar manner, by considering the asymptotics of the outgoing Whittaker functions $W_{-\alpha(s),\nu_k}(2i\omega z)$ we can derive the advanced correlation function for each helicity label $s$:
\be
{\cal G}_A(\alpha(s),\nu_k) = e^{i\pi \nu_k}\frac{\Gamma(-2\nu_k)  \Gamma\left(\frac{1}{2} + \alpha(s) + \nu_k   \right)}{\Gamma(2\nu_k) \Gamma\left(\frac{1}{2} + \alpha(s) - \nu_k   \right)}(2\omega L_{(2)})^{2\nu_k}\,.
\ee
It is interesting to note that
\be 
\frac{{\cal G}_R (\omega,\alpha(s),\nu_k)}{{\cal G}_A(\omega\,\alpha(s), \nu_k)} = 
\frac{e^{-2 i \pi \nu_k} - e^{- \sqrt{2}\pi q L_{(2)}}}{e^{2 i \pi \nu_k} - e^{- \sqrt{2}\pi q L_{(2)}}}\,,
\ee
where we have used the fact that for all of the helicity states of the gravitino we have $e^{2\pi i s} = -1$, a reflection that these are fermionic states. Quotients of ${\cal G}$ for two different helicities $s\neq s'$ are similar and can be related exactly to the above result by using the recursion identity for the Euler gamma function.

\bibliography{grav}{}

\providecommand{\href}[2]{#2}\begingroup\raggedright\begin{thebibliography}{10}

\bibitem{Gauntlett:2011mf}
J.~P. Gauntlett, J.~Sonner, and D.~Waldram, ``{Universal Fermionic Spectral
  Functions from String Theory},''
  \href{http://dx.doi.org/10.1103/PhysRevLett.107.241601}{{\em Phys. Rev.
  Lett.} {\bfseries 107} (2011) 241601},
\href{http://arxiv.org/abs/1106.4694}{{\ttfamily arXiv:1106.4694 [hep-th]}}.
%%CITATION = 1106.4694;%%.

\bibitem{Lee:2008xf}
S.-S. Lee, ``{A Non-Fermi Liquid from a Charged Black Hole: A Critical Fermi
  Ball},'' \href{http://dx.doi.org/10.1103/PhysRevD.79.086006}{{\em Phys. Rev.}
  {\bfseries D79} (2009) 086006},
\href{http://arxiv.org/abs/0809.3402}{{\ttfamily arXiv:0809.3402 [hep-th]}}.
%%CITATION = 0809.3402;%%.

\bibitem{Liu:2009dm}
H.~Liu, J.~McGreevy, and D.~Vegh, ``{Non-Fermi liquids from holography},''
  \href{http://dx.doi.org/10.1103/PhysRevD.83.065029}{{\em Phys. Rev.}
  {\bfseries D83} (2011) 065029},
\href{http://arxiv.org/abs/0903.2477}{{\ttfamily arXiv:0903.2477 [hep-th]}}.
%%CITATION = 0903.2477;%%.

\bibitem{Cubrovic:2009ye}
M.~Cubrovic, J.~Zaanen, and K.~Schalm, ``{String Theory, Quantum Phase
  Transitions and the Emergent Fermi-Liquid},''
  \href{http://dx.doi.org/10.1126/science.1174962}{{\em Science} {\bfseries
  325} (2009) 439--444},
\href{http://arxiv.org/abs/0904.1993}{{\ttfamily arXiv:0904.1993 [hep-th]}}.
%%CITATION = 0904.1993;%%.

\bibitem{Faulkner:2009wj}
T.~Faulkner, H.~Liu, J.~McGreevy, and D.~Vegh, ``{Emergent quantum criticality,
  Fermi surfaces, and AdS2},''
  \href{http://dx.doi.org/10.1103/PhysRevD.83.125002}{{\em Phys. Rev.}
  {\bfseries D83} (2011) 125002},
\href{http://arxiv.org/abs/0907.2694}{{\ttfamily arXiv:0907.2694 [hep-th]}}.
%%CITATION = 0907.2694;%%.

\bibitem{Iqbal:2009fd}
N.~Iqbal and H.~Liu, ``{Real-time response in AdS/CFT with application to
  spinors},'' \href{http://dx.doi.org/10.1002/prop.200900057}{{\em Fortsch.
  Phys.} {\bfseries 57} (2009) 367--384},
\href{http://arxiv.org/abs/0903.2596}{{\ttfamily arXiv:0903.2596 [hep-th]}}.
%%CITATION = 0903.2596;%%.

\bibitem{Albash:2009wz}
T.~Albash and C.~V. Johnson, ``{Holographic Aspects of Fermi Liquids in a
  Background Magnetic Field},''
  \href{http://dx.doi.org/10.1088/1751-8113/43/34/345405}{{\em J. Phys.}
  {\bfseries A43} (2010) 345405},
\href{http://arxiv.org/abs/0907.5406}{{\ttfamily arXiv:0907.5406 [hep-th]}}.
%%CITATION = 0907.5406;%%.

\bibitem{Basu:2009qz}
P.~Basu, J.~He, A.~Mukherjee, and H.-H. Shieh, ``{Holographic Non-Fermi Liquid
  in a Background Magnetic Field},''
  \href{http://dx.doi.org/10.1103/PhysRevD.82.044036}{{\em Phys. Rev.}
  {\bfseries D82} (2010) 044036},
\href{http://arxiv.org/abs/0908.1436}{{\ttfamily arXiv:0908.1436 [hep-th]}}.
%%CITATION = 0908.1436;%%.

\bibitem{Denef:2009yy}
F.~Denef, S.~A. Hartnoll, and S.~Sachdev, ``{Quantum oscillations and black
  hole ringing},'' \href{http://dx.doi.org/10.1103/PhysRevD.80.126016}{{\em
  Phys. Rev.} {\bfseries D80} (2009) 126016},
\href{http://arxiv.org/abs/0908.1788}{{\ttfamily arXiv:0908.1788 [hep-th]}}.
%%CITATION = 0908.1788;%%.

\bibitem{Kachru:2009xf}
S.~Kachru, A.~Karch, and S.~Yaida, ``{Holographic Lattices, Dimers, and
  Glasses},'' \href{http://dx.doi.org/10.1103/PhysRevD.81.026007}{{\em Phys.
  Rev.} {\bfseries D81} (2010) 026007},
\href{http://arxiv.org/abs/0909.2639}{{\ttfamily arXiv:0909.2639 [hep-th]}}.
%%CITATION = 0909.2639;%%.

\bibitem{Hung:2009qk}
L.-Y. Hung and A.~Sinha, ``{Holographic quantum liquids in 1+1 dimensions},''
  \href{http://dx.doi.org/10.1007/JHEP01(2010)114}{{\em JHEP} {\bfseries 01}
  (2010) 114},
\href{http://arxiv.org/abs/0909.3526}{{\ttfamily arXiv:0909.3526 [hep-th]}}.
%%CITATION = 0909.3526;%%.

\bibitem{Maity:2009zz}
D.~Maity, S.~Sarkar, N.~Sircar, B.~Sathiapalan, and R.~Shankar, ``{Properties
  of CFTs dual to Charged BTZ black-hole},''
  \href{http://dx.doi.org/10.1016/j.nuclphysb.2010.06.012}{{\em Nucl. Phys.}
  {\bfseries B839} (2010) 526--551},
\href{http://arxiv.org/abs/0909.4051}{{\ttfamily arXiv:0909.4051 [hep-th]}}.
%%CITATION = 0909.4051;%%.

\bibitem{Chen:2009pt}
J.-W. Chen, Y.-J. Kao, and W.-Y. Wen, ``{Peak-Dip-Hump from Holographic
  Superconductivity},''
  \href{http://dx.doi.org/10.1103/PhysRevD.82.026007}{{\em Phys. Rev.}
  {\bfseries D82} (2010) 026007},
\href{http://arxiv.org/abs/0911.2821}{{\ttfamily arXiv:0911.2821 [hep-th]}}.
%%CITATION = 0911.2821;%%.

\bibitem{Gubser:2009qt}
S.~S. Gubser and F.~D. Rocha, ``{Peculiar properties of a charged dilatonic
  black hole in $AdS_5$},''
  \href{http://dx.doi.org/10.1103/PhysRevD.81.046001}{{\em Phys. Rev.}
  {\bfseries D81} (2010) 046001},
\href{http://arxiv.org/abs/0911.2898}{{\ttfamily arXiv:0911.2898 [hep-th]}}.
%%CITATION = 0911.2898;%%.

\bibitem{Faulkner:2009am}
T.~Faulkner, G.~T. Horowitz, J.~McGreevy, M.~M. Roberts, and D.~Vegh,
  ``{Photoemission 'experiments' on holographic superconductors},''
  \href{http://dx.doi.org/10.1007/JHEP03(2010)121}{{\em JHEP} {\bfseries 03}
  (2010) 121},
\href{http://arxiv.org/abs/0911.3402}{{\ttfamily arXiv:0911.3402 [hep-th]}}.
%%CITATION = 0911.3402;%%.

\bibitem{Gubser:2009dt}
S.~S. Gubser, F.~D. Rocha, and P.~Talavera, ``{Normalizable fermion modes in a
  holographic superconductor},''
  \href{http://dx.doi.org/10.1007/JHEP10(2010)087}{{\em JHEP} {\bfseries 10}
  (2010) 087},
\href{http://arxiv.org/abs/0911.3632}{{\ttfamily arXiv:0911.3632 [hep-th]}}.
%%CITATION = 0911.3632;%%.

\bibitem{Hartnoll:2009kk}
S.~A. Hartnoll and D.~M. Hofman, ``{Generalized Lifshitz-Kosevich scaling at
  quantum criticality from the holographic correspondence},''
  \href{http://dx.doi.org/10.1103/PhysRevB.81.155125}{{\em Phys. Rev.}
  {\bfseries B81} (2010) 155125},
\href{http://arxiv.org/abs/0912.0008}{{\ttfamily arXiv:0912.0008
  [cond-mat.str-el]}}.
%%CITATION = 0912.0008;%%.

\bibitem{Hartnoll:2009ns}
S.~A. Hartnoll, J.~Polchinski, E.~Silverstein, and D.~Tong, ``{Towards strange
  metallic holography},'' \href{http://dx.doi.org/10.1007/JHEP04(2010)120}{{\em
  JHEP} {\bfseries 04} (2010) 120},
\href{http://arxiv.org/abs/0912.1061}{{\ttfamily arXiv:0912.1061 [hep-th]}}.
%%CITATION = 0912.1061;%%.

\bibitem{Albash:2010yr}
T.~Albash and C.~V. Johnson, ``{Landau Levels, Magnetic Fields and Holographic
  Fermi Liquids},''
  \href{http://dx.doi.org/10.1088/1751-8113/43/34/345404}{{\em J. Phys.}
  {\bfseries A43} (2010) 345404},
\href{http://arxiv.org/abs/1001.3700}{{\ttfamily arXiv:1001.3700 [hep-th]}}.
%%CITATION = 1001.3700;%%.

\bibitem{Faulkner:2010tq}
T.~Faulkner and J.~Polchinski, ``{Semi-Holographic Fermi Liquids},''
  \href{http://dx.doi.org/10.1007/JHEP06(2011)012}{{\em JHEP} {\bfseries 06}
  (2011) 012},
\href{http://arxiv.org/abs/1001.5049}{{\ttfamily arXiv:1001.5049 [hep-th]}}.
%%CITATION = 1001.5049;%%.

\bibitem{Gubser:2010dm}
S.~S. Gubser, F.~D. Rocha, and A.~Yarom, ``{Fermion correlators in non-abelian
  holographic superconductors},''
  \href{http://dx.doi.org/10.1007/JHEP11(2010)085}{{\em JHEP} {\bfseries 11}
  (2010) 085},
\href{http://arxiv.org/abs/1002.4416}{{\ttfamily arXiv:1002.4416 [hep-th]}}.
%%CITATION = 1002.4416;%%.

\bibitem{Albash:2010dr}
T.~Albash, ``{Non-Unitary Fermionic Quasinormal Modes at Zero Frequency},''
\href{http://arxiv.org/abs/1002.4431}{{\ttfamily arXiv:1002.4431 [hep-th]}}.
%%CITATION = 1002.4431;%%.

\bibitem{Ammon:2010pg}
M.~Ammon, J.~Erdmenger, M.~Kaminski, and A.~O'Bannon, ``{Fermionic Operator
  Mixing in Holographic p-wave Superfluids},''
  \href{http://dx.doi.org/10.1007/JHEP05(2010)053}{{\em JHEP} {\bfseries 05}
  (2010) 053},
\href{http://arxiv.org/abs/1003.1134}{{\ttfamily arXiv:1003.1134 [hep-th]}}.
%%CITATION = 1003.1134;%%.

\bibitem{Faulkner:2010da}
T.~Faulkner, N.~Iqbal, H.~Liu, J.~McGreevy, and D.~Vegh, ``{From black holes to
  strange metals},''
\href{http://arxiv.org/abs/1003.1728}{{\ttfamily arXiv:1003.1728 [hep-th]}}.
%%CITATION = 1003.1728;%%.

\bibitem{Hartman:2010fk}
T.~Hartman and S.~A. Hartnoll, ``{Cooper pairing near charged black holes},''
  \href{http://dx.doi.org/10.1007/JHEP06(2010)005}{{\em JHEP} {\bfseries 06}
  (2010) 005},
\href{http://arxiv.org/abs/1003.1918}{{\ttfamily arXiv:1003.1918 [hep-th]}}.
%%CITATION = 1003.1918;%%.

\bibitem{Mross:2010rd}
D.~F. Mross, J.~McGreevy, H.~Liu, and T.~Senthil, ``{A controlled expansion for
  certain non-Fermi liquid metals},'' {\em Phys. Rev.} {\bfseries B82} (2010)
  045121,
\href{http://arxiv.org/abs/1003.0894}{{\ttfamily arXiv:1003.0894
  [cond-mat.str-el]}}.
%%CITATION = 1003.0894;%%.

\bibitem{Cai:2010tr}
R.-G. Cai, Z.-Y. Nie, B.~Wang, and H.-Q. Zhang, ``{Quasinormal Modes of Charged
  Fermions and Phase Transition of Black Holes},''
\href{http://arxiv.org/abs/1005.1233}{{\ttfamily arXiv:1005.1233 [gr-qc]}}.
%%CITATION = 1005.1233;%%.

\bibitem{Benini:2010qc}
F.~Benini, C.~P. Herzog, and A.~Yarom, ``{Holographic Fermi arcs and a d-wave
  gap},''
\href{http://arxiv.org/abs/1006.0731}{{\ttfamily arXiv:1006.0731 [hep-th]}}.
%%CITATION = 1006.0731;%%.

\bibitem{Larsen:2010jt}
F.~Larsen and G.~van Anders, ``{Holographic Non-Fermi Liquids and the Luttinger
  Theorem},''
\href{http://arxiv.org/abs/1006.1846}{{\ttfamily arXiv:1006.1846 [hep-th]}}.
%%CITATION = 1006.1846;%%.

\bibitem{Hung:2010te}
L.-Y. Hung, D.~P. Jatkar, and A.~Sinha, ``{Non-relativistic metrics from
  back-reacting fermions},''
  \href{http://dx.doi.org/10.1088/0264-9381/28/1/015013}{{\em Class. Quant.
  Grav.} {\bfseries 28} (2011) 015013},
\href{http://arxiv.org/abs/1006.3762}{{\ttfamily arXiv:1006.3762 [hep-th]}}.
%%CITATION = 1006.3762;%%.

\bibitem{Sachdev:2010um}
S.~Sachdev, ``{Holographic metals and the fractionalized Fermi liquid},''
  \href{http://dx.doi.org/10.1103/PhysRevLett.105.151602}{{\em Phys. Rev.
  Lett.} {\bfseries 105} (2010) 151602},
\href{http://arxiv.org/abs/1006.3794}{{\ttfamily arXiv:1006.3794 [hep-th]}}.
%%CITATION = 1006.3794;%%.

\bibitem{Gubankova:2010ny}
E.~Gubankova, ``{Particle-hole instability in the $AdS_4$ holography},''
\href{http://arxiv.org/abs/1006.4789}{{\ttfamily arXiv:1006.4789 [hep-th]}}.
%%CITATION = 1006.4789;%%.

\bibitem{Vegh:2010fc}
D.~Vegh, ``{Fermi arcs from holography},''
\href{http://arxiv.org/abs/1007.0246}{{\ttfamily arXiv:1007.0246 [hep-th]}}.
%%CITATION = 1007.0246;%%.

\bibitem{Benini:2010pr}
F.~Benini, C.~P. Herzog, R.~Rahman, and A.~Yarom, ``{Gauge gravity duality for
  d-wave superconductors: prospects and challenges},''
  \href{http://dx.doi.org/10.1007/JHEP11(2010)137}{{\em JHEP} {\bfseries 11}
  (2010) 137},
\href{http://arxiv.org/abs/1007.1981}{{\ttfamily arXiv:1007.1981 [hep-th]}}.
%%CITATION = 1007.1981;%%.

\bibitem{Hartnoll:2010gu}
S.~A. Hartnoll and A.~Tavanfar, ``{Electron stars for holographic metallic
  criticality},'' \href{http://dx.doi.org/10.1103/PhysRevD.83.046003}{{\em
  Phys. Rev.} {\bfseries D83} (2011) 046003},
\href{http://arxiv.org/abs/1008.2828}{{\ttfamily arXiv:1008.2828 [hep-th]}}.
%%CITATION = 1008.2828;%%.

\bibitem{Kachru:2010dk}
S.~Kachru, A.~Karch, and S.~Yaida, ``{Adventures in Holographic Dimer
  Models},'' \href{http://dx.doi.org/10.1088/1367-2630/13/3/035004}{{\em New J.
  Phys.} {\bfseries 13} (2011) 035004},
\href{http://arxiv.org/abs/1009.3268}{{\ttfamily arXiv:1009.3268 [hep-th]}}.
%%CITATION = 1009.3268;%%.

\bibitem{Yamamoto:2010yp}
S.~J. Yamamoto and Q.~Si, ``{Global Phase Diagram of the Kondo Lattice: From
  Heavy Fermion Metals to Kondo Insulators},''
  \href{http://dx.doi.org/10.1007/s10909-010-0221-4}{{\em J. Low. Temp. Phys.}
  {\bfseries 161} (2010) 233--262},
\href{http://arxiv.org/abs/1006.4868}{{\ttfamily arXiv:1006.4868
  [cond-mat.str-el]}}.
%%CITATION = 1006.4868;%%.

\bibitem{Edalati:2010ww}
M.~Edalati, R.~G. Leigh, and P.~W. Phillips, ``{Dynamically Generated Mott Gap
  from Holography},''
  \href{http://dx.doi.org/10.1103/PhysRevLett.106.091602}{{\em Phys. Rev.
  Lett.} {\bfseries 106} (2011) 091602},
\href{http://arxiv.org/abs/1010.3238}{{\ttfamily arXiv:1010.3238 [hep-th]}}.
%%CITATION = 1010.3238;%%.

\bibitem{Gulotta:2010cu}
D.~R. Gulotta, C.~P. Herzog, and M.~Kaminski, ``{Sum Rules from an Extra
  Dimension},'' \href{http://dx.doi.org/10.1007/JHEP01(2011)148}{{\em JHEP}
  {\bfseries 01} (2011) 148},
\href{http://arxiv.org/abs/1010.4806}{{\ttfamily arXiv:1010.4806 [hep-th]}}.
%%CITATION = 1010.4806;%%.

\bibitem{Arsiwalla:2010bt}
X.~Arsiwalla, J.~de~Boer, K.~Papadodimas, and E.~Verlinde, ``{Degenerate Stars
  and Gravitational Collapse in AdS/CFT},''
  \href{http://dx.doi.org/10.1007/JHEP01(2011)144}{{\em JHEP} {\bfseries 01}
  (2011) 144},
\href{http://arxiv.org/abs/1010.5784}{{\ttfamily arXiv:1010.5784 [hep-th]}}.
%%CITATION = 1010.5784;%%.

\bibitem{Gubankova:2010rc}
E.~Gubankova {\em et al.}, ``{Holographic fermions in external magnetic
  fields},''
\href{http://arxiv.org/abs/1011.4051}{{\ttfamily arXiv:1011.4051 [hep-th]}}.
%%CITATION = 1011.4051;%%.

\bibitem{Parente:2010fs}
V.~Parente and R.~Roychowdhury, ``{A Study on Charged Neutron Star in
  $AdS_5$},'' \href{http://dx.doi.org/10.1007/JHEP04(2011)111}{{\em JHEP}
  {\bfseries 04} (2011) 111},
\href{http://arxiv.org/abs/1011.5362}{{\ttfamily arXiv:1011.5362 [hep-th]}}.
%%CITATION = 1011.5362;%%.

\bibitem{Edalati:2010ge}
M.~Edalati, R.~G. Leigh, K.~W. Lo, and P.~W. Phillips, ``{Dynamical Gap and
  Cuprate-like Physics from Holography},''
  \href{http://dx.doi.org/10.1103/PhysRevD.83.046012}{{\em Phys. Rev.}
  {\bfseries D83} (2011) 046012},
\href{http://arxiv.org/abs/1012.3751}{{\ttfamily arXiv:1012.3751 [hep-th]}}.
%%CITATION = 1012.3751;%%.

\bibitem{Cubrovic:2010bf}
M.~Cubrovic, J.~Zaanen, and K.~Schalm, ``{Constructing the AdS dual of a Fermi
  liquid: AdS Black holes with Dirac hair},''
\href{http://arxiv.org/abs/1012.5681}{{\ttfamily arXiv:1012.5681 [hep-th]}}.
%%CITATION = 1012.5681;%%.

\bibitem{Faulkner:2011tm}
T.~Faulkner, N.~Iqbal, H.~Liu, J.~McGreevy, and D.~Vegh, ``{Holographic
  non-Fermi liquid fixed points},''
\href{http://arxiv.org/abs/1101.0597}{{\ttfamily arXiv:1101.0597 [hep-th]}}.
%%CITATION = 1101.0597;%%.

\bibitem{Guarrera:2011my}
D.~Guarrera and J.~McGreevy, ``{Holographic Fermi surfaces and bulk dipole
  couplings},''
\href{http://arxiv.org/abs/1102.3908}{{\ttfamily arXiv:1102.3908 [hep-th]}}.
%%CITATION = 1102.3908;%%.

\bibitem{Wu:2011bx}
J.-P. Wu, ``{Holographic fermions in charged Gauss-Bonnet black hole},''
\href{http://arxiv.org/abs/1103.3982}{{\ttfamily arXiv:1103.3982 [hep-th]}}.
%%CITATION = 1103.3982;%%.

\bibitem{Huijse:2011hp}
L.~Huijse and S.~Sachdev, ``{Fermi surfaces and gauge-gravity duality},'' {\em
  Phys. Rev.} {\bfseries D84} (2011) 026001,
\href{http://arxiv.org/abs/1104.5022}{{\ttfamily arXiv:1104.5022 [hep-th]}}.
%%CITATION = 1104.5022;%%.

\bibitem{Iizuka:2011hg}
N.~Iizuka, N.~Kundu, P.~Narayan, and S.~P. Trivedi, ``{Holographic Fermi and
  Non-Fermi Liquids with Transitions in Dilaton Gravity},''
\href{http://arxiv.org/abs/1105.1162}{{\ttfamily arXiv:1105.1162 [hep-th]}}.
%%CITATION = 1105.1162;%%.

\bibitem{Jensen:2011su}
K.~Jensen, S.~Kachru, A.~Karch, J.~Polchinski, and E.~Silverstein, ``{Towards a
  holographic marginal Fermi liquid},''
\href{http://arxiv.org/abs/1105.1772}{{\ttfamily arXiv:1105.1772 [hep-th]}}.
%%CITATION = 1105.1772;%%.

\bibitem{Hartnoll:2011dm}
S.~A. Hartnoll, D.~M. Hofman, and D.~Vegh, ``{Stellar spectroscopy: Fermions
  and holographic Lifshitz criticality},''
\href{http://arxiv.org/abs/1105.3197}{{\ttfamily arXiv:1105.3197 [hep-th]}}.
%%CITATION = 1105.3197;%%.

\bibitem{Iqbal:2011in}
N.~Iqbal, H.~Liu, and M.~Mezei, ``{Semi-local quantum liquids},''
\href{http://arxiv.org/abs/1105.4621}{{\ttfamily arXiv:1105.4621 [hep-th]}}.
%%CITATION = 1105.4621;%%.

\bibitem{Cubrovic:2011xm}
M.~Cubrovic, Y.~Liu, K.~Schalm, Y.-W. Sun, and J.~Zaanen, ``{Spectral probes of
  the holographic Fermi groundstate: dialing between the electron star and AdS
  Dirac hair},''
\href{http://arxiv.org/abs/1106.1798}{{\ttfamily arXiv:1106.1798 [hep-th]}}.
%%CITATION = 1106.1798;%%.

\bibitem{Edalati:2011yv}
M.~Edalati, K.~W. Lo, and P.~W. Phillips, ``{Neutral Order Parameters in
  Metallic Criticality in d=2+1 from a Hairy Electron Star},''
\href{http://arxiv.org/abs/1106.3139}{{\ttfamily arXiv:1106.3139 [hep-th]}}.
%%CITATION = 1106.3139;%%.

\bibitem{Rangamani:2011ae}
M.~Rangamani and B.~Withers, ``{Fermionic probes of local quantum criticality
  in one dimension},''
\href{http://arxiv.org/abs/1106.3210}{{\ttfamily arXiv:1106.3210 [hep-th]}}.
%%CITATION = 1106.3210;%%.

\bibitem{Hoyos:2011us}
C.~Hoyos, T.~Nishioka, and A.~O'Bannon, ``{A Chiral Magnetic Effect from
  AdS/CFT with Flavor},''
\href{http://arxiv.org/abs/1106.4030}{{\ttfamily arXiv:1106.4030 [hep-th]}}.
%%CITATION = 1106.4030;%%.

\bibitem{Belliard:2011qq}
R.~Belliard, S.~S. Gubser, and A.~Yarom, ``{Absence of a Fermi surface in
  classical minimal four- dimensional gauged supergravity},''
\href{http://arxiv.org/abs/1106.6030}{{\ttfamily arXiv:1106.6030 [hep-th]}}.
%%CITATION = 1106.6030;%%.

\bibitem{Gauntlett:2007ma}
J.~P. Gauntlett and O.~Varela, ``{Consistent Kaluza-Klein Reductions for
  General Supersymmetric AdS Solutions},''
  \href{http://dx.doi.org/10.1103/PhysRevD.76.126007}{{\em Phys. Rev.}
  {\bfseries D76} (2007) 126007},
\href{http://arxiv.org/abs/0707.2315}{{\ttfamily arXiv:0707.2315 [hep-th]}}.
%%CITATION = 0707.2315;%%.

\bibitem{gswunpub}
J.~Gauntlett, Sonner, J., and D.~Waldram. Unpublished.

\bibitem{Bah:2010yt}
I.~Bah, A.~Faraggi, J.~I. Jottar, R.~G. Leigh, and L.~A. Pando~Zayas,
  ``{Fermions and D=11 Supergravity On Squashed Sasaki-Einstein Manifolds},''
  \href{http://dx.doi.org/10.1007/JHEP02(2011)068}{{\em JHEP} {\bfseries 02}
  (2011) 068},
\href{http://arxiv.org/abs/1008.1423}{{\ttfamily arXiv:1008.1423 [hep-th]}}.
%%CITATION = 1008.1423;%%.

\bibitem{Denef:2009tp}
F.~Denef and S.~A. Hartnoll, ``{Landscape of superconducting membranes},''
  \href{http://dx.doi.org/10.1103/PhysRevD.79.126008}{{\em Phys. Rev.}
  {\bfseries D79} (2009) 126008},
\href{http://arxiv.org/abs/0901.1160}{{\ttfamily arXiv:0901.1160 [hep-th]}}.
%%CITATION = 0901.1160;%%.

\bibitem{Donos:2011ut}
A.~Donos and J.~P. Gauntlett, ``{Superfluid black branes in $AdS_4\times
  S^7$},'' \href{http://dx.doi.org/10.1007/JHEP06(2011)053}{{\em JHEP}
  {\bfseries 06} (2011) 053},
\href{http://arxiv.org/abs/1104.4478}{{\ttfamily arXiv:1104.4478 [hep-th]}}.
%%CITATION = 1104.4478;%%.

\bibitem{Donos:2010ax}
A.~Donos, J.~P. Gauntlett, N.~Kim, and O.~Varela, ``{Wrapped M5-branes,
  consistent truncations and AdS/CMT},''
  \href{http://dx.doi.org/10.1007/JHEP12(2010)003}{{\em JHEP} {\bfseries 12}
  (2010) 003},
\href{http://arxiv.org/abs/1009.3805}{{\ttfamily arXiv:1009.3805 [hep-th]}}.
%%CITATION = 1009.3805;%%.

\bibitem{Lebedev:1989rz}
V.~V. Lebedev and A.~V. Smilga, ``{Supersymmetric sound},''
\href{http://dx.doi.org/10.1016/0550-3213(89)90636-6}{{\em Nucl. Phys.}
  {\bfseries B318} (1989) 669--704}.
%%CITATION = NUPHA,B318,669;%%.

\bibitem{Kovtun:2003vj}
P.~Kovtun and L.~G. Yaffe, ``{Hydrodynamic fluctuations, long-time tails, and
  supersymmetry},'' \href{http://dx.doi.org/10.1103/PhysRevD.68.025007}{{\em
  Phys. Rev.} {\bfseries D68} (2003) 025007},
\href{http://arxiv.org/abs/hep-th/0303010}{{\ttfamily arXiv:hep-th/0303010}}.
%%CITATION = HEP-TH/0303010;%%.

\bibitem{Kratzert:2003cr}
K.~Kratzert, ``{Finite-temperature supersymmetry: The Wess-Zumino model},''
  \href{http://dx.doi.org/10.1016/S0003-4916(03)00143-X}{{\em Ann. Phys.}
  {\bfseries 308} (2003) 285--310},
\href{http://arxiv.org/abs/hep-th/0303260}{{\ttfamily arXiv:hep-th/0303260}}.
%%CITATION = HEP-TH/0303260;%%.

\bibitem{Policastro:2008cx}
G.~Policastro, ``{Supersymmetric hydrodynamics from the AdS/CFT
  correspondence},''
  \href{http://dx.doi.org/10.1088/1126-6708/2009/02/034}{{\em JHEP} {\bfseries
  02} (2009) 034},
\href{http://arxiv.org/abs/0812.0992}{{\ttfamily arXiv:0812.0992 [hep-th]}}.
%%CITATION = 0812.0992;%%.

\bibitem{Iqbal:2010eh}
N.~Iqbal, H.~Liu, M.~Mezei, and Q.~Si, ``{Quantum phase transitions in
  holographic models of magnetism and superconductors},''
  \href{http://dx.doi.org/10.1103/PhysRevD.82.045002}{{\em Phys. Rev.}
  {\bfseries D82} (2010) 045002},
\href{http://arxiv.org/abs/1003.0010}{{\ttfamily arXiv:1003.0010 [hep-th]}}.
%%CITATION = 1003.0010;%%.

\bibitem{Freedman:1976aw}
D.~Z. Freedman and A.~K. Das, ``{Gauge Internal Symmetry in Extended
  Supergravity},''
\href{http://dx.doi.org/10.1016/0550-3213(77)90041-4}{{\em Nucl. Phys.}
  {\bfseries B120} (1977) 221}.
%%CITATION = NUPHA,B120,221;%%.

\bibitem{Fradkin:1976xz}
E.~S. Fradkin and M.~A. Vasiliev, ``{Model of Supergravity with Minimal
  Electromagnetic Interaction},''. LEBEDEV-76-197.

\bibitem{Kovtun:2005ev}
P.~K. Kovtun and A.~O. Starinets, ``{Quasinormal modes and holography},''
  \href{http://dx.doi.org/10.1103/PhysRevD.72.086009}{{\em Phys. Rev.}
  {\bfseries D72} (2005) 086009},
\href{http://arxiv.org/abs/hep-th/0506184}{{\ttfamily arXiv:hep-th/0506184}}.
%%CITATION = HEP-TH/0506184;%%.

\bibitem{Leaver:1990zz}
E.~W. Leaver, ``{Quasinormal modes of Reissner-Nordstrom black holes},''
\href{http://dx.doi.org/10.1103/PhysRevD.41.2986}{{\em Phys. Rev.} {\bfseries
  D41} (1990) 2986--2997}.
%%CITATION = PHRVA,D41,2986;%%.

\bibitem{Denef:2009kn}
F.~Denef, S.~A. Hartnoll, and S.~Sachdev, ``{Black hole determinants and
  quasinormal modes},''
  \href{http://dx.doi.org/10.1088/0264-9381/27/12/125001}{{\em Class. Quant.
  Grav.} {\bfseries 27} (2010) 125001},
\href{http://arxiv.org/abs/0908.2657}{{\ttfamily arXiv:0908.2657 [hep-th]}}.
%%CITATION = 0908.2657;%%.

\bibitem{Brattan:2010pq}
D.~K. Brattan and S.~A. Gentle, ``{Shear channel correlators from hot charged
  black holes},'' \href{http://dx.doi.org/10.1007/JHEP04(2011)082}{{\em JHEP}
  {\bfseries 04} (2011) 082},
\href{http://arxiv.org/abs/1012.1280}{{\ttfamily arXiv:1012.1280 [hep-th]}}.
%%CITATION = 1012.1280;%%.

\bibitem{Gauntlett:2009zw}
J.~P. Gauntlett, S.~Kim, O.~Varela, and D.~Waldram, ``{Consistent
  Supersymmetric Kaluza--Klein Truncations with Massive Modes},''
  \href{http://dx.doi.org/10.1088/1126-6708/2009/04/102}{{\em JHEP} {\bfseries
  04} (2009) 102},
\href{http://arxiv.org/abs/0901.0676}{{\ttfamily arXiv:0901.0676 [hep-th]}}.
%%CITATION = 0901.0676;%%.

\bibitem{Gauntlett:2009dn}
J.~P. Gauntlett, J.~Sonner, and T.~Wiseman, ``{Holographic superconductivity in
  M-Theory},'' \href{http://dx.doi.org/10.1103/PhysRevLett.103.151601}{{\em
  Phys. Rev. Lett.} {\bfseries 103} (2009) 151601},
\href{http://arxiv.org/abs/0907.3796}{{\ttfamily arXiv:0907.3796 [hep-th]}}.
%%CITATION = 0907.3796;%%.

\bibitem{Gauntlett:2009bh}
J.~P. Gauntlett, J.~Sonner, and T.~Wiseman, ``{Quantum Criticality and
  Holographic Superconductors in M- theory},''
  \href{http://dx.doi.org/10.1007/JHEP02(2010)060}{{\em JHEP} {\bfseries 02}
  (2010) 060},
\href{http://arxiv.org/abs/0912.0512}{{\ttfamily arXiv:0912.0512 [hep-th]}}.
%%CITATION = 0912.0512;%%.

\bibitem{Grassi:2000dm}
P.~A. Grassi and P.~van Nieuwenhuizen, ``{No van Dam-Veltman-Zakharov
  discontinuity for supergravity in AdS space},''
  \href{http://dx.doi.org/10.1016/S0370-2693(01)00023-5}{{\em Phys. Lett.}
  {\bfseries B499} (2001) 174--178},
\href{http://arxiv.org/abs/hep-th/0011278}{{\ttfamily arXiv:hep-th/0011278}}.
%%CITATION = HEP-TH/0011278;%%.

\end{thebibliography}\endgroup
\bibliographystyle{utphys}

\end{document}